\documentclass[11pt]{article}
\usepackage{amsfonts,amsmath,amssymb}
\usepackage{enumerate}
\usepackage{hyperref}
\usepackage{bbm}
\usepackage{nicefrac}
\usepackage[all]{xy}
\usepackage{graphicx}
\usepackage{bm}
\usepackage{makecell}
\usepackage[table]{xcolor}
\usepackage{longtable}		
\usepackage{subcaption}		
\usepackage{tikz}			

\usepackage{upgreek}

\usepackage{float}			
\usepackage{lscape}                 

\usepackage{booktabs}

\def\sst#1{{\scriptscriptstyle #1}}

\def\0{{\sst{(0)}}}
\def\1{{\sst{(1)}}}
\def\2{{\sst{(2)}}}
\def\3{{\sst{(3)}}}
\def\4{{\sst{(4)}}}
\def\5{{\sst{(5)}}}
\def\6{{\sst{(6)}}}
\def\7{{\sst{(7)}}}
\def\8{{\sst{(8)}}}
\def\9{{\sst{(9)}}}
\def\ten{{\sst{(10)}}}

\newcommand{\cN}{{\cal N}}

\allowdisplaybreaks  

\usepackage{multirow}
\usepackage{rotating}

\allowdisplaybreaks

\newcommand{\lgrav}{\text{LGRAV}}
\newcommand{\lgino}{\text{LGINO}}
\newcommand{\lvec}{\text{LVEC}}
\newcommand{\sgrav}{\text{SGRAV}}
\newcommand{\sgino}{\text{SGINO}}
\newcommand{\svec}{\text{SVEC}}
\newcommand{\mgrav}{\text{MGRAV}}

\newcommand{\mvec}{\text{MVEC}}

\usepackage{lipsum}

\newlength\Colsep
\begin{document}

\makeatletter
\renewcommand{\theequation}{\thesection.\arabic{equation}}
\@addtoreset{equation}{section}
\makeatother

\begin{titlepage}

\begin{flushright}
IFT-UAM/CSIC-21-101\\
LCTP-21-23
%
%
\end{flushright}

\vspace{5pt}

   \begin{center}
   \baselineskip=16pt

   \begin{Large}\textbf{
\hspace{-18pt} 
The spectrum of marginally-deformed ${\cal N} = 2$ CFTs  \\[4pt] 
with AdS$_4$ S-fold duals of type IIB    \\[8pt]
}
   \end{Large}

\vspace{25pt}

{\large  Mattia Ces\`aro$^{1,2}$, Gabriel Larios$^{1,3}$ \,and \,  Oscar Varela$^{1,4}$}
		
\vspace{25pt}

	\begin{small}

	{\it $^{1}$ Departamento de F\'\i sica Te\'orica and Instituto de F\'\i sica Te\'orica UAM/CSIC , \\
   Universidad Aut\'onoma de Madrid, Cantoblanco, 28049 Madrid, Spain}  \\

	\vspace{15pt}
	
	{\it $^{2}$ Department of Applied Science and Technology, Politecnico di Torino,\\
     Corso Duca degli Abruzzi 24, I-10129 Torino, Italy}  \\

	\vspace{15pt}
	
	{\it $^{3}$ Leinweber Center for Theoretical Physics, University of Michigan, Ann Arbor, MI 48109, USA}  \\

	\vspace{15pt}
	
	{\it $^{4}$ Department of Physics, Utah State University, Logan, UT 84322, USA}     \\	
		
	\end{small}

\vskip 50pt

\end{center}

\begin{center}
\textbf{Abstract}
\end{center}

\begin{quote}

A holographic duality was recently established between an ${\cal N} =4$ non-geometric AdS$_4$ solution of type IIB supergravity in the so-called S-fold class, and a three-dimensional conformal field theory (CFT) defined as a limit of ${\cal N} =4$ super-Yang-Mills at an interface. Using gauged supergravity, the ${\cal N} =2$ conformal manifold (CM) of this CFT has been assessed to be two-dimensional. Here, we holographically characterise the large-$N$ operator spectrum of the marginally-deformed CFT. We do this by, firstly, providing the algebraic structure of the complete Kaluza-Klein (KK) spectrum on the associated two-parameter family of AdS$_4$ solutions. And, secondly, by computing the ${\cal N} =2$ supermultiplet dimensions at the first few KK levels on a lattice in the CM, using new exceptional field theory techniques. Our KK analysis also allows us to establish that, at least at large $N$, this ${\cal N} =2$ CM is topologically a non-compact cylindrical Riemann surface bounded on only one side.

\end{quote}

\vfill

\end{titlepage}

\tableofcontents



\section{Introduction} \label{sec:introduction}


In general, determining the spectrum of operators of a strongly interacting quantum field theory is a very hard task, even for holographic conformal field theories (CFTs) with well-known anti-de Sitter (AdS) string theory duals. The specific subclass of AdS/CFT dualities in which the AdS solution arises upon consistent uplift of an AdS vacuum of a lower-dimensional maximal supergravity is very special in this regard. For this type of AdS solutions, new powerful techniques \cite{Malek:2019eaz,Malek:2020yue,Cesaro:2020soq} based on exceptional field theory (ExFT) \cite{Hohm:2013pua,Godazgar:2014nqa} have been introduced to compute the spectrum of Kaluza-Klein (KK) excitations, dual to the spectrum of operators of the corresponding CFT. These new ExFT-based KK spectral techniques have now been employed in a variety of cases \cite{Malek:2020mlk,Varela:2020wty,Guarino:2020flh,Eloy:2020uix,Bobev:2020lsk,Giambrone:2021zvp,Cesaro:2021haf}. 

The specific AdS/CFT instances with an associated maximal gauged supergravity interpretation are few and far between and, for that reason, must be treasured. These include, for example, the well-known cases of \cite{Maldacena:1997re,Aharony:2008ug} or the more recent \cite{Guarino:2015jca}. The proposed holographic duality between the $\cN=4$ AdS$_4$ solution of type IIB supergravity constructed in \cite{Inverso:2016eet} and the three-dimensional CFT described in \cite{Assel:2018vtq} is also of this concrete type. The CFT arises as an  $\cN=4$ infrared fixed point of the $\textrm{T} [ \textrm{U} (N) ] $ field theory of \cite{Gaiotto:2008sd}, enhanced with an adjoint Chern-Simons term at level $k$, and with its $\textrm{U}(N) \times \textrm{U}(N) $ global symmetry gauged with an $\cN=4$ vector multiplet. This field theory can be also thought to arise as a limit of four-dimensional $\cN=4$ super-Yang-Mills at a co-dimension one interface \cite{Assel:2011xz,Assel:2012cj}. In turn, the type IIB dual is non-geometric, of the form $\textrm{AdS}_4 \times S^5 \times S^1$, with non-trivial $ \textrm{SL} (2 , \mathbb{Z} )$ S-duality monodromy on $S^1$, and with $S^5$ and $S^1$ radii related to $N$ and $k$. This AdS$_4$ solution can be thought as a limit of a Janus solution of type IIB \cite{DHoker:2007zhm,DHoker:2007hhe}, compatible with the interface interpretation of the CFT. 

More interestingly for our purposes here, the AdS$_4$ type IIB dual \cite{Inverso:2016eet} also enjoys a maximal gauged supergravity interpretation. Type IIB supergravity admits a consistent truncation on $S^5 \times S^1$ \cite{Inverso:2016eet} down to $D=4$ $\cN=8$ supergravity with dyonic $ [\textrm{SO} (6) \times \textrm{SO} (1,1) ] \ltimes \mathbb{R}^{12}$ gauge group \cite{DallAgata:2011aa,DallAgata:2014tph}. The $D=4$ gauge couplings $g$ and $m \equiv gc$ are related to $N$ and $k$. Due to the consistency of the truncation, the vacua of this gauged supergravity (all of which are AdS, see \cite{DallAgata:2011aa,Gallerati:2014xra,Guarino:2019oct,Guarino:2020gfe,Bobev:2021yya,Guarino:2021hrc} for examples) give rise to (non-geometric) $\textrm{AdS}_4 \times S^5 \times S^1$ solutions of type IIB, with the $S^5$ possibly fibred trivially over the $S^1$. The above $D=4$ $\cN=8$ supergravity has an $\cN=4$, SO(4)-invariant critical point \cite{Gallerati:2014xra} that uplifts to the $\cN=4$ type IIB S-fold solution of \cite{Inverso:2016eet}. The $\cN=8$ gauged supergravity also has a two-parameter family of $\cN=2$ AdS vacua \cite{Bobev:2021yya} continuously connected to the $\cN=4$ point, with the same cosmological constant as the latter. These features led the authors of \cite{Bobev:2021yya} to put forward the interpretation of this family of AdS$_4$ solutions as the holographic realisation of the (necessarily $\cN=2$ \cite{Cordova:2016xhm}) conformal manifold (CM) of the $\cN=4$ CFT of \cite{Assel:2018vtq}. We review the holographic construction of the CM in section \ref{eq:ConfMan}, where we also make new observations about its global properties and establish its non-compactness.

The existence of a maximal gauged supergravity description of the AdS$_4$/CFT$_3$ dualities at hand allows one to apply the ExFT-based KK spectral methods of \cite{Malek:2019eaz,Malek:2020yue,Cesaro:2020soq} in the present case. This is what we set out to do in this paper: we characterise, in section \ref{sec:KKSpectra}, the operator spectrum on the $\cN=2$ CM of the $\cN=4$ CFT$_3$ of \cite{Assel:2018vtq}. We do this by a 
combination of traditional group theory arguments and these new ExFT techniques. The former allow us to obtain the algebraic structure of the complete KK spectrum across the entire CM, while the latter give us access to the explicit calculation of the mass (or equivalently, $\cN=2$ supermultiplet dimension) eigenvalues. We provide closed-form, analytic expressions for the multiplet dimensions of the complete spectrum at specific loci, and for specific multiplets at all points in the CM. Diagonalising analytically the KK mass matrices on the CM in full generality requires extraordinary computer power. Instead, we have resorted to numerics in order to determine the supermultiplets spectrum on a lattice of points in the CM. Rather than cluttering the appendices with endless tables, we include ancillary files containing a database with the first few KK levels of the $\cN=2$ multiplet spectrum on our grid.

A powerful aspect of these new ExFT spectral methods \cite{Malek:2019eaz,Malek:2020yue,Cesaro:2020soq} is that they can be applied to solutions that are only known as vacua of $D=4$ $\cN=8$ gauged supergravity, even if their fully uplifted counterparts are not known --provided, of course, that the latter exist. However, knowledge of the higher-dimensional solutions may still be useful to clarify features of the KK spectra. In the present case, for example, the latter carries information about the topology of the CM. The gauged supergravity is blind to these global features, which therefore must be instilled upon the spectra by the uplifted S-folds themselves. Similar observations have already been made in \cite{Giambrone:2021zvp}. In order to give further evidence of the relation of the global features of the CM and the uplifted S-folds we determine, in section \ref{eq:FamilyIIISfold}, the type IIB uplift of a notable subset of vacua in the class at hand. Section \ref{sec:Concs} concludes and further supplementary information is contained in the appendices.


\section{The holographic conformal manifold} \label{eq:ConfMan}


We are interested in the $\cN=2$ CM of the three-dimensional $\cN=4$ CFT at large-$N$ described in \cite{Assel:2018vtq}. This CM is dual to a certain two-parameter family of $\cN=2$ AdS$_4$ solutions of type IIB supergravity. With some exceptions, these type IIB duals are only known as AdS vacua of $D=4$ $\cN=8$ supergravity with $ [\textrm{SO} (6) \times \textrm{SO} (1,1) ] \ltimes \mathbb{R}^{12}$ gauging --by the consistency of the IIB truncation to the above $D=4$ $\cN=8$ gauging \cite{Inverso:2016eet}, all such vacua are guaranteed to uplift to ten-dimensional solutions. More concretely, the two-parameter family of $\cN=2$ AdS vacua of $D=4$ $\cN=8$ supergravity under discussion was recently obtained in \cite{Bobev:2021yya} building on \cite{Guarino:2020gfe}. Distinct one-parameter subfamilies of this holographic CM were previously constructed in \cite{Guarino:2020gfe} within this $D=4$ $\cN=8$ gauging and in \cite{Arav:2021gra} by other methods. Here we will review the holographic construction of the CM following  \cite{Guarino:2020gfe,Bobev:2021yya} in order to fix our conventions. Our main new observation drawing on the KK analysis of section \ref{sec:KKSpectra} is that the large-$N$ CM is a topological cylinder. 

A convenient subsector of the $D=4$ $\cN=8$ gauged supergravity was constructed in \cite{Guarino:2020gfe} containing seven scalars, $\varphi_i$, and seven pseudoscalars, $\chi_i$, $i= 1 , \ldots , 7$, that parameterise an $\left( \textrm{SL}(2, \mathbb{R}) / \textrm{SO}(2) \right)^7$ submanifold of $\textrm{E}_{7(7)} / \textrm{SU}(8)$. A one-parameter family of $\cN=2$ vacua was identified in \cite{Guarino:2020gfe} (and referred to as Family I in \cite{Bobev:2021yya}) located, in our conventions, at
\begin{eqnarray} \label{eq:LocationFamilyI}
& c^{-1} e^{-\varphi_1} = c^{-1} e^{-\varphi_2} =  e^{-\varphi_6} =e^{-\varphi_7} = \tfrac{1}{\sqrt{2}}  \; , \quad c^{-1} e^{-\varphi_3} = e^{-\varphi_4} = e^{-\varphi_5} = 1  \; , \nonumber \\
&  \chi_1 =   \chi_2 = c \,  \chi   \; , \quad \chi_3 = \chi_4 = \chi_5 =  0 \; , \quad \chi_6 = -\chi_ 7  =  \tfrac{1}{\sqrt{2}}   \; .
\end{eqnarray}
The free parameter here is the pseudoscalar $\chi$. A second one-parameter family of $\cN=2$ vacua was found in \cite{Bobev:2021yya}, where it was named Family II. This occurs at the locus 
\begin{eqnarray} \label{eq:LocationFamilyII}
& \varphi_1 = \varphi_2 = \varphi  \; , \quad e^{-\varphi_3} = c \; , \quad  e^{-\varphi_6} =e^{-\varphi_7} = \tfrac{1}{\sqrt{2}}   \; , \quad  e^{-\varphi_4} =e^{-\varphi_5} = \tfrac{c}{\sqrt{2}} \, e^{\varphi} \;,   \nonumber \\
&  \chi_1 =   \chi_2 =  \chi_3 = 0 \; , \quad \chi_6 = -\chi_ 7  =  \tfrac{1}{\sqrt{2}}   \; , \quad   \chi_4^2 = \chi_5^2 =  1- \tfrac12 \, c^2 \,  e^{2\varphi}  \; , 
\end{eqnarray}
parameterised by the scalar $\varphi$, and contains the $\cN=4$ point at $\varphi = \chi =0$. In (\ref{eq:LocationFamilyI}), (\ref{eq:LocationFamilyII}), $c=m/g\neq0$, with $g$ and $m$ the electric and magnetic gauge couplings of the parent $D=4$ $\cN=8$ supergravity. We henceforth set $c=1$ without loss of generality. A series of dualities can be performed on the $\textrm{E}_{7(7)} / \textrm{SU}(8)$ coset representative corresponding to the vacua (\ref{eq:LocationFamilyI}), (\ref{eq:LocationFamilyII}), in order to generate a larger set of vacua with both parameters $(\varphi , \chi)$ turned on  \cite{Bobev:2021yya}. This larger family is still $\cN=2$ (with supersymmetry enhancement at the $\cN=4$ point) and generically lies outside the $\left( \textrm{SL}(2, \mathbb{R}) / \textrm{SO}(2) \right)^7$ submanifold of \cite{Guarino:2020gfe}. Our parameters are related to those in \cite{Bobev:2021yya} as $\chi_{\textrm{here}} = \chi_{\textrm{there}}$ and $e^{-2\varphi_{\textrm{here}}} = \tfrac{1}{2} ( 1 +   \varphi^2_{\textrm{there}} )$.

This local family of AdS vacua parameterised by $( \varphi , \chi ) $ was proposed in \cite{Bobev:2021yya} as the holographic CM of the $\cN=4$ CFT of \cite{Assel:2018vtq} at large $N$. When restricted to this two-dimensional surface, the $\cN=8$ non-linear sigma model on $\textrm{E}_{7(7)} / \textrm{SU}(8)$ gives rise to the leading contribution to the Zamolodchikov metric on the CM \cite{Bobev:2021yya}. This metric is K\"ahler and reads, with our parameterisation,
\begin{equation}	\label{eq:cmmetric}
	ds^2=(4-e^{2\varphi})\Big[(2-e^{2\varphi})^{-1}\,d\varphi^2+d\chi^2\Big]\, .
\end{equation}
The corresponding Riemann tensor and Ricci scalar are 
\begin{equation}	\label{eq:confmanRicci}
R_{mnpq}=- R \, g_{m[p}g_{q]n} \; , \qquad
	R=\frac{2\, e^{2 \varphi} \left(e^{4 \varphi}-12\, e^{2\varphi}+16\right)}{(4-e^{2\varphi})^3} \; .
\end{equation}

The local $D=4$ $\cN=8$ supergravity scalars originally range on the entire real line, but we find the CM construction to be only well defined if the parameters are restricted as:
\begin{equation}	\label{eq: cmband}
	0 < e^{2\varphi} \leq 2 \quad , \qquad
	\textrm{$0 \leq \chi   <  \tfrac{2\pi}{T}$, and periodic: $\chi \sim \chi + \tfrac{2\pi}{T}$ } \; .
\end{equation}
Here, $T$ is the inverse radius of the $S^1$ factor of the associated type IIB S-fold solutions (and is related to the Chern-Simons level $k$ of the dual CFT \cite{Assel:2018vtq,Inverso:2016eet}).
Within the intervals (\ref{eq: cmband}), both the metric (\ref{eq:cmmetric}) and the curvature (\ref{eq:confmanRicci}) are smooth and finite. The Ricci scalar is in fact bounded, $-2 \leq R \leq \tfrac{10}{27}$, and the Riemann tensor vanishes at $e^{2\varphi}=2 (3 - \sqrt{5})$ and in the limit $e^{2\varphi} \rightarrow 0$,   with $\chi$ arbitrary within its allowed interval. 

The range of $\varphi$ specified in (\ref{eq: cmband}) must be enforced already at the gauged supergravity level, so that the solution (\ref{eq:LocationFamilyII}) (with $c=1$) is well defined and singularity-free. This is further confirmed by the KK analysis of section \ref{sec:KKSpectra}: only within the range (\ref{eq: cmband}) for $\varphi$ are the KK spectra on the CM free from tachyonic modes, as required by supersymmetry. The periodicity in $\chi$ cannot be seen at the $D=4$ gauged supergravity level: it is an intrinsically higher-dimensional feature of the corresponding type IIB S-folds, see section~\ref{eq:FamilyIIISfold}. The periodicity of $\chi$ is already present in the KK spectra, as discussed in \cite{Giambrone:2021zvp} and section~\ref{sec:KKSpectra} below. Thus, the large-$N$ CM has one closed one-dimensional boundary, corresponding to the circumference parameterised by $\chi$ located at the higher endpoint of the $\varphi$ range in (\ref{eq: cmband}). This upper boundary corresponds to Family I, (\ref{eq:LocationFamilyI}), of \cite{Guarino:2020gfe,Bobev:2021yya}:
\begin{equation} \label{eq:FamIandIII}
\textrm{Family I (upper boundary)} : \quad  e^{2\varphi} = 2 \; , \quad  \textrm{$0 \leq \chi   <  \tfrac{2\pi}{T}$, and periodic: $\chi \sim \chi + \tfrac{2\pi}{T}$. }
\end{equation}
The circumference parameterised by $\chi$ located at the lower endpoint of the $\varphi$ range in (\ref{eq: cmband}) is only reached asymptotically as $e^{2\varphi} \rightarrow 0$. The lower inequality in (\ref{eq: cmband}) is strict, and the singular locus $e^{2\varphi} = 0$ does not belong to the CM. The large-$N$ CM is thus non-compact, with infinite volume w.r.t.~the leading contribution (\ref{eq:cmmetric}) to the Zamolodchikov metric.

The CM is generically $\cN=2$ and $\textrm{U}(1)_F \times \textrm{U}(1)_R$--invariant, except at the locations specified below. Here, $\textrm{U}(1)_F \times \textrm{U}(1)_R$ is the subgroup of $\textrm{SO}(6) \sim \textrm{SU}(4)$ (the isometry of the internal $S^5$ in type IIB, or the R-symmetry of the parent dual $\cN=4$ super-Yang-Mills) defined by
\begin{equation} \label{eq:Embedding1}
\textrm{SO}(6) \sim \textrm{SU}(4) \; \supset \; \textrm{SO}(4) \sim \textrm{SU}(2)_1 \times \textrm{SU}(2)_2 \; \supset \;
\textrm{U}(1)_1 \times \textrm{U}(1)_2 \; , 
\end{equation}
with SO(4) the real subgroup of SU(4), $\textrm{SU}(2)_i   \supset \textrm{U}(1)_i$, $i=1,2$, and $\textrm{U}(1)_R$ and $\textrm{U}(1)_F$ respectively corresponding to the diagonal and antidiagonal combinations of $\textrm{U}(1)_1$ and $\textrm{U}(1)_2$. Alternatively, $\textrm{U}(1)_F \times \textrm{U}(1)_R$ is equivalently defined through 
\begin{equation} \label{eq:Embedding2}
\textrm{SO}(6) \sim \textrm{SU}(4) \; \supset \; \textrm{SU}(3) \times \textrm{U}(1)_b\; \supset \; \textrm{SU}(2)_F \times \textrm{U}(1)_a \times \textrm{U}(1)_b  \; \supset \;
\textrm{U}(1)_F \times \textrm{U}(1)_R \; , 
\end{equation}
with $\bm{3} \rightarrow \bm{2}$ under $\textrm{SU}(3)   \supset \textrm{SU}(2)_F$; then, $\textrm{SU}(2)_F   \supset \textrm{U}(1)_F$ and $\textrm{U}(1)_a \times \textrm{U}(1)_b \supset \textrm{U}(1)_R$, so that if $p$, $q$, $y_0$ are $\textrm{U}(1)_a$, $\textrm{U}(1)_b$, $\textrm{U}(1)_R$ charges, then $ y_0 = \tfrac13(p - q)$. The SO(6) in (\ref{eq:Embedding1}) and (\ref{eq:Embedding2}) is in turn embedded inside the SU(8) compact subgroup of the $\cN=8$ supergravity scalar manifold as the SO$(6)_v$ subgroup of the real subgroup SO(8) of SU(8) \cite{Giambrone:2021zvp,Guarino:2021kyp}. The commutant of SO$(6)_v$ inside SO(8) will be denoted below as SO(2): this is the group that rotates the internal $S^1$ in the type IIB geometry. The labels $F$ and $R$ in the U(1) and SU(2) groups above refer to the flavour and R-symmetry of the dual CFTs. An $R$ label could also be added to the SO(4) in (\ref{eq:Embedding1}), but is omitted for notational simplicity. Note, for later reference, that the embeddings (\ref{eq:Embedding1}), (\ref{eq:Embedding2}) are globally defined and independent of the $D=4$ supergravity scalars.

\begin{figure}
	\begin{subfigure}{0.45\textwidth}
		\centering
		\includegraphics[width=1.0\linewidth]{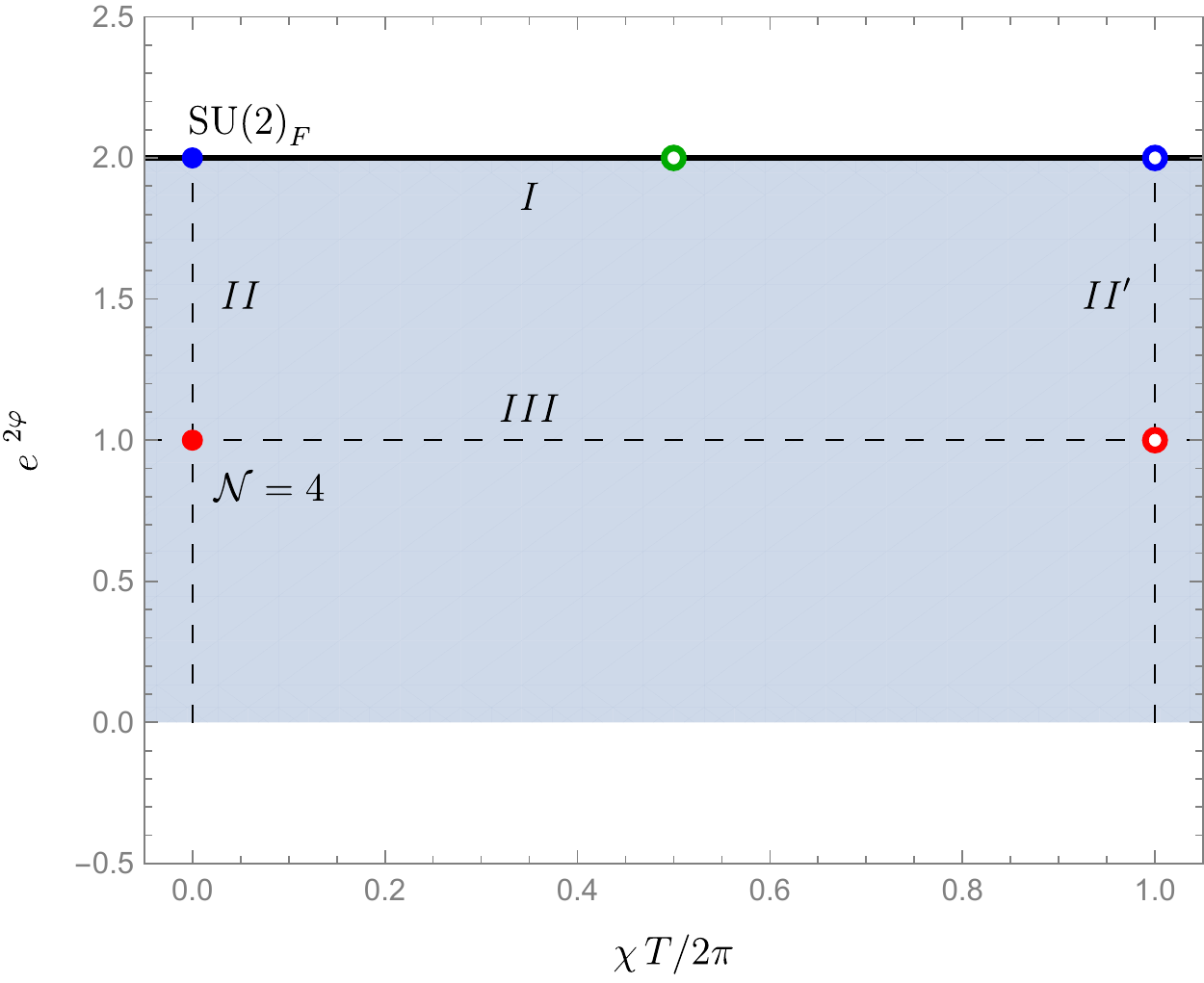}
	\end{subfigure}%
	\qquad
	\begin{subfigure}{0.45\textwidth}
		\centering
		\includegraphics[width=1.0\linewidth]{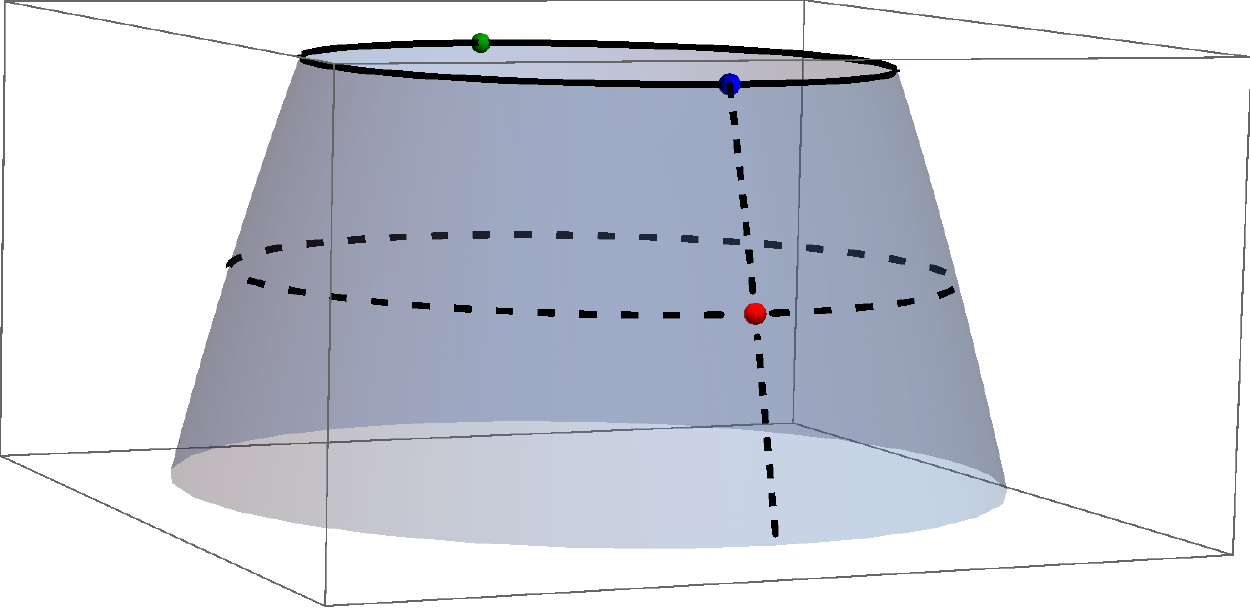}
	\end{subfigure}
	\caption{\footnotesize{The large-$N$ holographic CM. The left plot indicates the location of the upper boundary (solid black line), Family I (\ref{eq:FamIandIII}), along with the (super)symmetry enhanced points: (\ref{eq:N=4SO4point}) in solid ($n^\prime =0$) and hollow ($n^\prime =1$) red, (\ref{eq:N=2SU2xU1point1}) in solid ($n^\prime =0$) and hollow ($n^\prime =1$) blue, and (\ref{eq:N=2SU2xU1point2}) in hollow green ($n^\prime =1$). The dashed lines marked as II and II$^\prime$ correspond to Family II, (\ref{eq:FamII}), at $n^\prime = 0$ and $n^\prime = 1$, respectively. These two lines are identified per the periodicity (\ref{eq: cmband}) of $\chi$, rendering the topological cylinder on the right plot. Family III, (\ref{eq:FamIII}), in the interior is also indicated. The locus $e^{2\varphi} = 0$ lies outside the CM and is at infinite distance of the upper boundary w.r.t.~the metric (\ref{eq:cmmetric}).}  	
	\label{fig:conformalmanif} }
\end{figure}

The CM exhibits symmetry or supersymmetry enhancements at specific points. The $\cN=4$ SO(4)-invariant vacuum \cite{Gallerati:2014xra} of the $\cN=8$ supergravity, which uplifts to the AdS$_4 \times S^5 \times S^1$ type IIB S-fold solution of \cite{Inverso:2016eet} with the CFT dual of \cite{Assel:2018vtq} is attained in our parameterisation at
\begin{equation} \label{eq:N=4SO4point}
\textrm{$\cN=4$ \; SO(4) point} \; : \quad \varphi = 0 \; ,  \qquad \chi = \tfrac{2\pi}{T} n^\prime \; , \; n^\prime = 0 , \pm 1,  \pm 2 , \ldots  
\end{equation}
Strictly speaking, only the $n^\prime = 0$ (super)symmetry enhancement to $\cN=4$ SO(4) can be seen at the gauged supergravity level: the periodicity for $|n^\prime | \geq 1$ will be shown in sections \ref{sec:KKSpectra} and \ref{eq:FamilyIIISfold}. The SO(4) symmetry group in (\ref{eq:N=4SO4point}) is the same that appears in the branching (\ref{eq:Embedding1}). All other points in the CM are $\cN=2$, with generic $\textrm{U}(1)_F \times \textrm{U}(1)_R$ symmetry. The latter is enhanced to the $\textrm{SU}(2)_F \times \textrm{U}(1)_R$ defined in (\ref{eq:Embedding2}) at two specific locations up to periodicity. The first such symmetry enhancement occurs, in our parameterisation, for
\begin{equation} \label{eq:N=2SU2xU1point1}
\textrm{$\cN=2$ \; $\textrm{SU}(2)_F \times \textrm{U}(1)_R$   point 1} \; : \quad e^{2\varphi} = 2 \; ,  \qquad \chi = \tfrac{2\pi}{T} n^\prime \; , \; n^\prime = 0 , \pm 1, \pm 2 , \ldots 
\end{equation}
Again, only the $n^\prime = 0$ realisation is visible in gauged supergravity, and corresponds to the $\cN=2$ $\textrm{SU}(2) \times \textrm{U}(1)$ critical point found in \cite{Guarino:2020gfe}; the symmetry enhancement for $|n^\prime | \geq 1$ can only be seen with a KK analysis \cite{Giambrone:2021zvp}. The second such enhancement occurs at 
\begin{equation} \label{eq:N=2SU2xU1point2}
\textrm{$\cN=2$ \; $\textrm{SU}(2)_F \times \textrm{U}(1)_R$   point 2} \; : \quad e^{2\varphi} = 2 \; ,  \qquad \chi = \tfrac{\pi}{T} n^\prime \; , \; n^\prime = \pm 1, \pm 3 , \ldots 
\end{equation}
and has no counterpart in gauged supergravity \cite{Giambrone:2021zvp} (note the different ranges of $n^\prime$ in (\ref{eq:N=2SU2xU1point2}) and (\ref{eq:N=2SU2xU1point1})). Of course, the generic $\textrm{U}(1)_F \times \textrm{U}(1)_R$ symmetry group of the CM is a subgroup of both enhanced symmetry groups SO(4) and $\textrm{SU}(2)_F \times \textrm{U}(1)_R$ as indicated in (\ref{eq:Embedding1}) and (\ref{eq:Embedding2}), but the latter  $\textrm{SU}(2)_F \times \textrm{U}(1)_R$ is not a subgroup of the former SO(4).

There are no (super)symmetry enhancements across the CM other than (\ref{eq:N=4SO4point}), (\ref{eq:N=2SU2xU1point1}) and (\ref{eq:N=2SU2xU1point2}). The upper boundary, Family I (\ref{eq:FamIandIII}), is a geodesic of the metric (\ref{eq:cmmetric}). A couple of other notable loci within the CM are the following one-parameter families. The following locus parameterised by $\varphi$,
\begin{equation} \label{eq:FamII}
\textrm{Family II}  \; : \; \big( 0 < e^{2\varphi} \leq 2 \; , \;  \chi = \tfrac{2\pi}{T} n^\prime \big)  \; , \; n^\prime = 0 , \pm 1,  \pm2 , \ldots
\end{equation}
was discussed for $n^\prime = 0$ in \cite{Bobev:2021yya} (see also \cite{Arav:2021gra}) and, for this value of $n^\prime$, corresponds to the gauged supergravity solution (\ref{eq:LocationFamilyII}). Within the CM, Family II is the geodesic of the metric (\ref{eq:cmmetric}) that passes through the $\cN=4$ SO(4) point (\ref{eq:N=4SO4point}) and ends at the $\cN=2$ $\textrm{SU}(2)_F \times \textrm{U}(1)_R$ point 1, (\ref{eq:N=2SU2xU1point1}), with zero winding number on the cylindrical CM. Family II also provides a useful way to visualise the global aspects of the CM. If the latter is first represented as a rectangle in $\mathbb{R}^2$ with sides defined by (\ref{eq: cmband}), the cylinder is constructed by identifying the geodesics corresponding to Family II at $\chi=0$ and $\chi=\frac{2\pi}{T}$. Finally, the following circumference, parameterised by $\chi$, in the interior of the CM is also interesting, even though it des not correspond to a geodesic of (\ref{eq:cmmetric}):
\begin{equation} \label{eq:FamIII}
\textrm{Family III} : \quad  e^{2\varphi} = 1 \; , \quad  \textrm{$0 \leq \chi   <  \tfrac{2\pi}{T}$, and periodic: $\chi \sim \chi + \tfrac{2\pi}{T}$. }
\end{equation}
As we will see in section~\ref{sec:FamIIISpecttrum}, the complete KK spectrum on this locus can be given in closed form. See figure \ref{fig:conformalmanif} for a visual summary of the CM.

\section{KK towers on the two-parameter S-fold family} \label{sec:KKSpectra} 


We now turn to discuss the KK spectrum for the two-parameter family of AdS$_4$ solutions reviewed in section \ref{eq:ConfMan}. The spectrum can be labelled by two independent KK levels, $\ell$ and $n$, respectively associated with the internal $S^5$ and $S^1$ of the IIB S-folds. These range as 
\begin{equation} \label{eq:KKlevels}
\ell = 0, 1 , 2 , \ldots \qquad n = 0 , \, \pm 1 ,  \, \pm 2 , \, \ldots 
\end{equation}
At generic $\cN=2$ points in this two-dimensional holographic CM, the KK spectrum organises itself in representations of $\textrm{OSp}(4|2) \times \textrm{U}(1)_F$, with $\textrm{U}(1)_R \subset \textrm{OSp}(4|2)$ and $\textrm{U}(1)_F$ defined by either branching rule (\ref{eq:Embedding1}) or (\ref{eq:Embedding2}). At the $\cN=2$ points (\ref{eq:N=2SU2xU1point1}) and (\ref{eq:N=2SU2xU1point2}) with enhanced flavour symmetry, the KK spectrum lies in representations of $\textrm{OSp}(4|2) \times \textrm{SU}(2)_F$, with the latter factor defined in (\ref{eq:Embedding2}). Finally, at the $\cN=4$ point the KK spectrum is organised in OSp$(4|4)$ multiplets, with R-symmetry given by the SO(4) group defined in (\ref{eq:Embedding1}). Multiplets of these supergroups whose superconformal primary has dimension $E_0$ and $\textrm{U}(1)_R$ or SO(4) R-charges $y_0$ or $(\ell_1 , \ell_2)$ will be labelled as
\begin{eqnarray} \label{eq:OSpMultiplets}
\textrm{OSp}(4|2) \times \textrm{U}(1)_F & : &  \text{MULT}_2 \left[E_0,\,y_0;\;f\right] \; , \nonumber \\
\textrm{OSp}(4|2) \times \textrm{SU}(2)_F & : & \text{MULT}_2\left[E_0,\,y_0\right]\otimes [k] \; ,  \\
\textrm{OSp}(4|4) & : &  \text{MULT}_4 \left[E_0,\,\ell_1 , \ell_2  \right] \; , \nonumber
\end{eqnarray}
with $f$ and $k$ the additional $\textrm{U}(1)_F$ charge and $\textrm{SU}(2)_F$ (half-integer) spin, common to all states in a given OSp$(4|2)$ multiplet $\text{MULT}_2$. The subindices in $\text{MULT}_2$ and $\text{MULT}_4$ are used to distinguish $\cN=2$ and $\cN=4$ multiplets. For the former, we follow the notation and conventions of appendix A of \cite{Klebanov:2008vq}. See also that reference for their state contents. For the $\cN=4$ multiplets, we record some relevant aspects in appendix \ref{OSp44SuperMult}. See also appendix \ref{sec:KKmaterials} for more details on the setup and calculations behind the results reported in this section. Previous results on the spectra of these solutions may be found in \cite{Gallerati:2014xra,Dimmitt:2019qla,Guarino:2020gfe,Giambrone:2021zvp,Bobev:2021yya}.

\subsection{Algebraic structure of the complete spectrum} \label{sec:N=4Specttrum}

It is useful to start our discussion of the KK spectrum on the holographic $\cN=2$ CM under consideration by reviewing the spectrum at the parent $\cN=4$ point first. The reason is that the algebraic structure of the complete spectrum at all points in the CM, including the protected spectrum, is inherited from that at the $\cN=4$ point. The KK spectrum at this point was given for lowest KK levels $\ell = n = 0$ in \cite{Gallerati:2014xra} and was extended to all higher levels in \cite{Giambrone:2021zvp} (see also \cite{Dimmitt:2019qla} for previous partial results).

At fixed $\textrm{SO}(6)_v \times \textrm{SO}(2)$ KK levels $(\ell , n)$ ranging as in (\ref{eq:KKlevels}), the KK spectrum at the $\cN=4$ point is composed of a number\footnote{In (\ref{eq:NoMultN=4}), $H(x)$ is the Heaviside step function, with $H(x) = 0$ for $x\leq 0$ and $H(x) = 1$ for $x >0$. Also, $[ \ldots ]$ here and in (\ref{eq:GravBranching}) denotes integer part.},
\begin{equation} \label{eq:NoMultN=4}
\big( 1 + H ( |n| )  \big) \, \big( \ell + 1 -  \big[ \tfrac{\ell}{2} \big] \big) \big( 1 +  \big[ \tfrac{\ell}{2} \big] \big)   \; , 
\end{equation}
of OSp$(4|4)$ long graviton multiplets
\begin{equation} \label{eq:LGRAV4}
\lgrav_4\big[E_0,\ell_1,\ell_2\big] \; ,
\end{equation} 
whose scalar superconformal primaries have SO(4) Dynkin labels and dimensions specified as follows. The Dynkin labels correspond to all possible pairs $(\ell_1 , \ell_2)$ that appear on the r.h.s.~of the following branching under the first inclusion in the chain (\ref{eq:Embedding1}), namely,
\begin{equation} \label{eq:GravBranching}
	[0,\ell,0]  \; \rightarrow \; \bigoplus_{a=0}^{[\ell/2]} \bigoplus_{k=0}^{\ell-2a}(\ell-2a-k,k) \; .
\end{equation}	
At fixed $\ell$, each of these $\big( \ell + 1 -  \big[ \tfrac{\ell}{2} \big] \big) \big( 1 +  \big[ \tfrac{\ell}{2} \big] \big)$ pairs of integers $(\ell_1 , \ell_2)$ defines a multiplet (\ref{eq:LGRAV4}) present in the spectrum if $n=0$, or two if  $n \neq 0$, corresponding to the two signs\footnote{\label{fn:nDep} This is the only effect of the $S^1$ KK level $n$ in the algebraic structure of the $\cN=4$ spectrum, and the origin of the factor $\big( 1 + H ( |n| )  \big)$ in (\ref{eq:NoMultN=4}). The spectrum, though, does not come in $\textrm{OSp}(4|4) \times \textrm{SO}(2)$ representations  (\ref{eq:LGRAV4}) with definite SO(2) charge $2n$, because different states in a given $\textrm{OSp}(4|4)$ multiplet carry different charges under the (broken) SO(2), see {\it e.g.}~table~\ref{tab:so6xso2KKirreps} in appendix~\ref{sec:KKmaterials}. The $S^1$ level $n$ also affects the spectrum through the dimensions $E_0$, see (\ref{eq:GravDimensions}), with degeneracy for both signs of $n$ at all other quantum numbers held equal. On the rest of the CM, similar remarks apply about the dependence of the algebraic structure of the multiplet spectrum with $n$. The dimensions also acquire an $n$ dependence, and the sign degeneracy is lifted for flavoured multiplets.} of $n$. The conformal dimension for each of these depends on the KK levels $\ell$, $n$ and on the SO(4) Dynkin labels $\ell_1$, $\ell_2$, restricted as in (\ref{eq:GravBranching}), through the formula
\begin{equation} \label{eq:GravDimensions}
E_0 = -\tfrac12 + \sqrt{ \tfrac94  + \tfrac12 \ell (\ell +4 ) +\ell_1 (\ell_1 +1 ) +\ell_2 (\ell_2 +1 ) +\tfrac12 \big( \tfrac{2\pi n}{T} \big)^2 } \; . 
\end{equation}
The l.h.s.~in (\ref{eq:GravBranching}) corresponds to the $\textrm{SO}(6)_v $ representations of the putative graviton states discussed in appendix \ref{sec:PutativeSO6SO2}. The dimensions (\ref{eq:GravDimensions}), computed in \cite{Giambrone:2021zvp} using ExFT methods, agree with those that follow from the individual KK graviton masses found in (3.9) of \cite{Dimmitt:2019qla} (with $n_\textrm{here} = j_\textrm{there}$). See appendix  \ref{OSp44SuperMult} for the state content of the $\cN=4$ multiplets (\ref{eq:LGRAV4}).

For specific values of the quantum numbers some of the multiplets (\ref{eq:LGRAV4}) in the spectrum become short, and split into a $\sgrav_4$ (or $\mgrav_4$ for $\ell = 0$) and a $\sgino_4$ via \eqref{eq: recombN4}. Specifically, this happens for \cite{Dimmitt:2019qla}
\begin{equation} \label{eq:ShorteningQNs}
n=0 \; , \qquad \ell_1 = \ell_2 = \tfrac12 \ell  \; , \; \textrm{with $\ell$ even} \; ,
\end{equation}
a combination of $\ell_1$, $\ell_2$ and $\ell$ allowed by (\ref{eq:GravBranching}). Indeed, when (\ref{eq:ShorteningQNs}) holds, the dimension (\ref{eq:GravDimensions}) saturates the $\cN=4$ unitarity bound, (\ref{eq: unitaritybound}) with $s_0 = 0$.

The algebraic structure of the complete KK spectrum across the entire CM turns out to be determined by the spectrum at the $\cN=4$ point, through the branching (\ref{eq: N4toN2branching}) of the multiplets (\ref{eq:LGRAV4}) under
\begin{equation} \label{eq:OSp44OSp22U(1)F}
\textrm{OSp}(4|4) \supset \textrm{OSp}(4|2) \times \textrm{U}(1)_F \; .
\end{equation}
More concretely, at fixed $\ell$ and $n$, the spectrum  at an arbitrary point $( \varphi , \chi)$ in the CM contains $\big( 1 + H ( |n| )  \big)$ contributions of the form
{\setlength\arraycolsep{0pt}
\begin{eqnarray}	\label{eq:KKFamIIIellneq0nneq0General}
	 &&\bigoplus_{m_1=-\ell_1}^{\ell_1} \bigoplus_{m_2=-\ell_2}^{\ell_2}
	\Big\{
	 \lgrav_2\big[E_{m_1m_2}^\1  ,y_{m_1m_2}; f_{m_1 m_2} \big] \nonumber\\
     &\oplus &\lgino_2\big[ E_{m_1m_2}^\2 , y_{m_1m_2} ; f_{m_1 m_2} +1 \big] 
     \oplus \lgino_2\big[ E_{m_1m_2}^\3 , y_{m_1m_2} ; f_{m_1 m_2}  -1\big] \nonumber \\
&\oplus&\lgino_2\big[ E_{m_1m_2}^\4 , y_{m_1m_2} ; f_{m_1 m_2} +1\big] 
\oplus \lgino_2\big[ E_{m_1m_2}^\5 , y_{m_1m_2} ; f_{m_1 m_2}  -1 \big] \nonumber \\
	&\oplus& 		\lvec_2\big[ E_{m_1m_2}^\6 ,y_{m_1m_2}; f_{m_1 m_2} \big]  \nonumber \\
	&\oplus& \lvec_2\big[ E_{m_1m_2}^\7, y_{m_1m_2} ; f_{m_1 m_2} +2 \big]
	\oplus \lvec_2\big[ E_{m_1m_2}^\8 , y_{m_1m_2} ; f_{m_1 m_2}  \big]\nonumber\\[5pt]
	&\oplus &\lvec_2\big[ E_{m_1m_2}^\9 ,y_{m_1m_2} ; f_{m_1 m_2} -2 \big]
	\oplus \lvec_2\big[ E_{m_1m_2}^{\ten} , y_{m_1m_2} ; f_{m_1 m_2} \big]
  \Big\} \; ,
\end{eqnarray}
for each of the $\big( \ell + 1 -  \big[ \tfrac{\ell}{2} \big] \big) \big( 1 +  \big[ \tfrac{\ell}{2} \big] \big)$ pairs of integers $(\ell_1 , \ell_2)$ defined by the r.h.s.~of (\ref{eq:GravBranching}). 
All of the multiplets in (\ref{eq:KKFamIIIellneq0nneq0General}) are typically long. In (\ref{eq:OSp44OSp22U(1)F}), the $\cN=2$ $\textrm{U}(1)_R \subset \textrm{OSp}(4|2)$ R-symmetry and the $\textrm{U}(1)_F$ flavour symmetry are embedded into the $\cN=4$ $\textrm{SO}(4) \subset \textrm{OSp}(4|4)$ R-symmetry as indicated in (\ref{eq:Embedding1}) and below that equation. As remarked in section \ref{eq:ConfMan}, the group embedding (\ref{eq:Embedding1}) (and also (\ref{eq:Embedding2})) is independent of the $D=4$ supergravity scalars. For this reason, the R- and flavour charges of the $\cN=2$ multiplets in the spectrum do not depend on the position on the CM. Indeed, the quantities $y_{m_1m_2}$ and $f_{m_1m_2}$ in  (\ref{eq:KKFamIIIellneq0nneq0General}) that govern these charges are simply given, in our conventions, by the integers
\begin{equation} \label{eqRFCharges}
y_{m_1m_2} = m_1 + m_2 \; , \qquad f_{m_1 m_2} = m_1 - m_2 \; .
\end{equation}
The dimensions $E_{m_1m_2}^\1$, etc., in (\ref{eq:KKFamIIIellneq0nneq0General}) do depend on the moduli 
$(\varphi , \chi)$ and, except for Family III, do not follow in any obvious way from the $\cN=4$ dimensions (\ref{eq:GravDimensions}). It is for the calculation of these dimensions that we have resorted to ExFT spectral techniques \cite{Malek:2019eaz,Malek:2020yue,Cesaro:2020soq}. It is difficult to establish in general the analytical functional dependence of the dimensions either on the modulus $\varphi$ or on the quantum numbers $\ell$, $\ell_1$, $\ell_2$ (or possibly others). However, the dependence on $\chi$ of the dimension of a multiplet with flavour charge $f$ arising at $S^1$ KK level $n$ is locked into the combination
\begin{equation} \label{eq:chiComb}
\big( \tfrac{2\pi n}{T}+ f \chi \big)^2 \; ,
\end{equation}
across the entire CM. This combination was noted in \cite{Giambrone:2021zvp} to hold for the KK spectrum on Family I. Here, we extend this behaviour to the spectrum on all other points in the CM.

There are two immediate consequences of the $\chi$-dependence (\ref{eq:chiComb}) of the multiplet dimensions. Firstly, a multiplet in the spectrum is flavour neutral if and only if its dimension is independent of the modulus $\chi$. Secondly, the dependence (\ref{eq:chiComb}) establishes the periodic behaviour of the multiplet dimensions in $\chi$ advertised in (\ref{eq: cmband}). Indeed, for all fixed $S^5$ KK level $\ell$, the dimension of any given multiplet with flavour $f$, evaluated at $\chi = \chi_0$ and $S^1$ KK level $n$, coincides with the dimension of the same multiplet evaluated at $\chi = \chi_0 + \frac{2\pi}{T}$ and $S^1$ level $n^\prime$, with 
\begin{equation} \label{eq:KKreshuffle}
n^\prime = n - f \; .
\end{equation}
Such integer $n^\prime$ always exists given $n$ and $f$ because, as (\ref{eq:KKFamIIIellneq0nneq0General}), (\ref{eqRFCharges}) show, the flavour charges are also integer (in our conventions). As noted in footnote \ref{fn:nDep}, only the dimensions, but not the multiplet content (\ref{eq:KKFamIIIellneq0nneq0General}) itself, depend on $n$. For this reason, the entire contribution (\ref{eq:KKFamIIIellneq0nneq0General}) to the spectrum at KK level $\ell$ goes back to itself as $\chi$ ranges from $0$ to $2\pi/T$. Only the $S^1$ KK level needs to be readjusted as $\chi$ reaches each endpoint of its cycle. As remarked in \cite{Giambrone:2021zvp}, this mixture of KK levels is reminiscent of the `space invaders scenario' described in \cite{Duff:1986hr} (see also \cite{Cesaro:2020piw} for a more recent instance of this phenomenon). The periodicity of $\chi$ cannot be seen in $D=4$ gauged supergravity, which has fixed $n=0$. Thus, this effect must be due to an intrinsic feature of the fully-fledged type IIB uplifted solutions that is captured by their KK spectrum. Evidence that this is the case is given in \cite{Giambrone:2021zvp} and in section \ref{eq:FamilyIIISfold} below from the type IIB  uplifts of specific submanifolds of the CM.

\begin{table}[]
\begin{center}
\begin{tabular}{lcl} 					 \hline
$\sgrav_2\big[\ell+2  , \, \pm \ell ; \, 0 \big]$ 	&  &  	$\svec_2\big[\ell+1  , \,  \pm \ell ; \, 0 \big]$       		\\[5pt] 
$\sgino_2\big[\ell+\tfrac52  , \, \pm (\ell + 1 ); \,  0 \big]$ 	&  &  	$\textrm{HYP}_2 \big[\ell+2  , \, \pm ( \ell + 2 ) ; \, 0 \big]$   
		\\[2pt]  \hline
\end{tabular}
\caption{\footnotesize{The protected (short, moduli independent) OSp$(4|2)$ spectrum on the CM at KK levels $n=0$ and $\ell \geq 0$ even. At $\ell = 0$, there is only one graviton and one vector multiplets, both of them massless.
}\normalsize}
\label{tab:ShortN=2Spectrum}
\end{center}
\end{table}

At particular points in the CM and for specific choices of quantum numbers, the dimension of some of the multiplets in (\ref{eq:KKFamIIIellneq0nneq0General}) might saturate the corresponding $\cN=2$ unitarity bounds. In those cases, these long multiplets may be formally written in terms of short $\cN=2$ multiplets. In general, though, these accidental saturations will not lead to multiplet protection: the dimensions will typically remain moduli dependent and the short multiplets will tend to recombine into long ones. For the concrete choice of quantum numbers
\begin{equation} \label{eq:ShorteningQNsCM}
n=0 \; , \qquad |m_1 + m_2| = 2\ell_1 = 2\ell_2 = \ell  \; , \; \textrm{with $\ell$ even} \; ,
\end{equation}
which encompasses the $\cN=4$ shortening condition (\ref{eq:ShorteningQNs}), some of the multiplet dimensions in (\ref{eq:KKFamIIIellneq0nneq0General}) both saturate the $\cN=2$ unitarity bound and become moduli independent. This series, labelled by even $\ell$, is protected in the sense that the multiplet dimensions are independent of the moduli. The series includes, at $\ell=0$, a $\mgrav_2$ and a $\mvec_2$, respectively dual to the energy-momentum tensor and the $\textrm{U}(1)_F$ flavour current of the CFT, as well as two $\sgino_2$'s and two $\textrm{HYP}_2$'s. The latter contain the two real moduli on the CM, dual to a superpotential deformation \cite{Bobev:2021yya}. For each $\ell = 2 ,4 , \ldots$, the protected series includes two of each of the possible short multiplets of OSp$(4|2)$, with $\ell$-dependent opposite R-charges. By (\ref{eq:chiComb}), all the protected multiplets are U$(1)_F$ flavour neutral (the converse is not true, though). See table \ref{tab:ShortN=2Spectrum} for a summary.

\subsection{Spectrum on the upper boundary} \label{sec:FamISpecttrum}

The complete KK spectrum on the upper boundary, Family I (\ref{eq:FamIandIII}), of the CM has already been determined in \cite{Giambrone:2021zvp} for all KK levels $\ell$ and $n$ (see also \cite{Guarino:2020gfe} for the $\ell = n = 0$ spectrum). Our presentation will therefore be brief.

The main new observation is that the KK spectrum on this locus follows the pattern laid out in section \ref{sec:N=4Specttrum}, which is valid across the CM on general grounds. At fixed $\ell$ and for all $n$, the spectrum of $\textrm{OSp}(4|2) \times \textrm{U}(1)_F$ multiplets on the upper boundary of the CM contains a number (\ref{eq:NoMultN=4}) of contributions of the form (\ref{eq:KKFamIIIellneq0nneq0General}), with R- and flavour charges controlled by (\ref{eqRFCharges}). Expressions may be found for the multiplet dimensions in terms of the quantum numbers, adapted to the branching (\ref{eq:Embedding1}), that appear in those expressions. For example, the dimension on the upper boundary of the $\lgrav_2\big[E_{m_1m_2}^\1  ,y_{m_1m_2}; f_{m_1 m_2} \big]$ in (\ref{eq:KKFamIIIellneq0nneq0General}) can be written, suppressing the subindices on the l.h.s.~for simplicity, as 
{\setlength\arraycolsep{0pt}
\begin{eqnarray} \label{eq:LGRAVDimFamI}
 E^\1 &=& \tfrac12 + \Big[ \tfrac{9}{4} +\ell(\ell+4)+\tfrac12(m_1+m_2)^2+ \big(\tfrac{2\pi n}{T}+(m_1-m_2)\chi \big)^2  \\
&& \hspace{-1pt}  -\tfrac12 \big( \lvert m_1\lvert+\lvert m_2\lvert \big) \big( \lvert m_1\lvert+\lvert m_2\lvert+ 2 \ell- 2 \ell_1 -2\ell_2+2 \big)-\tfrac12 \big(\ell-\ell_1-\ell_2 \big) \big(\ell-\ell_1-\ell_2+2 \big) \Big]^{\frac12} . \nonumber 
\end{eqnarray}
}As usual, the $\chi$ dependence is introduced by a non-zero flavour, {\it e.g.} $f_{m_1m_2}$ in (\ref{eqRFCharges}) for the $\lgrav_2$ dimension in (\ref{eq:LGRAVDimFamI}). This dimension saturates the relevant $\cN=2$ unitarity bound for the choice of quantum numbers (\ref{eq:ShorteningQNsCM}), and the multiplet becomes short as indicated in table \ref{tab:ShortN=2Spectrum}. For all other multiplets in (\ref{eq:KKFamIIIellneq0nneq0General}), we also find the protected shortening patterns of that table and, for generic points in this family with only $\textrm{U}(1)_F \times \textrm{U}(1)_R$ symmetry, we find no further shortenings beyond the protected ones in table \ref{tab:ShortN=2Spectrum}. 

At the points (\ref{eq:N=2SU2xU1point1}) and (\ref{eq:N=2SU2xU1point2}) on this boundary, the flavour symmetry is enhanced to $\textrm{SU}(2)_F$, and the spectrum accordingly recombines into representations of $\textrm{OSp}(4|2) \times \textrm{SU}(2)_F$ \cite{Giambrone:2021zvp}. The algebraic structure and the dimensions (in particular (\ref{eq:LGRAVDimFamI})) at these symmetry-enhanced points are the same as in the rest of the upper boundary, only with U$(1)_F$ charges now labelling SU$(2)_F$ representations. This reassembling into $\textrm{SU}(2)_F$ multiplets occurs at every fixed $S^5$ KK number $\ell$, with the same (at $\chi =0$) or possibly different (at $\chi = \pi/T$ and $\chi = 2\pi/T$) $S^1$ KK levels $n$ \cite{Giambrone:2021zvp}. Though all these three (up to periodicity) locations exhibit $\textrm{SU}(2)_F$ symmetry enhancement, only $\chi =0$ and $\chi = 2\pi/T$ have the same KK spectrum, and this differs from that at $\chi = \pi/T$ \cite{Giambrone:2021zvp}. These $\textrm{SU}(2)_F$ symmetry enhancements are somewhat peculiar from the point of view of the parent $\cN=4$ point of the CM in the sense, noted in section \ref{eq:ConfMan}, that $\textrm{SU}(2)_F$ is not a subgroup of its SO(4) R-symmetry group. The symmetry breaking from SO(4) to $\textrm{SU}(2)_F$ proceeds by first breaking the former into $\textrm{U}(1)_F \times \textrm{U}(1)_R$ via (\ref{eq:Embedding1}) and then recombining back up through (\ref{eq:Embedding2}). In fact, an alternate dimension formula adapted to the quantum numbers of the latter branching also exists \cite{Giambrone:2021zvp}. Similarly to the $\cN=4$ point, the spectrum on the SU$(2)_F$-enhanced points has (U$(1)_F$-charged) short multiplets \cite{Giambrone:2021zvp} besides the ones in table \ref{tab:ShortN=2Spectrum}: see the discussion around equation (\ref{eq:shorteningpattern}) in appendix \ref{OSp44SuperMult}.

\subsection{Spectrum on Family III} \label{sec:FamIIISpecttrum}

Before discussing, in section \ref{sec:SpecCMInt}, the KK spectrum at generic interior locations in the CM, we will first look for simplicity at the one-parameter Family III defined in (\ref{eq:FamIII}). This locus contains the $\cN=4$ SO(4) point at (\ref{eq:N=4SO4point}), and is parameterised by $\chi$ with fixed $\varphi = 0$. The $\ell = n = 0$ spectrum on this locus follows from the results of \cite{Bobev:2021yya}. Here we will give the complete KK spectrum on this family at all KK levels.

On Family III, the contributions (\ref{eq:KKFamIIIellneq0nneq0General}) to the spectrum at KK levels $\ell$ and $n$ take on the specific form:
{\setlength\arraycolsep{0pt}
\begin{eqnarray}	\label{eq:KKFamIIIellneq0nneq0}
	 &&\bigoplus_{m_1=-\ell_1}^{\ell_1}\bigoplus_{m_2=-\ell_2}^{\ell_2}
	\Big\{
	 \lgrav_2\big[1+ E_0^{f_{m_1 m_2}}  ,y_{m_1m_2}; f_{m_1 m_2} \big] \nonumber\\
     &\oplus&\lgino_2\big[ \tfrac12 + E_0^{f_{m_1 m_2} +1} , y_{m_1m_2} ; f_{m_1 m_2} +1 \big] 
     \oplus\lgino_2\big[ \tfrac12+ E_0^{f_{m_1 m_2}  -1} , y_{m_1m_2} ; f_{m_1 m_2}  -1\big] \nonumber \\
&\oplus&\lgino_2\big[\tfrac32 + E_0^{f_{m_1 m_2} +1} , y_{m_1m_2} ; f_{m_1 m_2} +1\big] 
\oplus\lgino_2\big[\tfrac32 + E_0^{f_{m_1 m_2}  -1}, y_{m_1m_2} ; f_{m_1 m_2}  -1 \big] \nonumber \\
	&\oplus& 		\lvec_2\big[E_0^{f_{m_1 m_2}} ,y_{m_1m_2}; f_{m_1 m_2} \big]  \nonumber \\
	&\oplus& \lvec_2\big[1+ E_0^{f_{m_1 m_2} +2}, y_{m_1m_2} ; f_{m_1 m_2} +2 \big]
	\oplus\lvec_2\big[ 1+ E_0^{f_{m_1 m_2}} , y_{m_1m_2} ; f_{m_1 m_2}  \big]\nonumber\\[5pt]
	&\oplus&\lvec_2\big[ 1+ E_0^{f_{m_1 m_2} -2} ,y_{m_1m_2} ; f_{m_1 m_2} -2 \big]
	\oplus\lvec_2\big[ 2+ E_0^{f_{m_1 m_2}} , y_{m_1m_2} ; f_{m_1 m_2} \big]
  \Big\}\; ,
\end{eqnarray}
}with $y_{m_1m_2}$ and $f_{m_1 m_2}$ given in (\ref{eqRFCharges}), and dimensions $E_{m_1m_2}^\1 \equiv 1+ E_0^{f_{m_1 m_2}}$, etc., specified as follows.  The quantity $E_0^f$ that determines the dimension of a multiplet in (\ref{eq:KKFamIIIellneq0nneq0}) with U$(1)_F$ flavour $f$ is simply obtained from the $\cN=4$ expression (\ref{eq:GravDimensions}) with the same $\ell, n , \ell_1 , \ell_2$ quantum numbers by replacing the contribution $\big( \frac{2\pi n}{T} \big)^2$ there with (\ref{eq:chiComb}), namely,
\begin{equation} \label{eq:FamIIIDimensions}
E_0^f = -\tfrac12 + \sqrt{ \tfrac94  + \tfrac12 \ell (\ell +4 ) +\ell_1 (\ell_1 +1 ) +\ell_2 (\ell_2 +1 ) +\tfrac12 \big( \tfrac{2\pi n}{T} + f \chi  \big)^2 } \; . 
\end{equation}
At $\chi = 0$, the contributions to the spectrum (\ref{eq:KKFamIIIellneq0nneq0}) with (\ref{eq:FamIIIDimensions}) straightforwardly recombine KK level by KK level into the contributions at the $\cN=4$ point, (\ref{eq:LGRAV4}) with (\ref{eq:GravDimensions}), via the branching (\ref{eq: N4toN2branching}) under the supergroup embedding (\ref{eq:OSp44OSp22U(1)F}). By the argument laid down in section \ref{sec:N=4Specttrum}, for $\chi = 2\pi/T$ the contributions (\ref{eq:KKFamIIIellneq0nneq0}), (\ref{eq:FamIIIDimensions}) also reduce to (\ref{eq:LGRAV4}), (\ref{eq:GravDimensions}) for the $\cN=4$ point via (\ref{eq: N4toN2branching}). In the latter case, this recombination takes place by scrambling the copies of each different multiplet with flavour $f$ at KK levels $n$ and $n^\prime$ via (\ref{eq:KKreshuffle}).

It is instructive to write the above expressions for a few particular cases. The lowest lying, $\ell = n = 0$, case contains simply $(\ell_1 , \ell_2) = (0,0)$, and  becomes
{\setlength\arraycolsep{0.5pt}
\begin{eqnarray} \label{eq:KKFamIIIell=n=0}
	&&\mgrav_2\big[2,\,0;\,0\big]
	\, \oplus \, \sgino_2\big[\tfrac52,\, \pm1 ;\,0\big]
	\nonumber\\
	&&\quad\,\oplus\,   \lgino_2  \big[\tfrac12\sqrt{9+2\chi^2},\,0;\,\pm1\big]
	\,\oplus\,  \lgino_2\big[1+\tfrac12\sqrt{9+ 2\chi^2},\,0;\,\pm1\big]\nonumber\\
	&&\qquad\,\oplus\,  \lvec_2\big[\tfrac12+\tfrac12\sqrt{9+8 \chi^2},\,0;\,\pm2\big]
	\,\oplus\,\lvec_2\big[2,\,0;\,0\big]\nonumber\\
	&&\quad\qquad\,\oplus\,\lvec_2\big[3,\,0;\,0\big]
	\,\oplus\,\mvec_2\big[1,\,0;\,0\big]
	\, \oplus \, \textrm{HYP}_2 \big[2,\,\pm2;\, 0 \big]
	\, , 
\end{eqnarray}
}%
after writing all possible long multiplets at the $\cN=2$ unitarity bounds in terms of short ones as explained in section \ref{sec:N=4Specttrum}. This agrees with the gauged supergravity result, (4.3), (4.4) of \cite{Bobev:2021yya} with $\varphi_\textrm{there} = 1$, after some dimensions there are square-completed. In  (\ref{eq:KKFamIIIell=n=0}) and elsewhere, a flavour or R-symmetry charge with $\pm$ sign indicates the existence of multiplets with both charges. All short multiplets in (\ref{eq:KKFamIIIell=n=0}) are protected in the sense discussed in section \ref{sec:N=4Specttrum} and, reciprocally, table \ref{tab:ShortN=2Spectrum} at $\ell=0$ exhausts all short multiplets here. In agreement with the general discussion, the dimensions of all flavoured multiplets develop a $\chi$ dependence and thus the multiplets remain necessarily long. Not all long multiplets are flavoured, though, and those that are not have $\chi$-independent dimensions.

\begin{figure}
\centering
	\includegraphics[height=0.90\textheight]{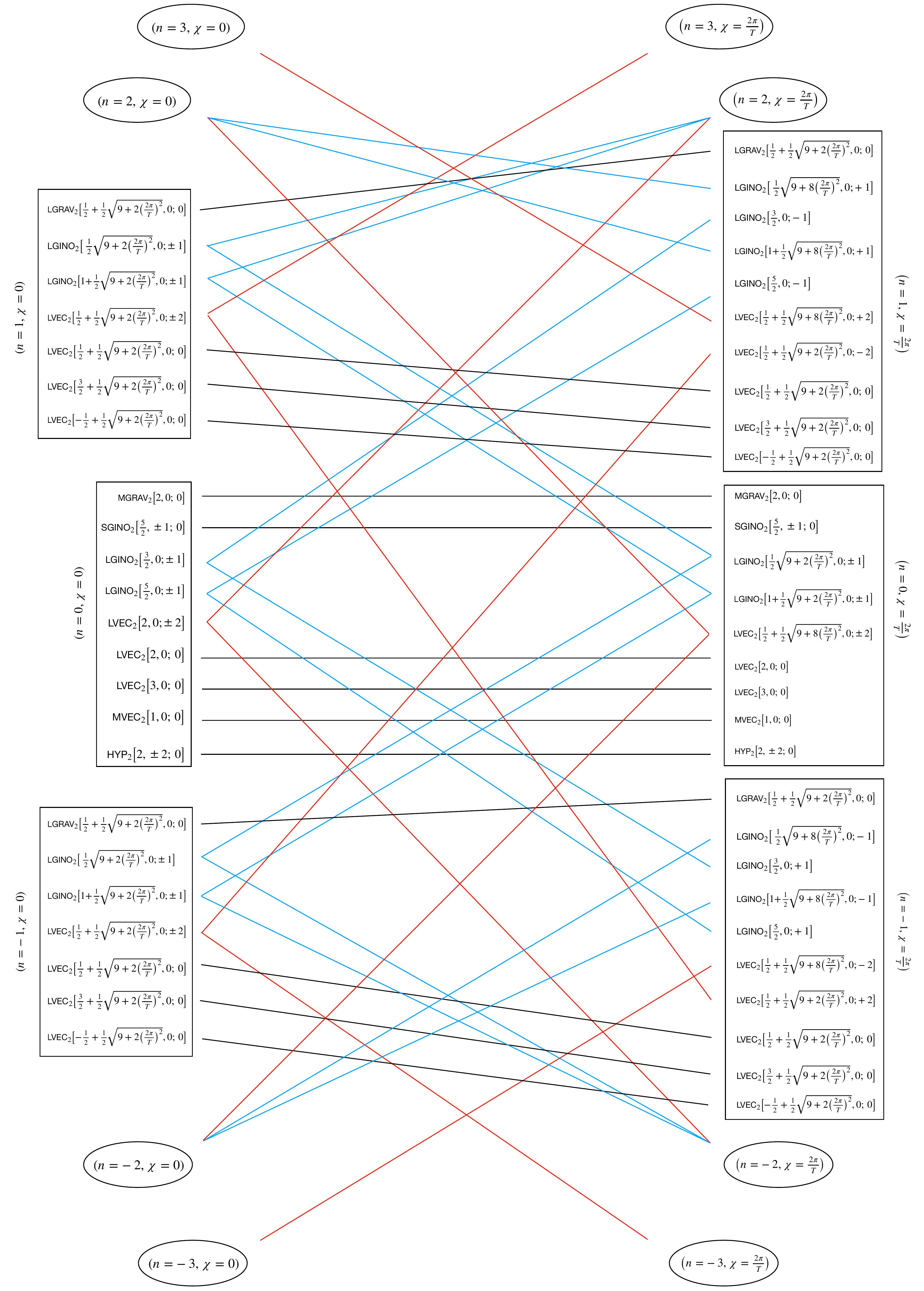}
		\caption{\footnotesize{`Space invasion' patterns for the reassembling of the $\textrm{OSp}(4|2) \times \textrm{U}(1)_F$ multiplets present in the KK spectrum on Family III, (\ref{eq:FamIII}), within the CM at KK levels $\ell = 0$, $n = 0 , \pm 1, \pm 2 , \ldots$, into $\textrm{OSp}(4|4)$ multiplets at the same $S^5$ level $\ell =0$ but possibly different $S^1$ level $n$, at $\chi = 0$ (left) and $\chi =2\pi /T$ (right). The boxes correspond to the multiplet content in (\ref{eq:KKFamIIIellneq0nneq0}) with $\ell_1 = \ell_2 = 0$ and $n$ fixed as indicated. Black lines connect $\chi$-independent, flavour-neutral $\cN=2$ multiplets. Blue and red lines respectively connect $\cN=2$ multiplets that need to be retrieved from one or two higher (or lower) $S^1$ KK levels.
		} \label{fig:InvasionNet} }
\end{figure}

At $\chi=0$, (\ref{eq:KKFamIIIell=n=0}) reduces to the $\ell = n = 0$ spectrum at the $\cN=4$ point \cite{Gallerati:2014xra}, branched out under (\ref{eq:OSp44OSp22U(1)F}) into $\cN=2$ representations through (\ref{eqMGRAV4and2}). At $\chi = 2\pi/T$ the $\cN=4$ spectrum at lowest KK levels is also reproduced, but with reshuffled $S^1$ levels. In order to make this more apparent, it is convenient to extract the KK tower with $\ell = 0$ and $n = 0,  \pm 1 , \pm 2 , \ldots$, from (\ref{eq:KKFamIIIellneq0nneq0}), (\ref{eq:FamIIIDimensions}). The result,
{\setlength\arraycolsep{0.5pt}
\begin{eqnarray} \label{eq:KKFamIIIell=0nneq0}
	&&\lgrav_2\big[\tfrac12+\tfrac12\sqrt{9+2\big(\tfrac{2\pi n}T\big)^2},\,0;\,0\big] \\
	&&\quad\,\oplus\,\lgino_2\big[\tfrac12\sqrt{9+2\big(\tfrac{2\pi n}T\pm\chi\big)^2},\,0;\,\pm1\big]
	\,\oplus\,\lgino_2\big[1+\tfrac12\sqrt{9+2\big(\tfrac{2\pi n}T\pm\chi\big)^2},\,0;\,\pm1\big]\nonumber\\
	&&\qquad\,\oplus\,\lvec_2\big[\tfrac12+\tfrac12\sqrt{9+2\big(\tfrac{2\pi n}T\pm2\chi\big)^2},\,0;\,\pm2\big]
	\,\oplus\,\lvec_2\big[\tfrac12+\tfrac12\sqrt{9+2\big(\tfrac{2\pi n}T\big)^2},\,0;\,0\big]\nonumber\\
	&&\quad\qquad\,\oplus\,\lvec_2\big[\tfrac32+\tfrac12\sqrt{9+2\big(\tfrac{2\pi n}T\big)^2},\,0;\,0\big]
	\,\oplus\,\lvec_2\big[-\tfrac12+\tfrac12\sqrt{9+2\big(\tfrac{2\pi n}T\big)^2},\,0;\,0\big]\, , \nonumber 
\end{eqnarray}
}reduces to (\ref{eq:KKFamIIIell=n=0}) at $n=0$ and extends that equation to all other $n$. Here and elsewhere, the presence in a multiplet of two labels with $\pm$ signs indicates the existence of two (not four) multiplets with correlated upper and lower signs (note incidentally that, at $ | n| \neq 0$ fixed, each of these appears twice like any other multiplet, once for each sign of $n$). All the multiplets present in (\ref{eq:KKFamIIIell=0nneq0}) are generically long, and the dimension of those with non-zero $\textrm{U}(1)_F$ charge develops a $\chi$ dependence, as usual. At $\chi = 0$, (\ref{eq:KKFamIIIell=0nneq0}) reproduces the $\ell =0$, $n=0 , \pm 1, \pm 2 , \ldots $ tower  at the $\cN=4$ point, (\ref{eq:LGRAV4}), (\ref{eq:GravBranching}) with $\ell = \ell_1 = \ell_2 =0$ therein, through the branching (\ref{eq: N4toN2branching}). At $\chi = 2\pi/T$, (\ref{eq:KKFamIIIell=n=0}), (\ref{eq:KKFamIIIell=0nneq0}) also recombine into $\cN=4$ multiplets through (\ref{eq: N4toN2branching}), possibly retrieved from different KK levels $n$. For example, the $\lgino_2\big[\tfrac12\sqrt{9+2\big(\tfrac{2\pi n}T\pm\chi\big)^2},\,0;\,\pm1\big]$  multiplets in (\ref{eq:KKFamIIIell=0nneq0}) are indeed long at $\chi =0$, but at $\chi = 2\pi/T$ become massless for KK levels $n = \mp 1$. For that value of $\chi$,  these join the flavour-neutral (and thus $\chi$-independent) $\text{MGRAV}_2[2,0;0]$ and $\text{MVEC}_2[1,0;0]$ that arise at level $n=0$ in (\ref{eq:KKFamIIIell=n=0}) into an $\text{MGRAV}_4[1,0,0]$ through (\ref{eqMGRAV4and2}). See figure \ref{fig:InvasionNet} for a graphical account of these `space invasion' patterns.

%
\begin{figure}
\centering
	\begin{subfigure}{.48\textwidth}
		\centering
		\includegraphics[width=1.0\linewidth]{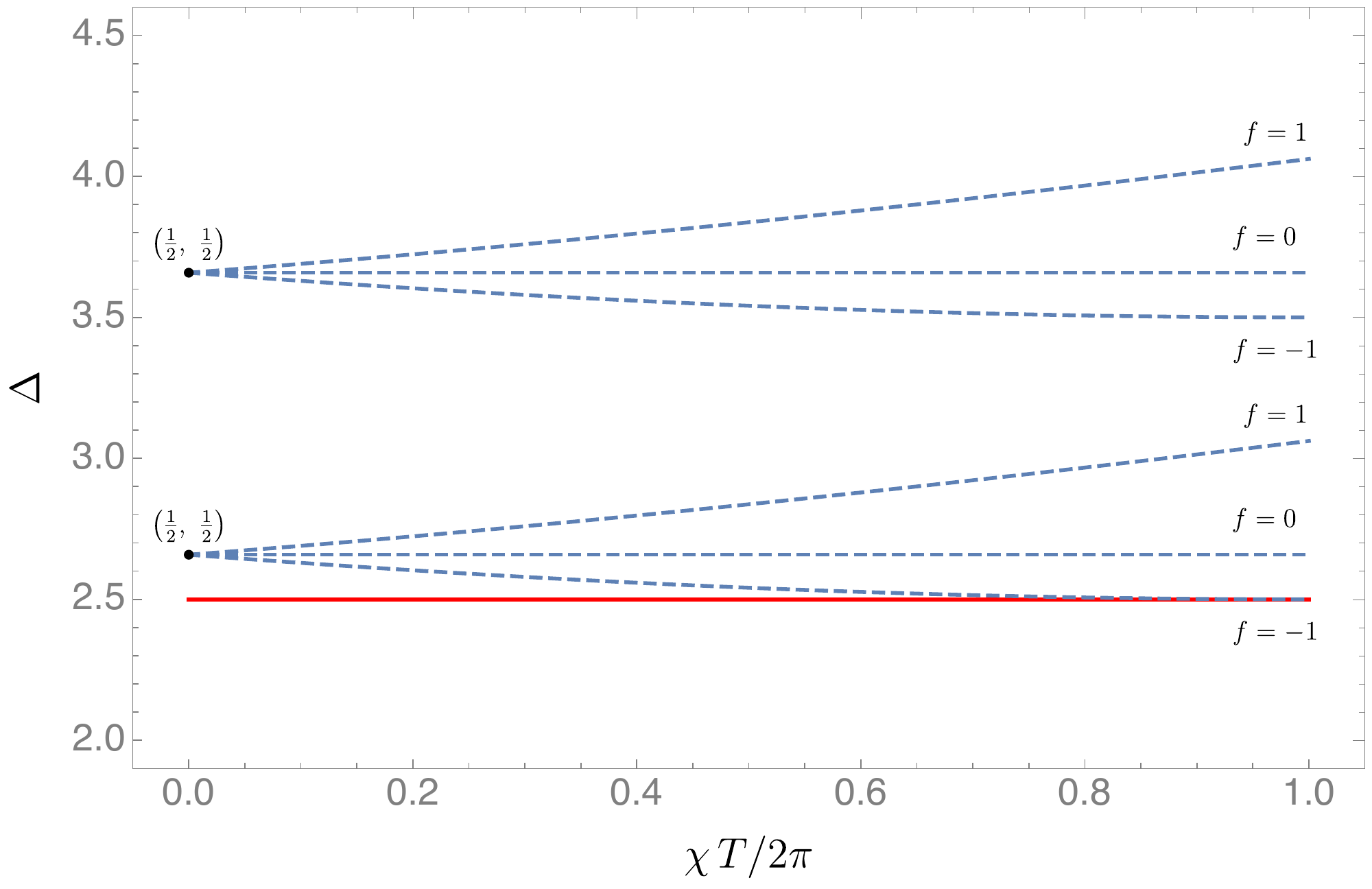}
  		\caption{\footnotesize{Gravitini}}
		\label{fig: FamIIginos01}
	\end{subfigure}%
	\quad
	\begin{subfigure}{.48\textwidth}
		\centering
		\includegraphics[width=1.0\linewidth]{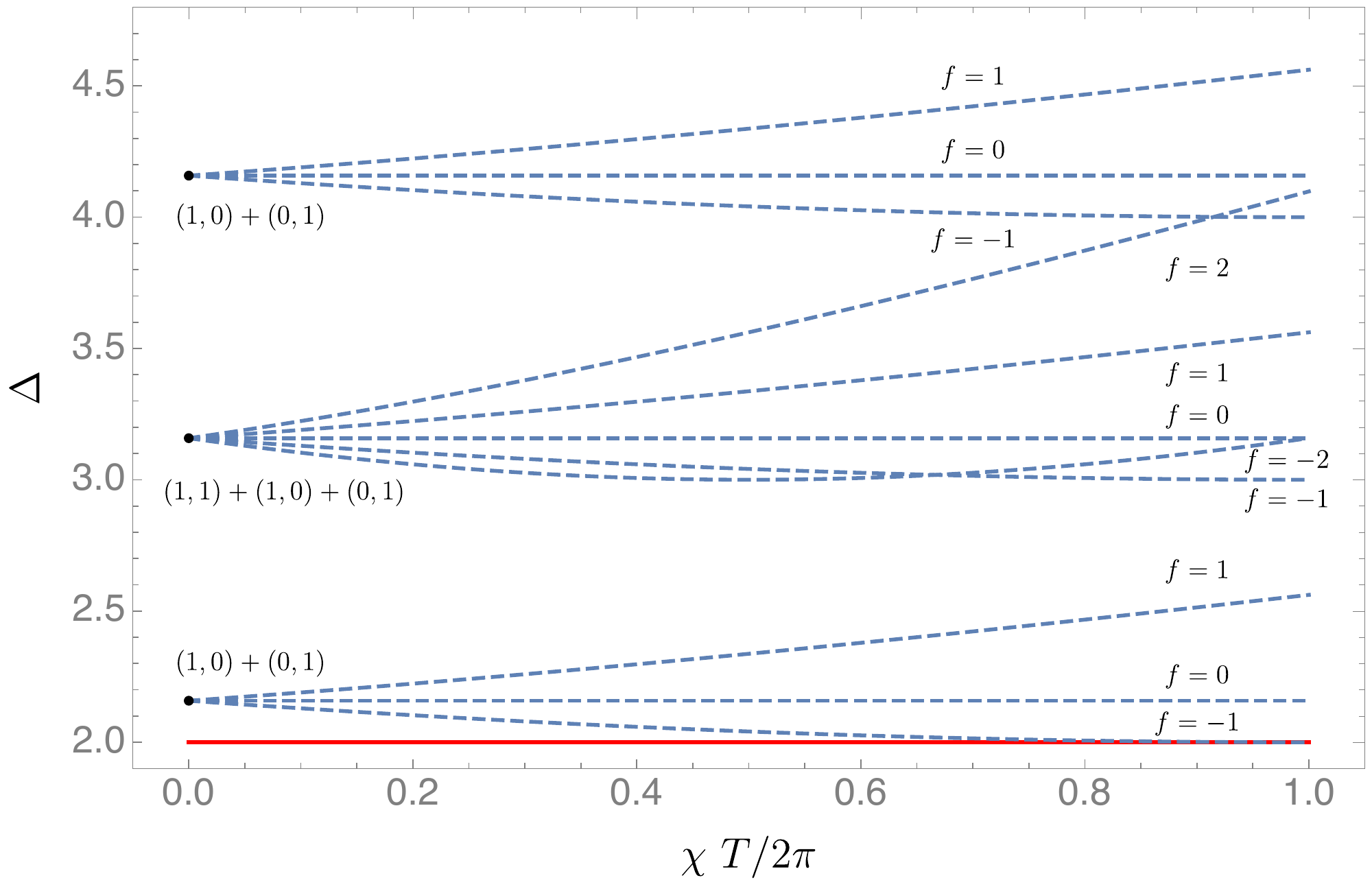}
  		\caption{\footnotesize{Vectors}}
		\label{fig: FamIIvecs01}
	\end{subfigure}
	\caption{\footnotesize{Dimensions $\Delta$ (dashed blue lines) of individual gravitino (left) and vector (right) states with flavour $f$ in the spectrum on Family III, at KK levels $\ell=0$, $n=1$, as functions of $\chi$. The solid red lines stand at the massless threshold. At $\chi=0$, the SO(4) representations from which the flavoured states branch down are shown.} \label{fig:spaceinvasion} }
\end{figure}

By the above analysis, the supermultiplets on Family III recombine into $\cN=4$ supermultiplets at both endpoints of the $\chi$ range (\ref{eq: cmband}). It is also informative to look at the individual states contained in those multiplets, and see how two gravitino states become `massless' (or rather, acquire AdS$_4$ mass $mL=1$ so that their dimension becomes $\Delta = \frac52$) at $\chi =2\pi/T$, thus enhancing the generic $\cN=2$ supersymmetry on Family III to $\cN=4$. Four KK vector states must also become massless, $\Delta=2$, in order for the bosonic symmetry to get enhanced from $\textrm{U}(1)_F \times \textrm{U}(1)_R$ to SO(4). The evolution with $\chi$ of the gravitino and vector mass eigenstates on Family III, as they arise from the diagonalisation of the $\ell=0$, $n=1$ mass matrices of \cite{Varela:2020wty,Cesaro:2020soq}, is depicted in figure~\ref{fig:spaceinvasion}. The left plot indeed identifies one gravitino with flavour $f = -1$ that branches out from the $( \frac12 , \frac12 )$ SO(4) mode with $\Delta = 1+\sqrt{\tfrac94+\tfrac{2\pi^2}{T^2}}$ at $\chi =0$ (and  $T=2\pi$ in the plot), and reaches $\chi = 2\pi/T$ with $\Delta = \tfrac52$. The other relevant gravitino, not depicted, has $f = 1$ and becomes massless at $n=-1$. A similar story unfolds for the vectors on the right plot. Two vector states (superimposed in the plot) with flavour $f = - 1$ branch out from the $(1,0) + (0,1)$ of SO(4) at $\chi = 0$ and become massless at $\chi = 2\pi/T$. Two more vectors, not depicted, with flavour $f=1$ become massless for $n=-1$, while the two vectors that gauge $\textrm{U}(1)_F \times \textrm{U}(1)_R$ stay massless all along.

\subsection{Spectrum at generic points on the interior}	\label{sec:SpecCMInt}

In the interior of the CM and at generic locations $(\varphi  , \chi)$, the KK spectrum again displays the algebraic structure laid down in section \ref{sec:N=4Specttrum}. Except for the origin in our parameterisation, corresponding to the $\cN=4$ point, there are no symmetry or supersymmetry enhancements. Thus, away from the origin, the spectrum remains organised for all $\ell$ and $n$ strictly in the collections (\ref{eq:KKFamIIIellneq0nneq0General}) of $\textrm{OSp}(4|2) \times \textrm{U}(1)_F$ multiplets with moduli-independent charges controlled by (\ref{eqRFCharges}). The multiplet dimensions typically depend on the modulus $\varphi$, and also on the periodic modulus $\chi$ through the combination (\ref{eq:chiComb}) for multiplets with flavour charge $f$. By the general discussion in section \ref{sec:N=4Specttrum}, the latter feature is responsible for the periodicity of $\chi$ for any value of $\varphi$. In particular, the spectrum on Family II, (\ref{eq:FamII}), at $\chi = 2\pi/T$ is mapped into the spectrum at $\chi =0$ level by level in $\ell$, but with $n$ levels reshuffled. Incidentally, the spectrum on Family II does not exhibit any noteworthy features other than this rearrangement of $S^1$ KK levels as $\chi$ crosses cycles.

Diagonalising analytically the KK mass matrices at generic locations of $\varphi$ and $\chi$ requires formidable computer power even at first $S^5$ KK level $\ell=1$. The tower with $\ell=0$ and $n$ arbitrary is still tractable analytically, and so are the first few KK levels of the graviton mass matrix. We report on these results in this section. More generally, we have resorted to numerics to obtain the multiplet spectrum on a (Euclidean) lattice on the CM, and we provide a database as an attachment: see appendix \ref{sec:attachment}. On the upper boundary and on Family III in the interior, our analytical results for generic $\varphi$, $\chi$ reduce to the results reported in sections \ref{sec:FamISpecttrum} and \ref{sec:FamIIISpecttrum}. Our numerics particularised to these loci also reproduce the results of the previous sections.

The spectrum at lowest, $\ell=n=0$, levels has already been computed from gauged supergravity at generic points in the CM \cite{Bobev:2021yya}:
\begin{eqnarray}
\label{eq: Twoparfaml0n0multiplets} 
&&\mgrav_2\big[2,\,0;\,0\big]
	\, \oplus \, \sgino_2\big[\tfrac52,\, \pm1 ;\,0\big]
	\nonumber\\[7pt]
&\oplus& \, \lgino_2\left[\tfrac{1}{2}-\tfrac{1}{2}\sqrt{2-e^{2\varphi}}+\tfrac12\sqrt{e^{-2\varphi}\left(2+e^{2\varphi}\right)^2+2e^{2\varphi}\chi^2},\,0;\; \pm1\right] \nonumber\\[7pt]
&\oplus& \,  \lgino_2\left[\tfrac{1}{2}+\tfrac{1}{2}\sqrt{2-e^{2\varphi}}+\tfrac12\sqrt{e^{-2\varphi}\left(2+e^{2\varphi}\right)^2+2e^{2\varphi}\chi^2},\,0;\; \pm1\right] \\[7pt]
&\oplus &\, \lvec_2\left[\tfrac{1}{2}+\sqrt{-\tfrac74+4e^{-2\varphi}+2e^{2\varphi}\chi^2},\,0;\; \pm2\right]\oplus \; \lvec_2\left[\tfrac{1}{2}+\sqrt{\tfrac14+2e^{2\varphi}},\, 0;\; 0\right]
\nonumber\\[7pt]
&\oplus& \lvec_2\left[\tfrac{1}{2}+\sqrt{\tfrac{33}{4}-2e^{2\varphi}},\,0;\; 0\right] \oplus\,\mvec_2\big[1,\,0;\,0\big]
	\, \oplus \, \textrm{HYP}_2 \big[2,\,\pm2;\, 0 \big]\;.\nonumber
\end{eqnarray}
This reduces to the $\ell=n=0$ spectra on the upper boundary, \cite{Giambrone:2021zvp,Guarino:2020gfe}, and on Family III, (\ref{eq:KKFamIIIell=n=0}). It also contains the protected multiplets of table \ref{tab:ShortN=2Spectrum} at $\ell=0$ and no other short multiplet. The dimensions of all the long multiplets depend on $\varphi$, and also on $\chi$ for flavour-charged multiplets. 

Still at $\ell =0$ but now at all $n$, (\ref{eq: Twoparfaml0n0multiplets}) extends into the following tower of generically long multiplets:
{\setlength\arraycolsep{.5pt}
\begin{eqnarray}
\label{eq: Twoparfaml0multiplets} 
&&\!\!\!\!\!\!\lgrav_2[\tfrac{1}{2}+\beta_1,\, 0;\; 0]\nonumber\\[7pt]
&\oplus& \; \lgino_2\left[\tfrac{1}{2}-\tfrac{1}{2}\sqrt{2-e^{2\varphi}}+\beta_2^+,\,0;\; +1\right]\oplus \; \lgino_2\left[\tfrac{1}{2}-\tfrac{1}{2}\sqrt{2-e^{2\varphi}}+\beta_2^-,\,0;\; -1\right] \nonumber\\[7pt]
&&\oplus\; \lgino_2\left[\tfrac{1}{2}+\tfrac{1}{2}\sqrt{2-e^{2\varphi}}+\beta_2^+,\,0;\; +1\right]\oplus\; \lgino_2\left[\tfrac{1}{2}+\tfrac{1}{2}\sqrt{2-e^{2\varphi}}+\beta_2^-,\,0;\; -1\right]	\nonumber\\[7pt]
&\oplus& \; \lvec_2\left[\tfrac{1}{2}+\beta_3^+,\,0;\; 0\right] \oplus \lvec_2\left[\tfrac{1}{2}+\beta_3^-,\,0;\; 0\right] \oplus \; \lvec_2\left[\tfrac{1}{2}+\beta_4,\, 0;\; 0\right]
\nonumber\\[7pt]
&&\oplus \; \lvec_2\left[\tfrac{1}{2}+\beta^+_5,\,0;\; +2\right]\oplus \; \lvec_2\left[\tfrac{1}{2}+\beta^-_5,\,0;\; -2\right] \;,
\end{eqnarray}
}%
where we have introduced the shorthands
{\setlength\arraycolsep{2pt}
\begin{eqnarray}	\label{eq: betas} 
	\beta_1^2&=&\tfrac{9}{4}+\tfrac{1}{2} e^{2\varphi} \big(\tfrac{2\pi n}{T}\big)^2\;,	\nonumber\\[5pt]
	(\beta_2^\pm)^2&=&\tfrac{1}{4} e^{-2\varphi} \left(2+e^{2\varphi}\right)^2+\tfrac{1}{2} e^{2\varphi} \left(\tfrac{2\pi n}{T}\pm\chi\right)^2\;,	\nonumber\\[5pt]
	(\beta_3^\pm)^2&=&\tfrac{17}{4}+\tfrac{1}{2} e^{2\varphi} \Big[\big(\tfrac{2\pi n}{T}\big)^2-2\Big]\pm\sqrt{(4-e^{2\varphi})^2+2\,e^{2\varphi}(2-e^{2\varphi})\big(\tfrac{2\pi n}{T}\big)^2}\;,	\nonumber\\[5pt]
	\beta_4^2&=&\tfrac{1}{4}+2e^{2\varphi}+\tfrac{1}{2} e^{2\varphi} \left(\tfrac{2\pi n}{T}\right)^2\;,	\nonumber\\[5pt]
	(\beta_5^\pm)^2&=&-\tfrac{7}{4}+4 \, e^{-2\varphi}+\tfrac{1}{2} e^{2\varphi} \left(\tfrac{2\pi n}{T}\pm2\chi\right)^2\;.
\end{eqnarray}
}At $n=0$, the multiplet content (\ref{eq: Twoparfaml0multiplets}) with (\ref{eq: betas}) reduces to (\ref{eq: Twoparfaml0n0multiplets}). It also reproduces the $\ell=0$, $n = 0 , \pm 1 , \pm 2 , \ldots$ towers at the upper boundary, (4.25) of \cite{Giambrone:2021zvp}, and on Family III, (\ref{eq:KKFamIIIell=0nneq0}), when $e^{2\varphi} = 2$ and $e^{2\varphi} = 1$, respectively. In particular, as $\varphi \rightarrow 0$, $\chi \rightarrow 0$, the multiplets in (\ref{eq: Twoparfaml0multiplets}), (\ref{eq: betas}) yield
{\setlength\arraycolsep{1.5pt}
\begin{eqnarray}	\label{eq: degsSO4}
\nonumber
\lgrav_2\left[\frac{1}{2}+\beta_1,\,0;\;0\right]&\rightarrow& \lgrav_2\left[\tfrac12+\tfrac12\sqrt{9+2\big(\tfrac{2\pi n}T\big)^2},\,0;\,0\right] ,	\\ [8pt]
\nonumber
\lgino_2\left[\tfrac{1}{2}-\tfrac{1}{2}\sqrt{2-e^{2\varphi}}+\beta_2^\pm,\,0;\; \pm1\right]&\rightarrow&
\lgino_2\left[\tfrac12\sqrt{9+2\big(\tfrac{2\pi n}T\big)^2},\,0;\,\pm1\right],\\ [8pt]
\nonumber
\lgino_2\left[\tfrac{1}{2}+\tfrac{1}{2}\sqrt{2-e^{2\varphi}}+\beta_2^\pm,\,0;\;\pm1\right]&\rightarrow&
\lgino_2\left[1+\tfrac12\sqrt{9+2\big(\tfrac{2\pi n}T\big)^2},\,0;\,\pm1\right],\\ [8pt]
\nonumber
\lvec_2\left[\tfrac{1}{2}+\beta_3^\pm,\,0;\;0\right]& \rightarrow &\lvec_2\left[\tfrac12\pm1+\tfrac12\sqrt{9+2\big(\tfrac{2\pi n}T\big)^2},\,0;\,0\right], \\ [8pt]
\nonumber
\lvec_2\left[\tfrac{1}{2}+\beta_4\,,0\;;0\right]
&\rightarrow&\lvec_2\left[\tfrac12+\tfrac12\sqrt{9+2\big(\tfrac{2\pi n}T\big)^2},\,0;\,0\right],\\ [6pt]
\lvec_2\left[\tfrac{1}{2}+\beta_5^\pm,\,0\;;\pm 2\right] 
&\rightarrow&\lvec_2\left[\tfrac12+\tfrac12\sqrt{9+2\big(\tfrac{2\pi n}T\big)^2},\,0;\,\pm2\right],
\end{eqnarray}
}%
and thus reproduce via (\ref{eq: N4toN2branching}) the $\ell=0$, $n=0 , \pm 1, \ldots$ tower at the $\cN=4$ point, 
(\ref{eq:GravBranching}) with $\ell = \ell_1 = \ell_2 =0$. When $e^{2\varphi} \rightarrow 2$, $\chi \rightarrow 0$, the multiplet content instead reproduces the $\ell=0$, $n=0 , \pm 1, \ldots$ tower at $\textrm{SU}(2)_F$ point 1, (\ref{eq:N=2SU2xU1point1}), in agreement with \cite{Giambrone:2021zvp}. This occurs through the recombinations 
{\setlength\arraycolsep{1.5pt}
\begin{eqnarray} \label{eq: degsSU2}
\nonumber
\lgrav_2\left[\tfrac{1}{2}+\beta_1,\,0;\;0\right]&\rightarrow& \lgrav_2\left[\tfrac12+\tfrac12\sqrt{9+2\big(\tfrac{2\pi n}T\big)^2},\,0\right]\otimes[0]\; ,\\ [6pt]
\left. \begin{matrix}
\nonumber
\lgino_2\left[\frac{1}{2}\pm\frac{1}{2}\sqrt{2-e^{2\varphi}}+\beta_2^+,\,0;\;+1\right]\\[8pt]
\lgino_2\left[\frac{1}{2}\pm\frac{1}{2}\sqrt{2-e^{2\varphi}}+\beta_2^-,\,0;\;-1\right]
\end{matrix} \right \}& \rightarrow & 2\times \lgino_2\left[\tfrac12+\sqrt{2+\left(\tfrac{2\pi n}{T}\right)^2},\,0\right] \otimes [\tfrac12]\;, \\[6pt]
\left. \begin{matrix}
\nonumber
\lvec_2\left[\tfrac{1}{2}+\beta_3^+,\,0;\;0\right]\\[8pt]
\lvec_2\left[\tfrac{1}{2}+\beta_4,\,0;\;0\right]
\end{matrix}\right\}& \rightarrow & 2\times\lvec_2\left[\tfrac{1}{2}+\sqrt{\tfrac{17}{4}+\big(\tfrac{2\pi n}T\big)^2},\,0\right]\otimes [0]\;,	\\ [6pt]
\left.\begin{matrix}
\lvec_2\left[\tfrac{1}{2}+\beta_5^+,\,0;\;+2\right]\\[8pt]
\lvec_2\left[\tfrac{1}{2}+\beta_3^-,\,0;\;0\right]\\[8pt]
\lvec_2\left[\tfrac{1}{2}+\beta_5^-,\,0;\;-2\right]
 \end{matrix}\right\} & \rightarrow &\lvec_2\left[\tfrac12+\sqrt{\tfrac{1}{4}+\big(\tfrac{2\pi n}T\big)^2},\,0\right] \otimes [1]\;, 
\end{eqnarray}
}%
in the notation of (\ref{eq:OSpMultiplets}), as usual. In our conventions, the U(1)$_F\subset$ SU$(2)_F$ charges are normalised to be integers so that, for example, the $[\tfrac12]$ of SU$(2)_F$ breaks into $\pm1$ U$(1)_F$ charges.

For the tower $\ell=1$, $n = 0 , \pm 1 , \pm 2, \ldots$, we can provide analytic expressions for the dimensions of the $\lgrav_2$ multiplets contained therein. By (\ref{eq:NoMultN=4}), there are $ 2 \big( 1 + H ( |n| )  \big) $ contributions at these KK levels of the form (\ref{eq:KKFamIIIellneq0nneq0General}), with SO(4) labels $(\ell_1,\ell_2)$ given by $(1,0)$ or $(0,1)$ according to (\ref{eq:GravBranching}). Altogether, there are the following graviton multiplets:
{\setlength\arraycolsep{.5pt}
\begin{eqnarray} \label{eq:LGRAVs10}
	&2&\times\lgrav_2\big[\tfrac{1}{2} + \gamma_1,\,0;0\big]		\nonumber\\[7pt]
	&\oplus&\; \lgrav_2\big[\tfrac{1}{2} + \gamma_2^+,\,+1;+1\big] \oplus \lgrav_2\big[\tfrac{1}{2} + \gamma_2^-,\,+1;-1\big]	\nonumber\\[7pt]
	&&\enspace\oplus\; \lgrav_2\big[\tfrac{1}{2} + \gamma_2^+,\,-1;+1\big] \oplus \lgrav_2\big[\tfrac{1}{2} + \gamma_2^-,\,-1;-1\big]\; .
\end{eqnarray}
}%
Here we have defined
\begin{equation} \label{eq:LGRAVs10Dim}
	(\gamma_1)^2=\tfrac{25}{4} +  \tfrac{e^{2\varphi}}{2}\big[(\tfrac{2 \pi n }{T})^2+ 1\big]\;,	\qquad
	(\gamma_2^\pm)^2=\tfrac{23}{4} + e^{-2\varphi}+\tfrac{ e^{2\varphi}}{2} (\tfrac{2 \pi n }{T}\pm\chi)^2\;. 
\end{equation}
It is again instructive to see how these expressions reduce to the known towers on the points with enhanced (super)symmetry. At the SO(4) point, all of the multiplets in (\ref{eq:LGRAVs10}) degenerate with dimension 
\begin{equation}
	E_0=\tfrac{1}{2} + \sqrt{\tfrac{27}{4} +  (\tfrac{2 \pi n }{T})^2}\; .
\end{equation}
This agrees with (\ref{eq:GravDimensions}), with $\ell=1$ and $(\ell_1,\ell_2)=(1,0)$ or $(0,1)$ there. At the SU(2) point $e^{2\varphi} = 2$, $\chi = 0$, the graviton multiplets (\ref{eq:LGRAVs10}), (\ref{eq:LGRAVs10Dim}) recombine as
{\setlength\arraycolsep{1.5pt}
\begin{eqnarray}
	\nonumber 
	\lgrav_2\big[\tfrac{1}{2} + \gamma_1,\,0;0\big]\; 
	&\rightarrow& \,\lgrav_2\big[\tfrac{1}{2} + \sqrt{\tfrac{29}{4} +  (\tfrac{2 \pi n }{T})^2},\,0\big] \otimes [0]\;,
	\\
	\left. \begin{matrix}
		\lgrav_2\big[\tfrac{1}{2} + \gamma_2^+,\,\pm1;+1\big]\;\\[8pt]
		\lgrav_2\big[\tfrac{1}{2} + \gamma_2^-,\,\pm1;-1\big]\;
	\end{matrix} \right\}
	&\rightarrow&\lgrav_2\big[\tfrac{1}{2} + \sqrt{\tfrac{25}{4} +  (\tfrac{2 \pi n }{T})^2},\,\pm1\big] \otimes[\tfrac12]\;,
\end{eqnarray}
}%
again matching the result in \cite{Giambrone:2021zvp}. 

Moving up in $S^5$ KK level, the multiplet content at levels $\ell =2$ and $n = 0 , \pm 1 , \pm 2,  \ldots$ includes $ 4 \big( 1 + H ( |n| )  \big) $ contributions of the form (\ref{eq:KKFamIIIellneq0nneq0General}) with possible $(\ell_1 , \ell_2)$ pairs $(0,0)$, $(1,1)$, $(2,0)$ and  $(0,2)$. All in all, there are the following $\lgrav_2$'s:
{\setlength\arraycolsep{.5pt}
\begin{eqnarray} \label{eq:LGRAVs20}
	&&\!\!\!\!\!\!\lgrav_2\big[\tfrac{1}{2} + \delta_1^+,\,0;0\big]	\oplus	\;\lgrav_2\big[\tfrac{1}{2} + \delta_1^-,\,0;0\big]	\nonumber \\[7pt]
 	&\oplus&\; 2\times\lgrav_2\big[\tfrac{1}{2} + \delta_2^+,\,0;0\big]\oplus\; \lgrav_2\big[\tfrac{1}{2} + \delta_2^-,\,+2;0\big]\oplus\;
	\lgrav_2\big[\tfrac{1}{2} + \delta_2^-,\,-2;0\big]		\nonumber \\[7pt]
	&\oplus&\,2\times\lgrav_2\big[\tfrac{1}{2} + \delta_3^{++},\,+1;+1\big]	\oplus\,
	2\times\lgrav_2\big[\tfrac{1}{2} + \delta_3^{+-},\,+1;-1\big]		\nonumber \\[4pt]
	&&\enspace\oplus\,2\times\lgrav_2\big[\tfrac{1}{2} +\delta_3^{-+},\,-1;+1\big]\oplus\,
	2\times\lgrav_2\big[\tfrac{1}{2} + \delta_3^{--},\,-1;-1\big]		\nonumber \\ [6pt]
	&\oplus&\; \lgrav_2\big[\tfrac{1}{2} + \delta_4^{-+},\,0;+2\big]\oplus\;
	\lgrav_2\big[\tfrac{1}{2} +\delta_4^{--},\,0;-2\big]			\nonumber \\[6pt]
	&&\enspace\oplus\; \lgrav_2\big[\tfrac{1}{2} + \delta_4^{++}\;;\,+2,+2\big]\oplus\;
	\lgrav_2\big[\tfrac{1}{2} + \delta_4^{+-},\,-2;-2\big] 			\nonumber \\[4pt]
	&&\qquad\oplus\; \lgrav_2\big[\tfrac{1}{2} + \delta_4^{++},\,-2;+2\big]\oplus\;
	\lgrav_2\big[\tfrac{1}{2} +\delta_4^{+-},\,+2;-2\big]\;,	
\end{eqnarray}
}where we have introduced
{\setlength\arraycolsep{1.5pt}
\begin{eqnarray} \label{eq:LGRAVs20Dim}
	\big(\delta_1^\pm)^2&=&\tfrac{33}{4}\pm2 +  \tfrac{e^{2\varphi}}{2}\big[\big(\tfrac{2\pi n}{T}\big)^2+4\big]\;,		\nonumber\\[7pt]
	\big(\delta_2^\pm)^2&=&\tfrac{53}{4}\pm1 +  \tfrac{e^{2\varphi}}{2}\big(\tfrac{2\pi n}{T}\big)^2\;,	\nonumber\\[7pt]
	\big(\delta_3^{(pq)}\big)^2&=&\tfrac{47}{4} + e^{-2\varphi} +p\sqrt{-e^{2\varphi}+2} +\tfrac{e^{2\varphi}}{2}\big[1+\left(\tfrac{2\pi n}{T}+q\chi\right)^2\big]\; ,	\nonumber\\[7pt]
	\big(\delta_4^{(pq)}\big)^2&=&\tfrac{37}{4}+p + 4e^{-2\varphi} + \tfrac{e^{2\varphi}}{2} \left(\tfrac{2\pi n}{T} +2q \chi \right)^2\; ,
\end{eqnarray}
}
with $p,q=\pm1$. At the $\varphi = \chi=0$ SO(4) point, these degenerate as 
{\setlength\arraycolsep{1.5pt}
\begin{eqnarray}
	\delta_1^+,\ \delta_2^-,\ \delta_3^{+\pm},\ \delta_4^{-\pm} &\rightarrow& \sqrt{\tfrac{49}{4} + \tfrac12\big(\tfrac{2\pi n}{T}\big)^2}\;,	\nonumber\\[6pt]
	\delta_2^+,\ \delta_3^{-\pm},\ \delta_4^{+\pm} &\rightarrow& \sqrt{\tfrac{57}{4} + \tfrac12\big(\tfrac{2\pi n}{T}\big)^2}\;,				\nonumber\\[6pt]
	\delta_1^{-} &\rightarrow& \sqrt{\tfrac{33}{4} + \tfrac12\big(\tfrac{2\pi n}{T}\big)^2}\;,		
\end{eqnarray}
}%

\begin{figure}
\centering
	\begin{subfigure}{.45\textwidth}
		\centering
		\includegraphics[width=1.0\linewidth]{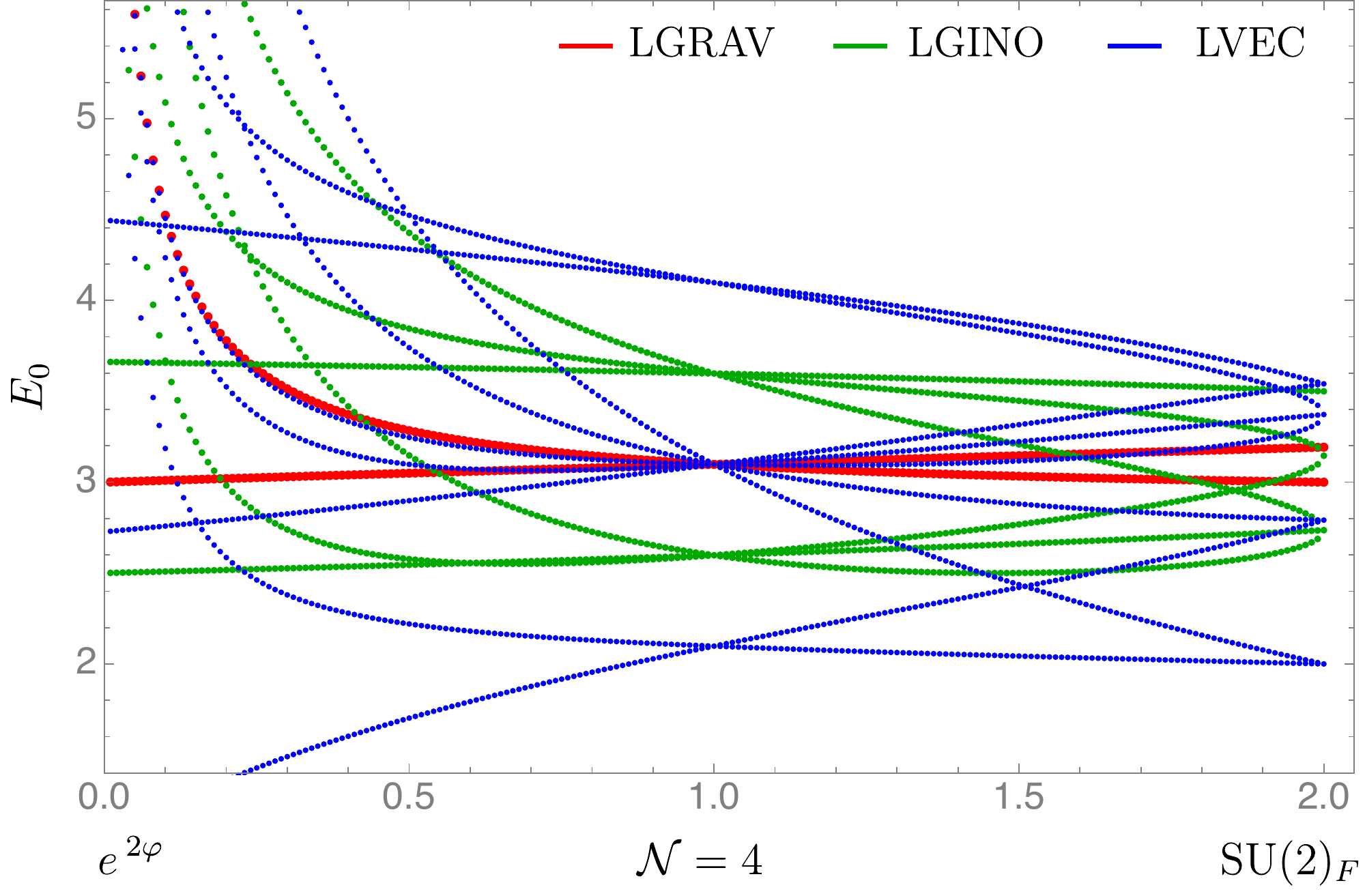}
  		\caption{\footnotesize{$n=0$}}
		\label{fig: dimsFamII10}
	\end{subfigure}%
	\quad
	\begin{subfigure}{.45\textwidth}
		\centering
		\includegraphics[width=1.0\linewidth]{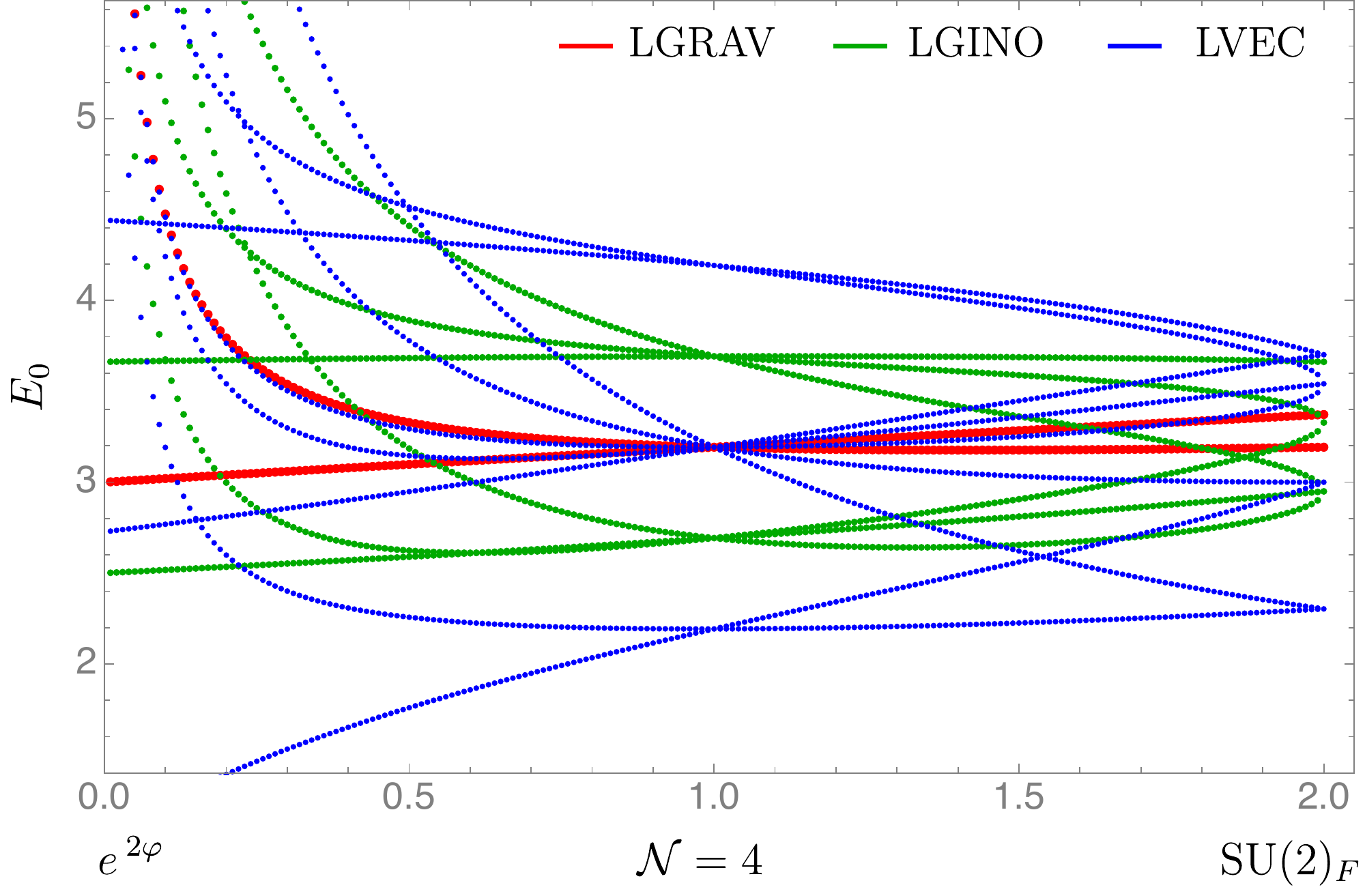}
  		\caption{\footnotesize{$n=1$}}
		\label{fig: dimsFamII11}
	\end{subfigure}\\[8pt]%
	\begin{subfigure}{.45\textwidth}
		\centering
		\includegraphics[width=1.0\linewidth]{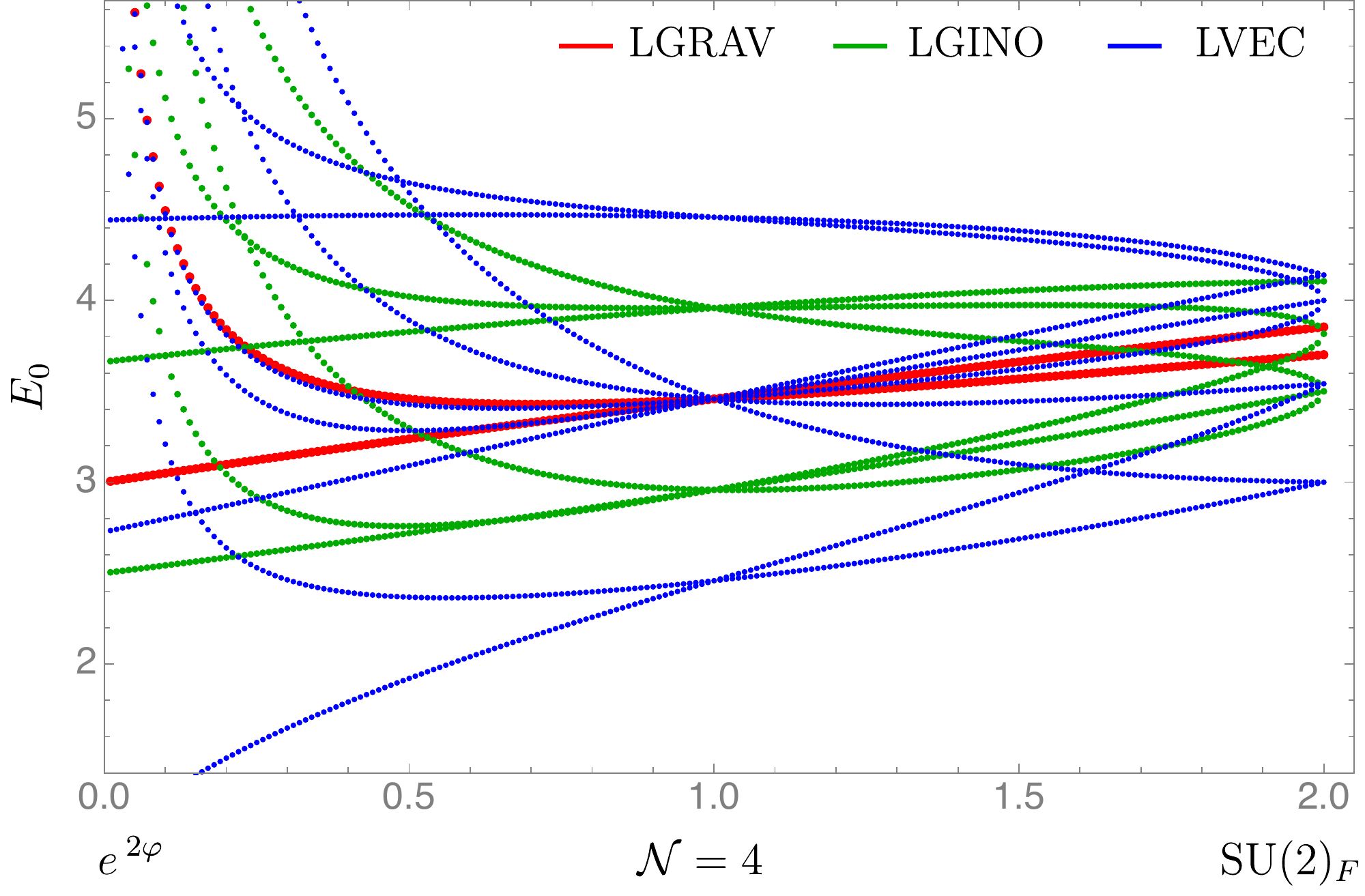}
  		\caption{\footnotesize{$n=2$}}
		\label{fig: dimsFamII12}
	\end{subfigure}%
	\quad
	\begin{subfigure}{.45\textwidth}
		\centering
		\includegraphics[width=1.0\linewidth]{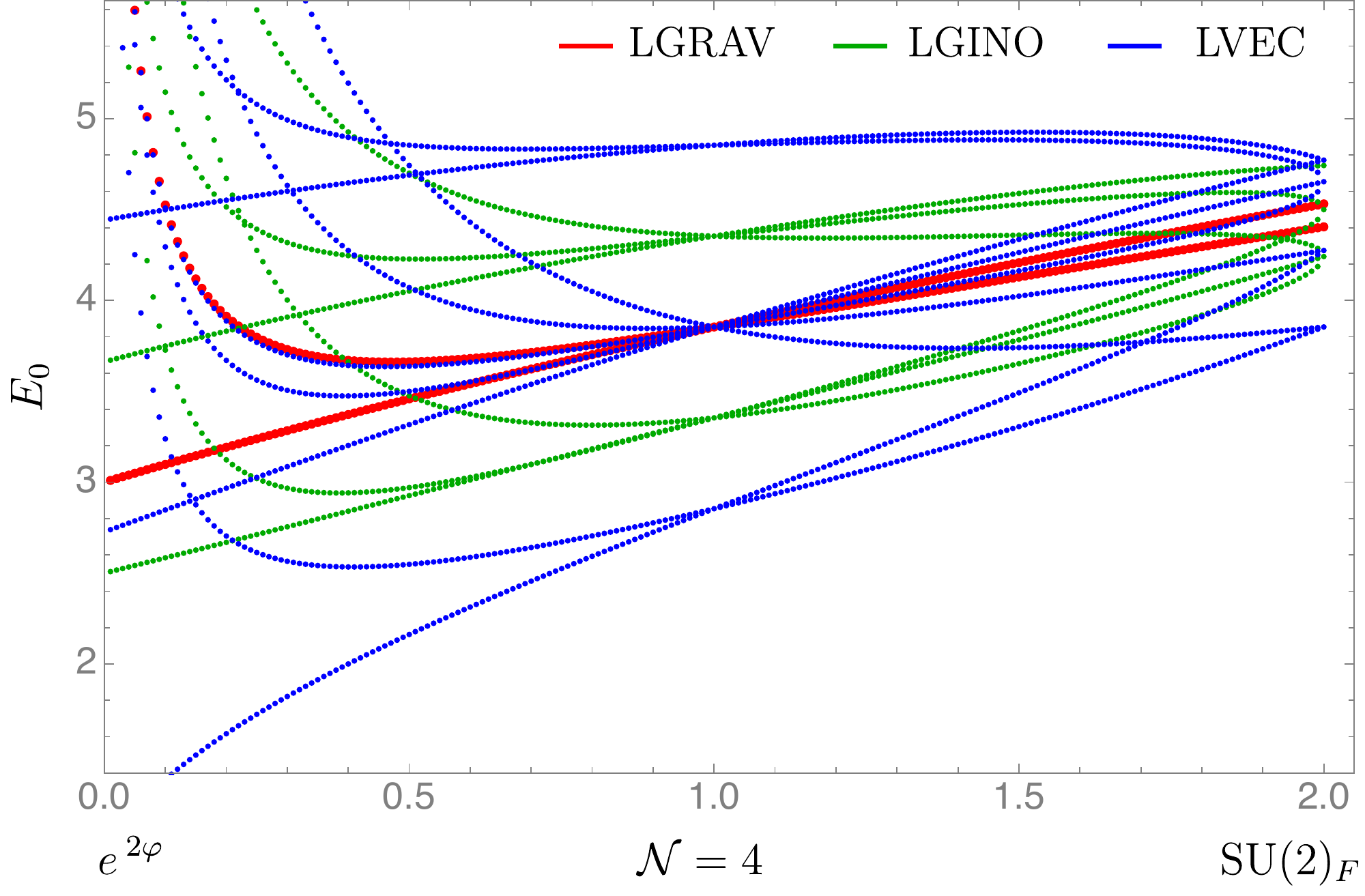}
  		\caption{\footnotesize{$n=3$}}
		\label{fig: dimsFamII13}
	\end{subfigure}
	\caption{\footnotesize{$\mathcal{N}=2$ multiplets at KK levels $\ell=1$ and $n=0 , \ldots , 3$ on Family II at $\chi = 0$, with $T= 2\pi$.} \label{fig:FamiliIIell1} }
\end{figure}

\noindent in agreement with the $\cN=4$ spectrum at the level both of the degeneracies and the dimensions (\ref{eq:GravDimensions}). At the SU(2) point $e^{2\varphi} = 2$, $\chi = 0$, the contributions (\ref{eq:LGRAVs20}) with (\ref{eq:LGRAVs20Dim}) recombine instead as 
{\setlength\arraycolsep{1.5pt}
\begin{eqnarray}
	\left.\begin{matrix}
		\lgrav_2\big[\tfrac{1}{2} + \delta_2^-,\,\pm2;0\big]\\[8pt]
		\lgrav_2\big[\tfrac{1}{2} + \delta_4^{++},\,\pm2;+2\big]\\[8pt]
		\lgrav_2\big[\tfrac{1}{2} + \delta_4^{+-},\,\pm2;-2\big]
	\end{matrix}\right\} 
	&\rightarrow& \lgrav_2\big[\tfrac{1}{2} + \sqrt{\tfrac{49}{4} +  \left(\tfrac{2\pi n}{T}\right)^2},\,\pm2\big] \otimes [1] \;,	\nonumber\\[8pt]
	\left.\begin{matrix}
		\lgrav_2\big[\tfrac{1}{2} + \delta_1^-,\,0;0\big]	\\[8pt]
		\lgrav_2\big[\tfrac{1}{2} + \delta_4^{-+},\,0;+2\big]	\\[8pt]
		\lgrav_2\big[\tfrac{1}{2} + \delta_4^{--},\,0;-2\big]
	\end{matrix}\right\} 
	&\rightarrow& \lgrav_2\big[\tfrac{1}{2} + \sqrt{\tfrac{41}{4} + \left(\tfrac{2\pi n}{T}\right)^2},\,0\big] \otimes [1]	\;,\nonumber\\[8pt]
	\left.\begin{matrix}
		\lgrav_2\big[\tfrac{1}{2} + \delta_3^{++},\,\pm1;+1\big] \\[8pt]
		\lgrav_2\big[\tfrac{1}{2} + \delta_3^{+-},\,\pm1;-1\big]
	\end{matrix} \right\} 
	&\rightarrow& \lgrav_2\big[\tfrac{1}{2} + \sqrt{\tfrac{53}{4} \left(\tfrac{2\pi n}{T}\right)^2},\,\pm1\big] \otimes [\tfrac12]	\;,\nonumber\\[8pt]
	\lgrav_2\big[\tfrac{1}{2} + \delta_1^+,\,0;0\big]
	&\rightarrow& \lgrav_2\big[\tfrac{1}{2} + \sqrt{\tfrac{57}{4} +  \left(\tfrac{2\pi n}{T}\right)^2},0\big] \otimes[0]	\;,		\nonumber\\[8pt]
	\lgrav_2\big[\tfrac{1}{2} + \delta_2^+,\,0;0\big]
	&\rightarrow& \lgrav_2\big[\tfrac{1}{2} + \sqrt{\tfrac{57}{4} +  \left(\tfrac{2\pi n}{T}\right)^2},0\big] \otimes[0]\; ,
\end{eqnarray}
}%
reproducing the results that derive from \cite{Giambrone:2021zvp}, at the level both of the dimensions and the SU$(2)_F$ representation content. Note also that the multiplets $\lgrav_2\big[\tfrac{1}{2} + \delta_2^-,\,\pm2;0\big]$ become short everywhere on the CM at $n=0$, in agreement with (\ref{eq:ShorteningQNsCM}) and table \ref{tab:ShortN=2Spectrum}.

\begin{figure}
\centering
	\begin{subfigure}{.45\textwidth}
		\centering
		\includegraphics[width=1.0\linewidth]{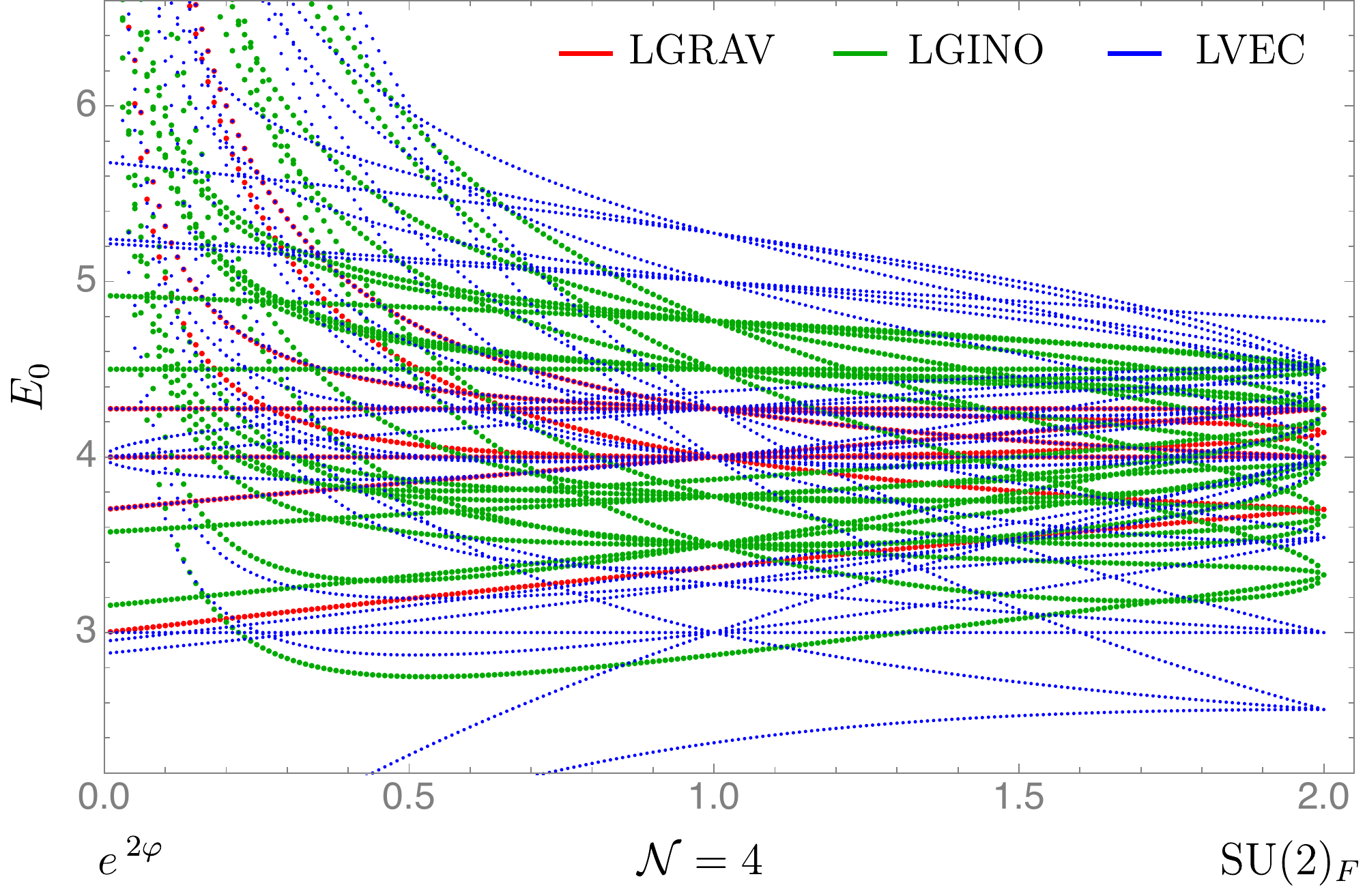}
  		\caption{\footnotesize{$n=0$}}
		\label{fig: dimsFamII20}
	\end{subfigure}%
	\quad
	\begin{subfigure}{.45\textwidth}
		\centering
		\includegraphics[width=1.0\linewidth]{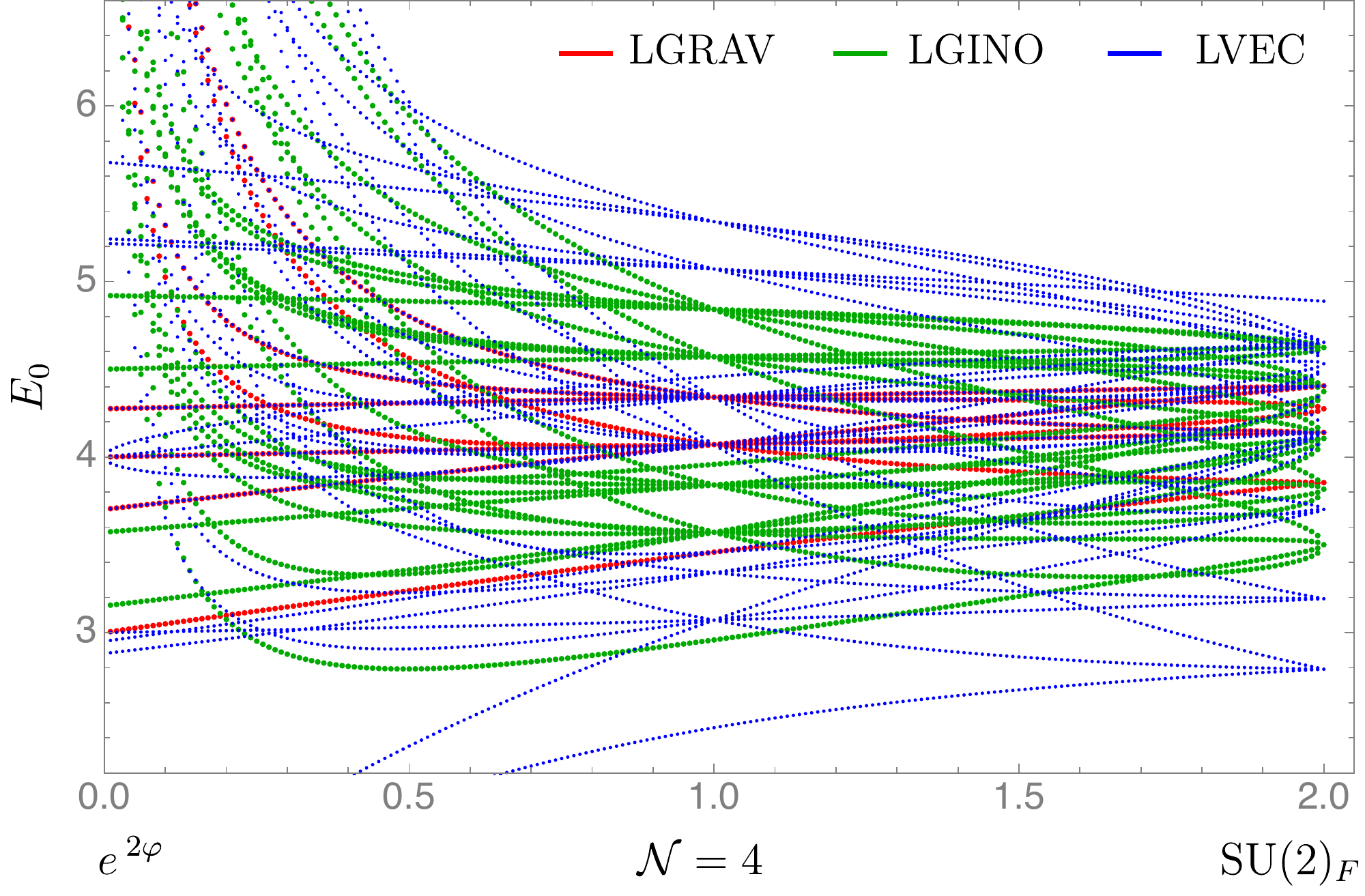}
  		\caption{\footnotesize{$n=1$}}
		\label{fig: dimsFamII21}
	\end{subfigure}\\[8pt]%
	\quad
	\begin{subfigure}{.45\textwidth}
		\centering
		\includegraphics[width=1.0\linewidth]{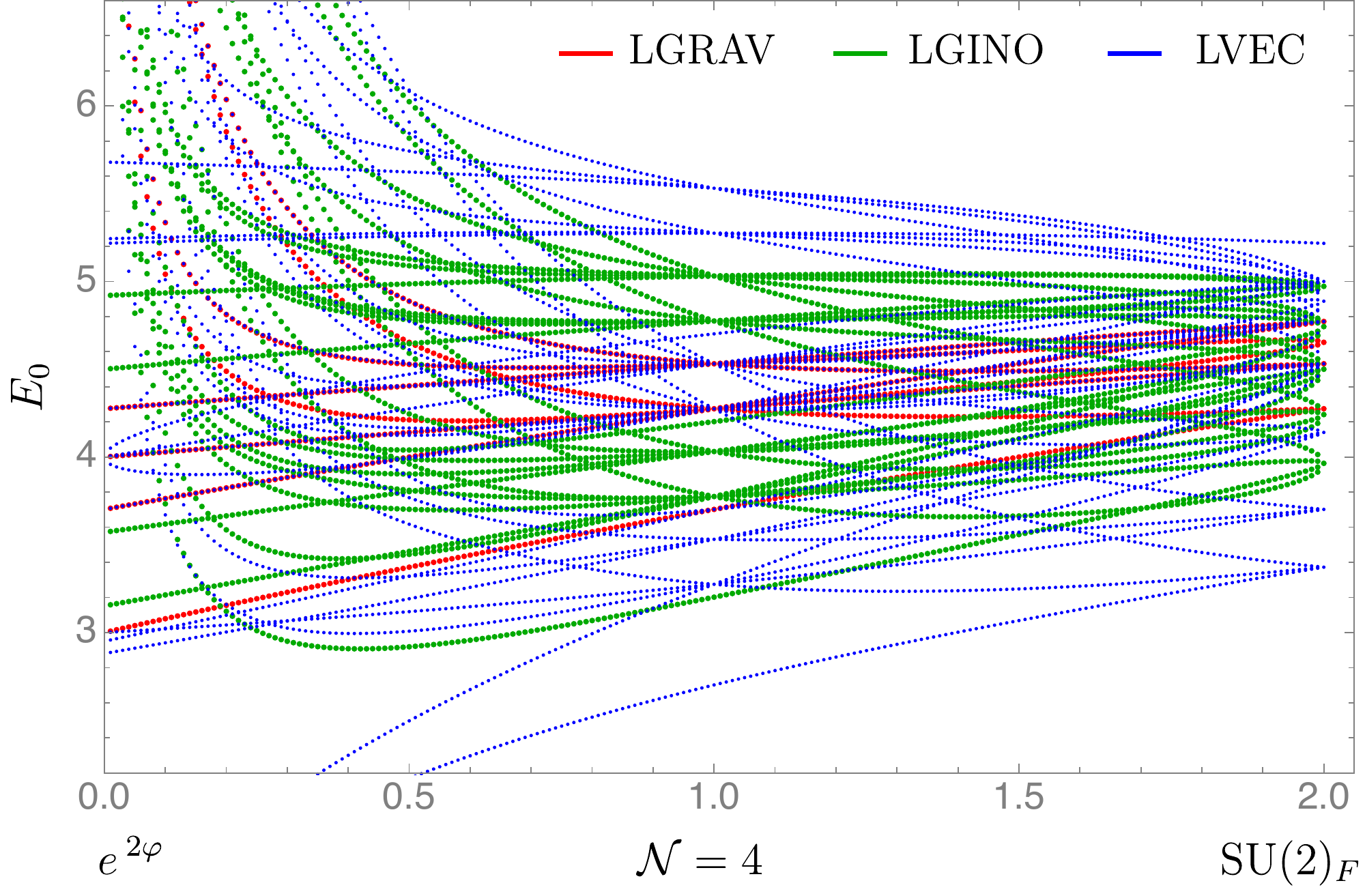}
  		\caption{\footnotesize{$n=2$}}
		\label{fig: dimsFamII22}
	\end{subfigure}%
	\quad
	\begin{subfigure}{.45\textwidth}
		\centering
		\includegraphics[width=1.0\linewidth]{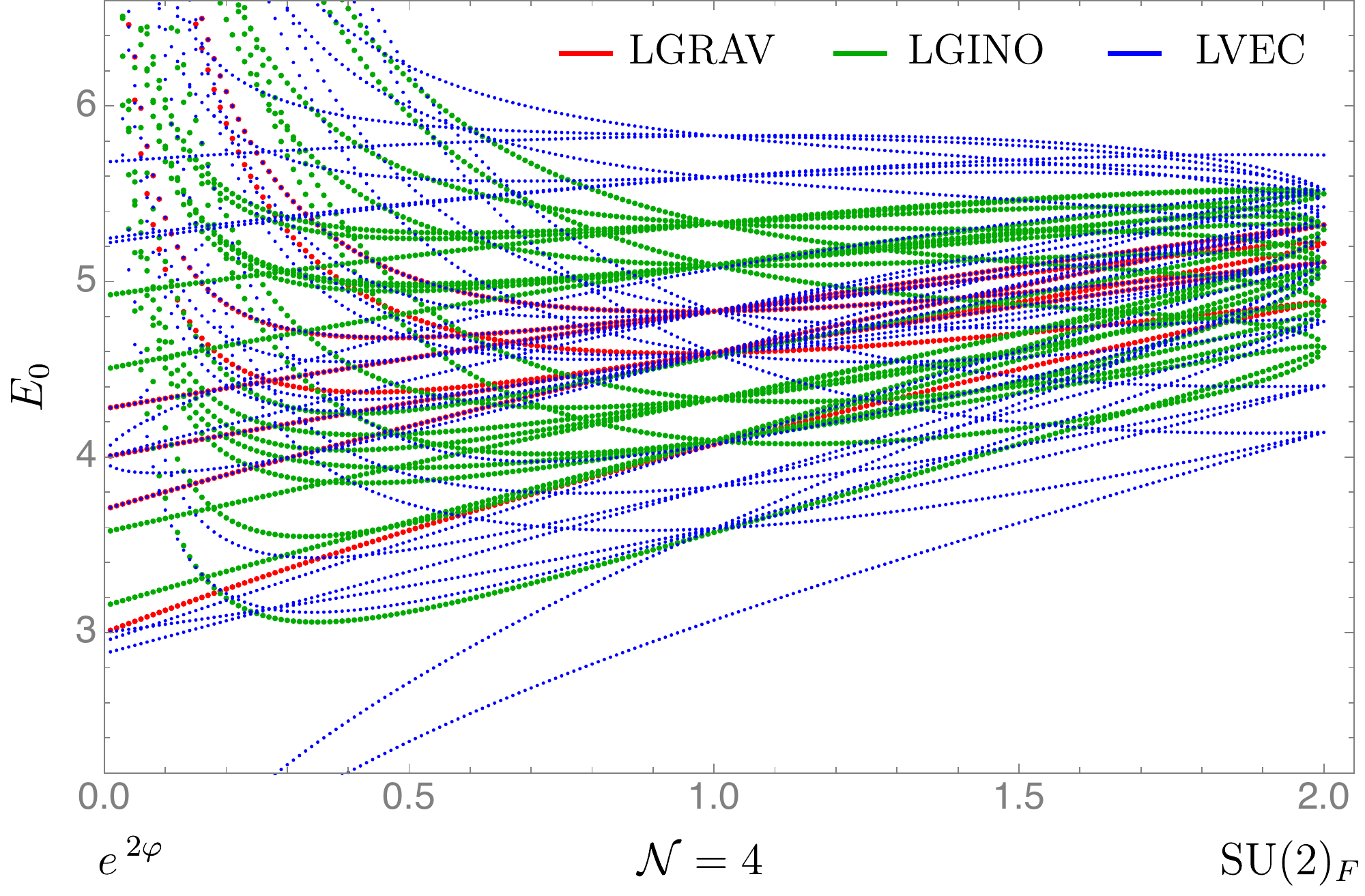}
  		\caption{\footnotesize{$n=3$}}
		\label{fig: dimsFamII23}
	\end{subfigure}
	\caption{\footnotesize{$\mathcal{N}=2$ multiplets at KK levels $\ell=2$ and $n=0 , \ldots , 3$ on Family II at $\chi = 0$, with $T= 2\pi$.} \label{fig:FamiliIIell2} }
\end{figure}

For low values of the KK levels up to $\ell = |n| =3$, we have recomputed numerically the spectrum of graviton multiplets at a grid of locations in the CM, and our results agree with the analytic expressions above. We have also determined numerically on this grid the remaining contributions, from $\lgino_2$'s and $\lvec_2$'s, to the KK spectrum at those levels. The complete results are presented in separate files, see appendix \ref{sec:attachment}. Here, we only provide figures \ref{fig:FamiliIIell1} and \ref{fig:FamiliIIell2} as graphical summaries of those calculations, on a representative one-parameter locus on the CM corresponding to Family II at $\chi =0$. These plots show the dependence on the modulus $\varphi$ of the dimensions of all long graviton, gravitino and vector OSp$(4|2)$ multiplets present in the spectrum at $S^5$ levels $\ell =1$ (in figure \ref{fig:FamiliIIell1}) and $\ell =2$ (in figure \ref{fig:FamiliIIell2}), for various choices of the $S^1$ KK level $n$.

Our numerical results across the interior of the holographic CM are compatible with the shortening patterns of table \ref{tab:ShortN=2Spectrum}. Reciprocally, we do not see any other accidental shortenings taking place, at least on our grid. At level $\ell=2$ we indeed see moduli-independent multiplet dimensions given, within numerical precision, by the integers specified in the table. Curiously, we also see other integer multiplet dimensions arising on certain loci of the CM, although these should not be regarded as particularly significant, as they were obtained for fixed $T=2\pi$. For example, at fixed $e^{2\varphi}=\tfrac32$ and all $\chi$, there is a (flavour neutral) $\lvec_2$ with dimension $E_0 =3$ that arises at KK levels $\ell = 0$ and $n=2$. This multiplet contains classically marginal, $\Delta = 3$, scalars which, however, cannot become exactly marginal because the multiplet lies above the unitarity bound and thus must be long. Also on this locus, and on the $e^{2\varphi}=\tfrac65$, $\chi$ free locus, there are $\lgrav_2$'s arising at $(\ell , n) =(2,2)$ and $(\ell , n) =(1,3)$, respectively, with $E_ 0 = 4$. The latter locus has an $\lvec_2$ with $E_0 = 6$ at  $(\ell, n)=(3,3)$, and there is also a $\lvec_2$ with $E_ 0 = 4$ at $(\ell, n)=(2,2)$ on the family $e^{2\varphi}=\tfrac85$ with $\chi$ free. The points $(e^{2\varphi},\chi)=(\tfrac65,0)$ and $(e^{2\varphi},\chi)=(\tfrac54,1)$ have $\lvec_2$'s with $E_0=5$ and $E_0=7$, arising in both cases at $(\ell, n)=(3,1)$. This list is presumably not exhaustive. Finally, in the region $1 \leq e^{2\varphi} \leq 2$ for all $\chi$, all relevant or marginal, $\Delta \leq 3$, scalars arise at KK levels up to $\ell=2$: at KK levels $\ell=3$, all scalars have dimensions $\Delta > 3$ for all $n$. For $0 < e^{2\varphi} <1$, there are $\Delta \leq 3$ scalars even at $\ell =3$. Our numerical calculations fix $T= 2\pi$ for simplicity, but the results do not differ qualitatively from those with the more realistic $k$-dependent choices for $T$ in \cite{Inverso:2016eet,Assel:2018vtq}.

Our numerics show that for all values of the parameters within the ranges specified in (\ref{eq: cmband}), the KK spectra are well behaved. As the singular limit $e^{2\varphi}=0$ with $\chi$ arbitrary is approached, the multiplet dimensions become independent of $\chi$ even for flavoured multiplets. The dimensions also become independent of the $S^1$ KK level $n$. For example, from (\ref{eq: Twoparfaml0multiplets}), the $\ell = 0$ spectrum on this asymptotic locus becomes, for all $n = 0 , \pm 1 , \pm 2, \ldots$,
{\setlength\arraycolsep{0.5pt}
\begin{eqnarray} \label{eq:Asymptoticell=0Alln}
	&&\mgrav_2\big[2,\,0;\,0\big]
	\, \oplus \, \sgino_2\big[\tfrac52,\, \pm1 ;\,0\big]
	\,\oplus\, 2 \times   \lgino_2  \big[ e^{-\varphi} +\ldots ,\,0;\,\pm1\big]
	\nonumber\\
	&&\qquad\,\oplus\,  \lvec_2\big[ 2 e^{-\varphi}  +\ldots ,\,0;\,\pm2\big]
	\,\oplus\,\lvec_2\big[\tfrac{1}{2}\big( 1 +\sqrt{33} \big) ,\,0;\,0\big]\nonumber\\
	&&\quad\qquad\,
	\,\oplus\,2 \times \mvec_2\big[1,\,0;\,0\big]
	\, \oplus \, 2 \times  \textrm{HYP}_2 \big[2,\,\pm2;\, 0 \big]
	\, , 
\end{eqnarray}
}after again writing all the multiplets at the $\cN=2$ unitarity bound as short. The dimensions of the flavour-neutral multiplets reduce to finite constants, but those with flavour $f$ appear to grow without bound as $| f|  \, e^{-\varphi}$. This is apparent from (\ref{eq:Asymptoticell=0Alln}) for the $\ell = 0$, $n = 0 , \pm 1 , \pm 2, \ldots$ tower (the ellipses in (\ref{eq:Asymptoticell=0Alln}) denote subleading terms with respect to that behaviour), and is further confirmed at higher $\ell$, at least for the graviton multiplets, by the $e^{2\varphi} \rightarrow 0$ limit of (\ref{eq:LGRAVs10}), (\ref{eq:LGRAVs20}). Note also that (\ref{eq:Asymptoticell=0Alln}) contains an infinite tower, $n = 0 , \pm 1 , \pm 2, \ldots$, of massless scalars, vectors and gravitons.


\section{Type IIB uplift of Family III} \label{eq:FamilyIIISfold}

Some aspects of the holographic CM, like the periodicity in $\chi$ and the symmetry enhancement (\ref{eq:N=2SU2xU1point2}), are intrinsically higher-dimensional. These cannot be seen at the level of the $D=4$ gauged supergravity, but are already present in the KK spectrum as discussed in \cite{Giambrone:2021zvp} and section \ref{sec:KKSpectra}  above. Determining the type IIB uplift of the entire two-parameter family of $D=4$ gauged supergravity vacua of \cite{Bobev:2021yya} is certainly beyond the scope of this work. Here, we will only give the uplift of Family III within this set of solutions. This will be enough to explain the periodicity in $\chi$ of the spectrum discussed in section \ref{sec:KKSpectra}, including the $\cN=4$ SO(4) (super)symmetry enhancements (\ref{eq:N=4SO4point}). Previously known uplifts into type IIB S-folds of points or loci in the holographic CM include that of the $\cN=4$ SO(4) point \cite{Inverso:2016eet}, the $\cN=2$ $\textrm{SU}(2)_F \times \textrm{U}(1)_R$ point \cite{Guarino:2020gfe}, and Family I \cite{Giambrone:2021zvp}. 

The type IIB uplift of any solution of $D=4$ $\cN=8$ $ [\textrm{SO} (6) \times \textrm{SO} (1,1) ] \ltimes \mathbb{R}^{12}$--gauged supergravity may be obtained using the ExFT formulae of \cite{Inverso:2016eet}. However, we have not followed this route to obtain the type IIB solutions corresponding to Family III. Instead, we have used the following reverse engineering approach. Firstly, we wrote an educated guess for the ten-dimensional metric, and confirmed it by reproducing the graviton sector of the KK spectrum of section \ref{sec:FamIIISpecttrum} using the formalism of \cite{Bachas:2011xa}. Secondly, we wrote ans\"atze for the remaining supergravity fields, and enforced the type IIB field equations on the full configuration. In retrospect, the successful reproduction of the graviton spectrum for Family III using \cite{Bachas:2011xa}, together with the fact that all KK modes close into $\textrm{OSp}(4|2) \times \textrm{U}(1)_F$ representations, provides a solid crosscheck on our implementation of the ExFT spectral techniques of \cite{Malek:2019eaz,Malek:2020yue,Cesaro:2020soq}.

In order to write the type IIB solutions, it is convenient to employ the same coordinates, $\eta$ on $S^1$ and $(r,\theta_i , \phi_i )$, $i=1,2$, on $S^5$, used in \cite{Inverso:2016eet} to express the $\cN=4$ SO(4)-invariant solution. These coordinates range as
\begin{equation} \label{eq:CoordRanges}
0 \leq \eta < T \; , \qquad 0 \leq r \leq 1 \; , \qquad  0 \leq \theta_i \leq \tfrac{\pi}{2}  , \qquad  0 \leq \phi_i < 2\pi   \; , \quad i =1,2 \; .
\end{equation}
In particular, $\eta$ and $\phi_i$ are  periodic with periods $T$ and $2\pi$,
\begin{equation} \label{eq:Periodicities}
\eta \sim \eta +T \; , \qquad \phi_i \sim \phi_i +2 \pi \; , \quad  i=1,2 \; .
\end{equation}
It is also helpful to introduce the following $\chi$-dependent one-
\begin{equation} \label{eq:OneFormsIIB}
\bm{e}_1 = d\phi_1 - \chi \, d\eta \; , \qquad \bm{e}_2 = d\phi_2 + \chi \, d\eta \; ,
\end{equation}
and two-forms
\begin{equation} \label{eq:TwoFormsIIB}
\bm{v}_1 = \tfrac{r^2}{1+2r^2} \, \sin{\theta_1}\,d\theta_1\wedge \bm{e}_1 \; , \qquad 
\bm{v}_2 = \tfrac{1-r^2}{3-2r^2} \, \sin{\theta_2}\,d\theta_2\wedge \bm{e}_2 \; .
\end{equation}
The relative signs in \eqref{eq:OneFormsIIB} have been chosen to match the flavour group in \eqref{eq:Embedding1}, but any other choice of signs will also yield a solution of the equations of motion.

With these definitions, the type IIB uplift of $D=4$ Family III of vacua can be written as follows. The metric reads
\newpage

{\setlength\arraycolsep{1pt}
\begin{eqnarray} \label{IIBMetricFamIII}
ds_{10}^2 &=&  L^{2} \Delta^{-1} \, \Big[ds^2 (\textrm{AdS}_4)  \\
&& +  2d\eta^{2}+ \frac{2dr^{2}}{1-r^{2}}+\frac{2r^{2}}{1+2r^{2}} \, \big[d\theta_1^2 + \sin^2\!\theta_1 \, \bm{e}_1^2\big]	+ \frac{2(1-r^{2})}{3-2r^{2}} \, \big[d\theta_2^2+\sin^2\!\theta_2 \, \bm{e}_2^2\big]\Big] \; , 
 \nonumber
\end{eqnarray}
}and the self-dual five-form is 
\begin{equation} \label{eq:F5IIB}
 F_\5= L^4\Big[-6 \, \textrm{vol} \big( \textrm{AdS}_4 \big)\wedge \Big( d\eta-\frac{4}{3}r dr \Big) +  \frac{6}{\sqrt{1-r^2}}\,dr \wedge \bm{v}_1 \wedge  \bm{v}_2 +\;8 r\sqrt{1-r^2} \,  \bm{v}_1 \wedge \bm{v}_2 \wedge d\eta\Big]   \, .
 \end{equation}
Here, $ds^2 (\textrm{AdS}_4)$ and $\textrm{vol} \big( \textrm{AdS}_4 ) $ are the metric and volume form on unit radius AdS$_4$ space and $L$ is related both to the electric gauge coupling $g$ of the $D=4$ $\cN=8$ supergravity as $L^2 \equiv \frac12 g^{-2}$, and to the dual gauge group rank $N$ as $L^4 \sim N$ upon flux quantisation. The warp factor depends only on the coordinate $r$,
\begin{equation}
\Delta= \Big((1+2r^2)(3-2r^2) \Big)^{-\frac14} \; ,
\end{equation}
while the dilaton and axion depend also on $\eta$:
\begin{eqnarray}
\label{eq:axion-dilaton}
e^{-\Phi}=\frac{\sqrt{2}\sqrt{(1+2r^2)(3-2r^2)}}{(3+2r^3)\cosh{2\eta}+4r^2\sinh{2\eta}}  \; ,  \quad 
C_\0=\frac{4r^2\cosh{2\eta}+(3+2r^3)\sinh{2\eta}}{(3+2r^3)\cosh{2\eta}+4r^2\sinh{2\eta}} . \;
\end{eqnarray}
Finally, the Neveu-Schwarz and Ramond-Ramond three-form field strengths are
{\setlength\arraycolsep{1pt}
\begin{eqnarray} \label{eq:ThreeFormsIIBFamIII}
H_\3 &=& 4L^2\Big[- 3^{-\frac{1}{4}}\, e^{-\eta} \, \left(\frac{(3+2r^2)}{(1+2r^2)}dr  -r d\eta\right) \wedge \bm{v}_1 	\nonumber \\[2pt]
&& \hspace{2cm}+ \, 3^{\frac{1}{4}} \, e^{\eta}  \left(\frac{(5-2r^2)}{(3-2r^2)}\frac{r}{\sqrt{1-r^2}} dr\wedge  \bm{v}_2 -\sqrt{1-r^2} \,  \bm{v}_2  \wedge d\eta\right)\Big]\, , \nonumber \\
F_\3 &=& \tilde{F}_\3-C_\0 H_\3 \; , 
\end{eqnarray}
}where
{\setlength\arraycolsep{1pt}
\begin{eqnarray} \label{eq:AuxThreeForm}
\tilde{F}_\3&=& 4L^2 \Big[ 3^{-\frac{1}{4}} \, e^{-\eta} \, \left(\frac{(3+2r^2)}{(1+2r^2)}dr -r\,  d\eta\right) \wedge \bm{v}_1 \nonumber\\[2pt]
&& \hspace{2cm}+\, 3^{\frac{1}{4}} \, e^{\eta}\, \left(\frac{(5-2r^2)}{(3-2r^2)}\frac{r}{\sqrt{1-r^2}} dr\wedge \bm{v}_2 -\sqrt{1-r^2} \, \bm{v}_2  \wedge d\eta\right)\Big]\, . 
\end{eqnarray}
}%
We also note the following expressions for the two-form potentials,
{\setlength\arraycolsep{1pt}
\begin{eqnarray} \label{eqTwoFormPots}
B_\2 &=& 4L^2\Big(- 3^{-\frac{1}{4}}\, e^{-\eta} r\, \bm{v}_1 \,- 3^{\frac{1}{4}}\, e^{\eta} \sqrt{1-r^2} \, \bm{v}_2 \Big) \ , \nonumber\\[2pt]
C_\2 &=& 4L^2\Big( 3^{-\frac{1}{4}}\, e^{-\eta}  r\, \bm{v}_1 \,-3^{\frac{1}{4}}\, e^{\eta}\,\sqrt{1-r^2}\, \bm{v}_2\Big) \, ,
\end{eqnarray}
}%
such that $H_\3 = dB_\2$ and  $\tilde{F}_\3 = dC_\2$. 

We have verified that (\ref{IIBMetricFamIII})--(\ref{eq:AuxThreeForm}) solve the equations of motion and Bianchi identities of type IIB supergravity, as given in {\it e.g.}~appendix A of \cite{Gauntlett:2010vu}. This configuration thus defines a one-parameter family, labelled by the constant $\chi$, of non-geometric S-fold solutions of type IIB supergravity. At both endpoints of the interval in (\ref{eq:CoordRanges}) for the $S^1$ coordinate $\eta$, the fields (\ref{eq:axion-dilaton}), (\ref{eq:ThreeFormsIIBFamIII}) charged under $\textrm{SL}(2 , \mathbb{R})$ (or $\textrm{SL}(2 , \mathbb{Z})$ in the full string theory), are related by an $\textrm{SL}(2 , \mathbb{R})$ (or $\textrm{SL}(2 , \mathbb{Z})$) S-duality transformation, exactly as in  \cite{Inverso:2016eet,Assel:2018vtq}. Further, as argued in the first of these references, supersymmetry is not upset by the uplifting process as long as such S-duality transformation lies in the hyperbolic $\textrm{SL}(2 , \mathbb{Z})$ conjugacy class. Thus, the type IIB solution (\ref{IIBMetricFamIII})--(\ref{eq:AuxThreeForm}) inherits the generic $\cN=2$ supersymmetry of the $D=4$ Family III solution it uplifts from. It also contains the $\cN=4$ SO(4) point at $\chi =0$ and at the other locations specified below.

The type IIB solution (\ref{IIBMetricFamIII})--(\ref{eq:AuxThreeForm}) depends on the parameter $\chi$ only through the one-forms (\ref{eq:OneFormsIIB}) (and the two-forms (\ref{eq:TwoFormsIIB}) via their dependence on the former). For all values of $\chi$ and the specified coordinate ranges (\ref{eq:CoordRanges}), the solution extends globally over $S^5 \times S^1$, with the $S^5$ trivially fibred over $S^1$. At $\chi = 0$, our solution reduces to the $\cN=4$ SO(4)-invariant solution on $S^5 \times S^1$, (3.35)--(3.41) of \cite{Inverso:2016eet}, upon identifying ${\cal Y}^p$, ${\cal Z}^p$, $p=1,2,3$ there as
\begin{eqnarray}
{\cal Y}^1 = r\,\cos{\theta_1} \, ,\hspace{0.9cm}\quad {\cal Y}^2 = r\,\sin{\theta_1}\cos{\phi_1} \,, \hspace{0.7cm} \quad  {\cal Y}^3 = r\,\sin{\theta_1}\sin{\phi_1}\;, \hspace{0.9cm} \nonumber \\[2pt]
 {\cal Z}^1= \sqrt{1-r^2}\,\cos{\theta_2} \, ,\quad {\cal Z}^2 = \sqrt{1-r^2}\,\sin{\theta_2}\cos{\phi_2} \,, \quad  {\cal Z}^3 = \sqrt{1-r^2}\,\sin{\theta_2}\sin{\phi_2} \, . \qquad 
\end{eqnarray}
This is the type IIB counterpart of the fact that the $D=4$ Family III reduces to the four-dimensional $\cN=4$ SO(4) vacuum of \cite{Gallerati:2014xra} when $\chi = 0$. For this value of $\chi$, the brackets in the internal portion of the ten-dimensional metric (\ref{IIBMetricFamIII}) become the round metrics on two two-spheres, $S_i^2$, $i=1,2$. In turn, the two-forms (\ref{eq:TwoFormsIIB}) become the volume forms $\textrm{vol} ( S_i^2)$, up to overall functions of $r$. The $\chi =0$ metric on $S^5$ is thus a deformation of the round, Einstein metric on the join of the two $S^2_i$, $i=1,2$, such that only the $\textrm{SO}(4) \sim \textrm{SU}(2)_1 \times \textrm{SU}(2)_2$ subgroup of SO(6) in (\ref{eq:Embedding1}) is preserved. Each $\textrm{SU}(2)_i$ rotates each $S^2_i$, for $i=1,2$. The SO(2) isometry of $S^1$ is broken by the supergravity fields.

When $\chi \neq 0$, the symmetry of the solution (\ref{IIBMetricFamIII})--(\ref{eq:AuxThreeForm}) generically reduces to the $\textrm{U}(1)_1 \times \textrm{U}(1)_2$ defined in (\ref{eq:Embedding1}), with U$(1)_i$ generated by $\partial_{\phi_i}$ for $i=1,2$. Equivalently, the generic symmetry when $\chi \neq 0$ is the $\textrm{U}(1)_R \times \textrm{U}(1)_F$ group generated by the diagonal and anti-diagonal combinations
\begin{equation}
\partial_R = \partial_{\phi_1} + \partial_{\phi_2} \; , \qquad
\partial_F = \partial_{\phi_1} - \partial_{\phi_2} \; ,
\end{equation}
as specified below (\ref{eq:Embedding1}). Interestingly, the change of coordinates
\begin{equation}	\label{eq: coordchange}
	\phi_1  \,  \longrightarrow \,  \phi_1^\prime =\phi_1  -\chi\eta \; , \qquad
	\phi_2  \,  \longrightarrow \,  \phi_2^\prime =\phi_2  +\chi\eta \; , 
\end{equation}
with $\eta$, $r$, $\theta_i$, $i=1,2$, untouched, can be used to eliminate $\chi$ locally from the solution. Generically, though, the change (\ref{eq: coordchange}) is not globally well defined, {\it i.e.} is not a diffeomorphism, and does not generically allow one to eliminate $\chi$ globally. For specific values of $\chi$, the change (\ref{eq: coordchange}) {\it is} globally well defined: these are the values that render $\phi_i^\prime$ periodic, $\phi_i^\prime \sim \phi_i^\prime +2 \pi$. Given the periods (\ref{eq:Periodicities}) of the original coordinates, this induces a periodic identification $\chi \sim \chi +2\pi/T$ such that, for $\chi = 2\pi n^\prime / T$, with $n^\prime$ integer, the solution (\ref{IIBMetricFamIII})--(\ref{eq:AuxThreeForm}) becomes diffeomorphic to the $\chi=0$, $\cN=4$ SO(4) solution. This explains the periodic symmetry enhancements of the spectrum on Family III observed in section \ref{sec:FamIIISpecttrum}. 

\newpage


\section{Final comments} \label{sec:Concs}


In this paper we have studied the structure of the KK spectrum above a class of AdS$_4$ solutions of type IIB string theory of non-geometric, S-fold type. We have done this by a combination of well established group theory methods and new KK spectral techniques \cite{Malek:2019eaz,Malek:2020yue,Cesaro:2020soq} derived from the ExFT \cite{Hohm:2013pua,Godazgar:2014nqa} reformulation of the higher-dimensional supergravities. For these group theory and ExFT methods to be applicable for this type of solutions, it is crucial that they arise upon consistent uplift of vacua of $D=4$ $\cN=8$ supergravity with a suitable gauging. The fully uplifted higher-dimensional solutions need not be explicitly known in order to extract their KK spectrum using these tools. In fact, the class of AdS solutions of interest in this paper is only known completely at the $D=4$ gauged supergravity level \cite{Bobev:2021yya}. Only the type IIB uplifts of particular subsets in this class are known: see
\cite{Inverso:2016eet,Guarino:2020gfe,Giambrone:2021zvp} and section \ref{eq:FamilyIIISfold} above. While the uplifted solutions are not needed in order to compute their spectrum, they are still helpful to explain certain features of these spectra that are invisible in gauged supergravity. These include the compactness of one of the parameters that characterise the family of solutions, which is inherited from the global properties of the type IIB uplifts.

In the case at hand, the relevant gauging of $D=4$ $\cN=8$ supergravity is that with dyonic $ [\textrm{SO} (6) \times \textrm{SO} (1,1) ] \ltimes \mathbb{R}^{12}$ \cite{DallAgata:2011aa,DallAgata:2014tph} gauge group. The scalar potential of this gauged supergravity has AdS vacua that tend to come in critical loci, rather than critical points, on the $\cN=8$ scalar manifold $\textrm{E}_{7(7)}/\textrm{SU}(8)$ \cite{Guarino:2019oct,Guarino:2020gfe,Bobev:2021yya}. Of course, these loci occur at fixed cosmological constant, but may display symmetry or supersymmetry enhancements at selected points. This feature of critical loci versus critical points distinguishes $ [\textrm{SO} (6) \times \textrm{SO} (1,1) ] \ltimes \mathbb{R}^{12}$ from other $\cN=8$ gaugings, like SO(8) \cite{deWit:1982ig} or ISO(7) \cite{Guarino:2015qaa}, with similar higher-dimensional origins. Since they have fixed cosmological constant, leading to fixed free energies on the boundary, these loci are amenable to holographic interpretation as CMs, with the $D=4$ scalars that parameterise the loci interpreted as the marginal couplings in the dual CFT. The present setting thus provides a rare instance where such CMs can be explored from lower-dimensional gauged supergravity and its (comparatively) simple uplift to higher-dimensions. For other gaugings like SO(8) or ISO(7), the CMs of the CFTs dual to critical points can still be assessed holographically, see~{\it e.g.}~\cite{Bobev:2021gza}, but using other methods \cite{Lunin:2005jy,Ashmore:2016oug}.

The two-parameter family of AdS vacua \cite{Bobev:2021yya} relevant to this paper contains the $\cN=4$ critical point \cite{Gallerati:2014xra} of the gauging at hand. Its type IIB S-fold uplift was constructed in \cite{Inverso:2016eet} and a dual CFT candidate was put forward in \cite{Assel:2018vtq}. This type of S-fold solutions also arises as a limiting cases of certain supersymmetric Janus solutions in the class of \cite{DHoker:2007zhm,DHoker:2007hhe} constructed in \cite{Bobev:2020fon}. The $\cN=4$ CFTs dual to this type of Janus solutions have been described in  \cite{Assel:2011xz,Assel:2012cj} building on \cite{Gaiotto:2008sd}, and their $\cN=2$ CMs have also been recently studied in \cite{Bachas:2017wva,Bachas:2019jaa}. It would be interesting to  track back the limiting process and determine what aspects of our KK spectra on the holographic $\cN=2$ CM of the $\cN=4$ S-fold CFT of \cite{Assel:2018vtq} extrapolate to the Janus CMs of \cite{Bachas:2017wva,Bachas:2019jaa}. Does, for example, the protected S-fold spectrum of table \ref{tab:ShortN=2Spectrum} extrapolate to the Janus CM? In this regard, it should be noted that our short spectrum is protected in the sense discussed in section \ref{eq:ConfMan} that it does not depend on the CM moduli. However, it is not {\it absolutely} protected in the sense of~{\it e.g.}~\cite{Cordova:2016xhm}: there are no genuinely short multiplets in our spectrum, all of them may recombine into long multiplets at the unitarity bound. Extending the superconformal index results of \cite{Garozzo:2019ejm,Beratto:2020qyk} into the large-$N$ regime corresponding to our supergravity results would be helpful to shed light on this issue.

We have provided extensive evidence that one of the directions, parameterised by $\chi$ in our conventions, of the large-$N$ conformal manifold is compact. However, the direction along the other coordinate, $\varphi$, is non-compact. Everywhere within the moduli ranges (\ref{eq: cmband}) is the family of AdS$_4$ solutions smooth and their KK spectra well behaved. The singular locus $e^{2\varphi}=0$ is at infinite distance of any interior point in the CM with respect to the leading order Zamolodchikov metric (\ref{eq:cmmetric}). Thus, at least at leading order in the gauge group rank $N$, the CM is non-compact. Further, our KK analysis is consistent with a strong coupling description of the dual CFT across the entire CM, with dimensions that nowhere reduce to free-field values. For these reasons, our results do not lend support to the `CFT distance conjecture' recently put forward in \cite{Perlmutter:2020buo}. According to the latter, the CMs of three-dimensional $\cN=2$ CFTs should be expected to be compact, or else become free at infinite geodesic distance with respect to the Zamolodchikov metric. Interestingly, the $e^{2\varphi} \rightarrow 0$ limit of the $\ell=0$ spectrum discussed in (\ref{eq:Asymptoticell=0Alln}) exhibits infinite towers of flavour-neutral massless scalars, vectors and gravitons,  along with flavoured modes becoming infinitely massive. Since the flavour charges are controlled by the $S^5$ quantum numbers and are independent of $S^1$, this suggests that the modulus $\varphi$ controls the radii of AdS$_4$, $S^5$ and $S^1$, in such a way that the $e^{2\varphi} \rightarrow 0$ limit corresponds to a decompactification of $S^1$ while AdS$_4$ and $S^5$ both become highly curved well into the stringy regime. In particular, the $S^1$ decompactification should be responsible for the existence in this limit of infinite towers of massless lower-spin fields, in agreement with the more general `swampland distance conjecture' of \cite{Ooguri:2006in}, but without the need to invoke the existence of the infinite towers of massless higher spin fields that are central to \cite{Perlmutter:2020buo}. 


\section*{Acknowledgements}


A previous preprint version of this article overlooked part of the range (\ref{eq: cmband}) above
for the parameter $\varphi$. We thank Nikolay Bobev, Jerome Gauntlett, Fri$\eth$rik Freyr Gautason and Jesse van Muiden for bringing this to our attention. We would also like to thank Thomas Dumitrescu, Leonardo Rastelli, Cumrun Vafa and Irene Valenzuela for correspondence, and Emilio Ambite for technical support on our cluster computations. The latter were performed on the Hydra cluster of IFT. MC is supported by a La Caixa Foundation (ID 100010434) predoctoral fellowship LCF/ BQ/DI19/11730027. GL is supported by an FPI-UAM predoctoral fellowship and a Spain-US Fulbright scholarship. OV is supported by the NSF grant PHY-2014163. All of us are partially sup\-por\-ted by grants RYC-2015-18741, SEV-2016-0597 and PGC2018-095976-B-C21 from MCIU/AEI/FEDER, UE.


\newpage

\appendix

\addtocontents{toc}{\setcounter{tocdepth}{1}}


\section{Ancillary files: numerical spectrum across the CM}	\label{sec:attachment}


This article comes with three companion Wolfram Mathematica files,
\begin{center}
\texttt{KKSpectrum.nb} \; , \quad
\texttt{KKSpectrum$\_$CM.wl}  \; ,  \quad
\texttt{KKSpectrum$\_$FamilyII.wl}  \; ,
\end{center}
which provide a numerical database for the first few KK levels of the spectrum at a grid of locations in the CM. The first of these files, with \texttt{nb} extension, provides a user interface, while the last two \texttt{wl} files contain data used by the former. All three files must be downloaded into the same local folder before the \texttt{nb} file can be executed. The data contained in the \texttt{wl} files must be loaded into memory by running the \texttt{Get} commands in the \texttt{nb} file.

The database can be accessed by executing from the \texttt{nb} file either of the following two functions, with syntax:
\begin{center}
\texttt{KKSpecCM}$[ e^{2\varphi} , \chi , \ell , n ]$ \; , \quad
\texttt{KKSpecFamII}$[ e^{2\varphi} , \ell , n ]$ \; .
\end{center}
These respectively provide the spectrum of OSp$(4|2)$ multiplets on the interior or boundary point  $(e^{2\varphi} , \chi)$ in the CM, or point $e^{2\varphi}$ on Family II, (\ref{eq:FamII}) with $n^\prime =0$ therein, at $S^5$ and $S^1$ KK levels $\ell$ and $n$. The KK levels must be contained in $\ell\in\{0,1,2,3\}$ and $n\in\{0,\pm1,\pm2,\pm3\}$. For the function \texttt{KKSpecCM}, the argument $e^{2\varphi}$ can be any number contained in $0.05\leq e^{2\varphi}\leq2$ in steps of $\Delta(e^{2\varphi})=0.05$, while $\chi$ must be in the range $0\leq \chi\leq1$, in steps of $\Delta\chi=0.05$. For the function \texttt{KKSpecFamII}, $e^{2\varphi}$ ranges in $0.01\leq e^{2\varphi}\leq2$ and the steps can be taken as short as $\Delta(e^{2\varphi})=0.01$. In both cases, the internal $S^1$ radius is fixed to $T=2\pi$.

The output of both functions is the list of OSp$(4|2)$ multiplets present in the KK spectrum at the specified KK levels and location on the CM, or Family II. More concretely, the functions provide the eigenvalues of the KK mass matrices at the requested KK level on the selected CM point, translated into conformal dimensions, and repacked into supermultiplets of OSp$(4|2)$. The functions do not keep track of the R- or flavour charges of these multiplets, and only tally up their multiplicities. For this reason, the output OSp$(4|2)$ multiplets are displayed as
\begin{center}
\texttt{MULT[$\mathtt{E_0}$,deg]} \ , 
\end{center}
rather than as in (\ref{eq:OSpMultiplets}). Here, \texttt{MULT} is one of the acronyms for the OSp$(4|2)$ multiplets specified in appendix A of \cite{Klebanov:2008vq}, which we also use in the main text, $\mathtt{E_0}$ is the dimension and \texttt{deg} the multiplicity. The spectrum at both the original, $(e^{2\varphi} =1, \chi =0 )$, and additional, $(e^{2\varphi} =1, \chi =1 )$, $\cN=4$ points is displayed in $\cN=2$ multiplets via the branchings (\ref{eq: N4toN2branching}). Also, the multiplet shortenings that occur at $\chi=0$ (but, for simplicity, not at $\chi = \frac12$ or $\chi = 1$) are explicitly indicated using the corresponding acronyms. Finally, if the input arguments do not meet the above specifications, an error message is printed. 

Our calculations were performed with Mathematica's default machine precision of fifteen decimal places. For simplicity of presentation, the database contained in the \texttt{wl} file is truncated to eight digits.



\section{Further details on the KK spectra} \label{sec:KKmaterials}


Two main elements intervene in the determination of the KK spectra presented in section \ref{sec:KKSpectra}: the underlying algebraic structure and the eigenvalue problem itself. Further details on both aspects are provided in appendices \ref{sec:PutativeSO6SO2} and \ref{sec:KKMassMat}, respectively. 


\subsection{Putative $\textrm{SO}(6)_v \times \textrm{SO}(2)$ structure of the KK spectra} \label{sec:PutativeSO6SO2} 


The vacua of $D=4$ $\cN=8$ supergravity with $[\textrm{SO} (6) \times \textrm{SO} (1,1) ] \ltimes \mathbb{R}^{12}$ gauging uplift to type IIB S-fold solutions of the form AdS$_4 \times S^5 \times S^1$. The SO(2) isometry of $S^1$ is broken and the metric on $S^5$ has an isometry group that contains the symmetry group $G \subset \textrm{SO}(6)_v$ of the corresponding $D=4$ vacuum. The KK spectra of these type IIB solutions is thus labelled by two independent KK levels $\ell$ and $n$, ranging as in (\ref{eq:KKlevels}), respectively related to the internal $S^5$ and $S^1$. The individual states of definite spin in these spectra come in representations of $G$. The modes lying at the bottom of the KK towers, $\ell=0$, $n=0$, correspond to the linearisation of the $D=4$ $\cN=8$ gauged supergravity fields. In particular, the $G$ representations present in the $\ell = n = 0$ spectrum are obtained by branching the $D=4$ $\cN=8$ fields under $\textrm{SU}(8) \supset \textrm{SO}(8) \supset \textrm{SO} (6)_v \times \textrm{SO} (2) \supset G$. See table \ref{tab:so6xso2KKirreps} (upper) for the list of intermediate $\textrm{SO} (6)_v \times \textrm{SO} (2)$ representations that appear in this branching. In turn, the actual masses can be computed by linearising the $D=4$ $\cN=8$ gauged supergravity around the vacuum under consideration.

 \begin{table}[h]

\centering

\resizebox{.85\textwidth}{!}{

\begin{tabular}{|lllll|}
\hline
 spin & &SO(6)$_v\times$SO(2) irrep& & SO(6)$_v\times$SO(2) Dynkin labels
            \\ \hline
$2$ && $\bm{1}_0$&& $[0,0,0]_0$ \\[4pt]
 $\frac32$ && $\bm{4}_1+\bar{\bm{4}}_{-1}$&& $[1,0,0]_1+[0,0,1]_{-1}$ \\[4pt]
 $1$ && $\bm{15}_0+\bm{1}_0+\bm{6}_2+\bm{6}_{-2}$&& $[1,0,1]_0 + [0,0,0]_0+[0,1,0]_2+[0,1,0]_{-2} $ \\[4pt]
 $\frac12$ && $\bar{\bm{20}}_{-1}+\bm{20}_1+\bm{4}_1+\bm{4}_{-3}+\bar{\bm{4}}_{3}+\bar{\bm{4}}_{-1}$&&  $[1,1,0]_{-1} + [0,1,1]_{1} + [1,0,0]_{1} + [1,0,0]_{-3} + [0,0,1]_{3} + [0,0,1]_{-1} $ \\[4pt]
 $0^+$ && $\bm{20'}_0+\bm{1}_4+\bm{1}_0+\bm{1}_{-4}$&& $[0,2,0]_0+[0,0,0]_4+[0,0,0]_0+[0,0,0]_{-4}$ \\[4pt]
 $0^-$ && $\bm{15}_0+\bm{10}_{-2}+\bar{\bm{10}}_{2}$&& $[1,0,1]_0+[2,0,0]_{-2}+[0,0,2]_2$   \\
 \hline
\end{tabular}
}
\bigskip

\resizebox{\textwidth}{!}{
\begin{tabular}{|lll|}
\hline
 spin  & & SO(6)$_v\times$SO(2) Dynkin labels
            \\ \hline
$2$ && $[0,\ell,0]_{2n}$ \\[6pt]
 $\frac32$ &&  $[1,\ell,0]_{2n+1}+[0,\ell-1,1]_{2n+1}+[0,\ell,1]_{2n-1}+[1,\ell-1,0]_{2n-1}$ \\[6pt]
 $1$ &&  $[1,\ell,1]_{2n} + [2,\ell-1,0]_{2n}+ [0,\ell-1,2]_{2n} +[1,\ell-2,1]_{2n} + [0,\ell,0]_{2n}$ \\[3pt]
 && \quad $+[0,\ell+1,0]_{2n+2} + [1,\ell-1,1]_{2n+2}+ [0,\ell-1,0]_{2n+2} +[0,\ell+1,0]_{2n-2} + [1,\ell-1,1]_{2n-2}+ [0,\ell-1,0]_{2n+2}$ \\[6pt]
 $\frac12$ &&   $[1,\ell+1,0]_{2n-1} + [0,\ell,1]_{2n-1}+ [2,\ell-1,1]_{2n-1} +[1,\ell-1,0]_{2n-1} + [1,\ell-2,2]_{2n-1} + [0,\ell-2,1]_{2n-1} $ \\[3pt]
 &&   \quad+ $[0,\ell+1,1]_{2n+1} + [1,\ell,0]_{2n+1}+ [1,\ell-1,2]_{2n+1} +[0,\ell-1,1]_{2n+1} + [2,\ell-2,1]_{2n+1} + [1,\ell-2,0]_{2n+1} $ \\[3pt]
  &&   \qquad+ $[1,\ell,0]_{2n-3} + [0,\ell-1,1]_{2n-3}+ [0,\ell,1]_{2n+3} +[1,\ell-1,0]_{2n+3} $ \\[6pt]
 $0^+$ &&  $[0,\ell+2,0]_{2n} + [1,\ell,1]_{2n}+ [0,\ell,0]_{2n} +[2,\ell-2,2]_{2n}+ [1,\ell-2,1]_{2n}+ [0,\ell-2,0]_{2n}+ [0,\ell,0]_{2n+4}+ [0,\ell,0]_{2n-4} $ \\[6pt]
 $0^-$ &&  $[2,\ell,0]_{2n-2} + [1,\ell-1,1]_{2n-2}+ [0,\ell-2,2]_{2n-2}+[0,\ell,2]_{2n+2} + [1,\ell-1,1]_{2n+2}+ [2,\ell-2,0]_{2n+2} $   \\
 \hline
\end{tabular}
}

\caption{\footnotesize{States in SO(6)$_v\times$SO(2) representations at KK levels $(\ell,n)=(0,0)$ (above) and $\ell=0 , 1, 2 , \ldots$, $n\in\mathbb{Z}$ (below) in the KK towers for AdS$_4$ solutions of type IIB that uplift from $D=4$ $\cN=8$ $[\textrm{SO} (6) \times \textrm{SO} (1,1) ] \ltimes \mathbb{R}^{12}$--gauged supergravity. SO(6)$_v$ representations are given in terms of SU(4) Dynkin labels, and SO(2) charges are given as subscripts. Representations with negative Dynkin labels are absent and need to be crossed out. For a solution with residual symmetry $G \subset \textrm{SO}(6)_v$, the spectrum organises itself in the representation of $G$ that branch from these SO(6)$_v$ representations. The tables exclude some $0^+$ states of $D=4$ supergravity that are always Higgsed away.}  \normalsize}
\label{tab:so6xso2KKirreps}
\end{table}

At higher levels $\ell \geq 0$, $\vert n\vert \geq 0$, the individual KK modes also come in representations of $G$ that still follow a rigid pattern. The KK gravitons at levels $(\ell , n)$ have SO(2) charge $2n$ and are in the $G$ representations that result from branching the order-$\ell$ symmetric traceless representation of SO$(6)_v$, $[0,\ell,0]$ (denoted here and elsewhere with SU(4) Dynkin labels) under $G$. More generally, the algebraic structure of the spectrum at KK levels $(\ell, n)$ is obtained by the following two-step process. Firstly, tensor the $\ell=n=0$ gauged supergravity fields (branched-out into $\textrm{SO}(6)_v \times \textrm{SO}(2)$ representations as above) with the $[0,\ell,0]_{2n}$ representation of $\textrm{SO}(6)_v \times \textrm{SO}(2)$ and remove Goldstones and Goldstinos. The resulting $\textrm{SO}(6)_v \times \textrm{SO}(2)$ content is summarised for convenience in table \ref{tab:so6xso2KKirreps} (lower). Secondly, branch under $\textrm{SO}(6)_v \supset G$. This algorithm has already been applied in the same context in \cite{Giambrone:2021zvp} and is analogue to that specified previously in \cite{Englert:1983rn,Varela:2020wty} to obtain the algebraic structure of the KK spectra of AdS$_4$ solutions that uplift from the SO(8) and ISO(7) gaugings of $D=4$ $\cN=8$ supergravity. 

At this stage, having obtained the content of the spectrum at KK levels $(\ell,n)$ in terms of $G$ representations, one must still compute the masses of the individual states. This can be done by diagonalising the KK mass matrices recently derived from ExFT \cite{Hohm:2013pua,Godazgar:2014nqa} in \cite{Malek:2019eaz,Malek:2020yue,Cesaro:2020soq}. The particularisation of these mass matrices to the present context is summarised in appendix~\ref{sec:KKMassMat}. Finally, for solutions preserving $\cN$ supersymmetries, these individual KK states must be further sorted into representations of $\textrm{OSp} (4|\cN) \times G^\prime$ with superconformal primary dimensions related to the relevant KK masses. The $\textrm{OSp} (4|\cN)$ factor here contains the subgroup of $G$ corresponding to the R-symmetry, while $G^\prime$ is identified with any leftover flavour symmetry. As noted in the text, for the cases at hand with $\cN=2$ and $\cN=4$, all states can be arranged in long supermultiplets of $\textrm{OSp} (4|\cN)$.

This lenghty process allows us to assess the KK spectrum above the class of AdS$_4$ solutions reviewed in section \ref{eq:ConfMan}, the results of which we have discussed in section \ref{sec:KKSpectra}. The supermultiplet structure of the spectrum reported in section \ref{sec:N=4Specttrum} is perfectly compatible with the multiplicities and values of the individual mass states that we have explicitly computed for the first few KK levels. We take this compatibility as a solid self-consistency test on our results, given the very different methods to obtain these two aspects of the KK spectra. The multiplet structure was determined by the group theory methods specified above, while the explicit mass states up the KK towers were obtained by diagonalising the first few levels of the KK mass matrices of \cite{Malek:2019eaz,Malek:2020yue,Cesaro:2020soq}. Next, we comment on our implementation of the latter techniques.


\subsection{KK mass matrices} \label{sec:KKMassMat}


To compute the masses of the individual states in the spectra we have diagonalised the ExFT-derived KK mass matrices as given in \cite{Cesaro:2020soq,Varela:2020wty}. These mass matrices feature the $D=4$ $\cN=8$ $\textrm{E}_{7(7)}/\textrm{SU}(8)$ coset representative and the embedding tensor for the gauging at hand, $ [\textrm{SO} (6) \times \textrm{SO} (1,1) ] \ltimes \mathbb{R}^{12}$, see \cite{DallAgata:2011aa} for the latter. All the information that the KK mass matrices carry about  the internal $S^5 \times S^1$ in type IIB is brought in by a collection of matrices $\mathcal{T}_{\underline{M}}$, labelled by an index $\underline{M} = 1 , \ldots , 56$ in the fundamental of E$_{7(7)}$, which encode the generators of $\textrm{SO}(6)_v \times \textrm{SO}(2)$ in the infinite-dimensional, reducible representation 
\begin{equation} \label{eq:SymTrac}
\oplus_{\ell=0}^\infty \oplus_{n= -\infty}^\infty \,  [0,\ell,0]_{2n} \; .
\end{equation}
Here as in the main text, we have used SU(4) Dynkin labels to denote the rank-$\ell$ symmetric traceless representation of SO$(6)_v$. Expressions for these $\mathcal{T}_{\underline{M}}$'s have been previously given in \cite{Giambrone:2021zvp}. In our conventions, these can be specified with the following double-index structure
\begin{equation} \label{eq:curlyTs}
	(\mathcal{T}_{\underline{M}})_{\Lambda c}{}^{\Sigma d}=(\mathcal{T}_{\underline{M}})_{\Lambda}{}^{\Sigma}\;\delta_{c}{}^{d}+\delta_{\Lambda}{}^{\Sigma}\;(\mathcal{T}_{\underline{M}})_{c}{}^d\; ,
\end{equation}
tailored to (\ref{eq:SymTrac}). The index $c=1,2$ is rotated by SO(2), while $\Lambda$ ranges in $\oplus_{\ell=0}^\infty \,  [0,\ell,0] $ so that, if $I=1 , \ldots , 6$ is a fundamental SO$(6)_v$ index, 
\begin{equation} \label{eq:SigmaIndex}
\Lambda =  \big( 1, I_1  \, , \,  \{I_1 I_2 \}  \, , \, \ldots ,  \{I_1 \ldots I_\ell  \} \, , \, \ldots  \big) \; , 
\end{equation}
where curly brackets denote traceless symmetrisation. The matrices on the r.h.s.~of (\ref{eq:curlyTs}) can in turn be defined as follows. Introducing SL$(8,\mathbb{R})$ fundamental indices $A=1 , \ldots , 8$, we have
\begin{equation} \label{eq:TsSplit}
({\cal T}_{\underline{M} })_\Lambda{}^\Sigma = \left( ({\cal T}_{AB})_\Lambda{}^\Sigma \; , \; ({\cal T}^{AB})_\Lambda{}^\Sigma \equiv 0 \right) \; , \qquad
({\cal T}_{\underline{M} })_c{}^d = \left( ({\cal T}_{AB})_c{}^d \equiv 0 \; , \; ({\cal T}^{AB})_c{}^d  \right) \; ,
\end{equation}
with $({\cal T}_{AB})_\Lambda{}^\Sigma =({\cal T}_{[AB]})_\Lambda{}^\Sigma $ specified by splitting the $\Lambda$, $\Sigma$ indices as in (\ref{eq:SigmaIndex}), as
\begin{equation} \label{eq:TABSymTr}
({\cal T}_{AB})_{K_1 \ldots K_\ell}{}^{L_1 \ldots L_\ell}   =   \ell \, (  {\cal T}_{AB})_{\{ K_1}{}^{ \{ L_1} \delta_{K_2}^{L_2} \ldots    \delta_{K_\ell\}}^{L_\ell\}} \; .
\end{equation}
Finally, the matrices $({\cal T}_{AB})_{  K}{}^{ L} =  ({\cal T}_{[AB]})_{  K}{}^{ L} $ in (\ref{eq:TABSymTr}) and $({\cal T}^{AB})_a{}^b = ({\cal T}^{[AB]})_a{}^b$ in (\ref{eq:TsSplit}) contain the generators of  $\textrm{SO}(6)_v$ and $\textrm{SO}(2)$ in the fundamental representation. Indeed, further splitting the indices $A=(I ,a)$ under $\textrm{SL}(8,\mathbb{R} ) \supset \textrm{SO}(8) \supset \textrm{SO}(6)_v \times \textrm{SO}(2)$, only the following matrices are non-zero,
\begin{equation}
( {\cal T}_{IJ})_{K}{}^{ L} \equiv 2 \, \delta_{[I}^{L} \delta_{J]K} \; , \qquad
( {\cal T}^{ab})_{c}{}^{ d} \equiv \tfrac{2\pi n}{T} \, \epsilon^{ab} \, \epsilon_c{}^d \; ,
\end{equation}
with all the rest, $( {\cal T}_{Ib})_{K}{}^{ L} $, $( {\cal T}_{ab})_{K}{}^{ L} $, $( {\cal T}^{IJ})_{c}{}^{ d}$ and $( {\cal T}^{Ib})_{c}{}^{ d}$, identically zero.

A powerful aspect of these ExFT spectral techniques is that explicit knowledge of the full higher dimensional AdS solution is not needed in order to obtain its KK spectrum. It is enough to know the solution as a vacuum of $D=4$ $\cN=8$ supergravity. We have explicitly constructed the KK mass matrices up to KK level $\ell=3$ and for all $n$ by bringing the coset representative $\textrm{E}_{7(7)}/\textrm{SU}(8)$ for the two-parameter family of $D=4$ vacua \cite{Bobev:2021yya} reviewed in section \ref{eq:ConfMan}, the relevant embedding tensor \cite{DallAgata:2011aa}, and the matrices ${\cal T}_{\underline{M}}$ specified above, to the KK mass matrices given in \cite{Cesaro:2020soq,Varela:2020wty}. We have obtained the mass spectrum for the individual states of definite spin. Translating these masses to conformal dimensions, we have been able to allocate these into $\textrm{OSp}(4|4)$, $\textrm{OSp}(4|2) \times \textrm{SU}(2)_F$ or $\textrm{OSp}(4|2) \times \textrm{U}(1)_F$ representations as appropriate, using the group representations computed in appendix \ref{sec:PutativeSO6SO2}. As emphasised above, the compatibility with the diagonalisation results with the group theory provides a reassuring self-consistency test.


\section{$\mathcal{N}=4$ supermultiplets} \label{OSp44SuperMult}


As explained in the main text, the algebraic structure of the KK spectrum across the holographic CM is inherited from that at the $\cN=4$ point. Thus, it is useful to collect some aspects of OSp$(4|4)$ representation theory. More concretely, in this appendix we give the explicit state content of the multiplets present in the $\cN=4$ spectrum reviewed in section \ref{sec:N=4Specttrum}. We also give some relevant shortening conditions and  branching rules under (\ref{eq:OSp44OSp22U(1)F}) into $\cN=2$ multiplets.

The general representation theory of OSp$(4|4)$ has been laid out in \cite{Cordova:2016emh}. The states that compose a given\footnote{As in the main text, we use the acronym  $\textrm{MULT}_4$ to refer to a generic multiplet of OSp$(4|4)$. We specifically denote long and short graviton and short gravitino multiplets as LGRAV$_4$, SGRAV$_4$ and SGINO$_4$. These respectively correspond to the multiplets denoted in \cite{Cordova:2016emh} as $L$ (with $j_{\text{there}}=0$), $A_2$ and $B_1$. Graviton and gravitino OSp$(4|4)$ multiplets have scalar, $s_0 = 0$, superconformal primaries, and gravitino multiplets are necessarily short.} $\textrm{MULT}_4$ representation of OSp$(4|4)$ carry definite SO(4) R-charges. In our conventions, these are labelled with half-integer Dynkin labels $(\ell_1,\,\ell_2)$. Unfortunately, the generic expressions for the OSp$(4|4)$ multiplet contents given in \cite{Cordova:2016emh} do not work well for scalar superconformal primaries or low values of $(\ell_1,\,\ell_2)$, where many states are actually absent and need to be sieved out. These are the cases relevant to our analysis. Here, we will determine the state content of the OSp$(4|4)$ long graviton multiplet (\ref{eq:LGRAV4}) for all possible values of the Dynkin labels on a case-by-case basis. Only multiplets with integer $(\ell_1,\,\ell_2)$ enter the KK spectra of interest in this paper. Once we got down to business though, it only took a finite amount of additional pain to get the strictly half-integer cases as well. Similar remarks apply to the OSp$(4|3)$ representation theory contained in \cite{Cordova:2016emh}: see appendix B of \cite{Varela:2020wty} for complete listings.

It is useful to start by listing the possible Lorentz spins, $[s] = 0 , \tfrac12 , 1 , \tfrac32 , 2$, and SO(4) Dynkin labels, $(\ell_1,\,\ell_2)$, that subsequent powers $Q^p$, $p=0, 1 , \ldots , 8$, of the OSp$(4|4)$ supercharge $Q$ may have. The result is:
{\setlength\arraycolsep{3.5pt}
\begin{eqnarray} \label{eq:MULT4BuildingBlock}
	1\ \&\ Q^8		&:&	[0] \qquad (0,0)\;,						\nonumber	\\[3mm]
	Q\ \&\ Q^7		&:&	[\tfrac12] \qquad \big(\tfrac12,\tfrac12\big)\;,	\nonumber	\\[3mm]
	Q^2\ \&\ Q^6	&:&	\begin{cases}	[1] \qquad (1,0)+(0,1)\;,					\\[1mm]
								[0] \qquad (1,1)+(0,0)\;,
					\end{cases}							\nonumber	\\[3mm]
	Q^3\ \&\ Q^5	&:&	\begin{cases}	[\tfrac32] \qquad \big(\tfrac12,\tfrac12\big)\;,		\\[1mm]
								[\tfrac12] \qquad \big(\tfrac32,\tfrac12\big)+\big(\tfrac12,\tfrac32\big)+\big(\tfrac12,\tfrac12\big)\;,
					\end{cases}										\\[3mm]
	Q^4			&:&	\begin{cases}	[2] \qquad (0,0)\;,						\\[1mm]
								[1] \qquad (1,1)+(1,0)+(0,1)\;,				\\[1mm]
								[0] \qquad (2,0)+(0,2)+(1,1)+(0,0) \; .
					\end{cases}							\nonumber
\end{eqnarray}
}%
Together with the fact that the action with $Q$ increases the conformal dimension by $\frac12$, the information summarised in (\ref{eq:MULT4BuildingBlock}) is the basic building block to find out the state content of the long graviton multiplets (\ref{eq:LGRAV4}) of OSp$(4|4)$. For a (scalar) superconformal primary with Lorentz and SO(4) spins $s_0=0$ and $(\ell_1 , \ell_2 )$ and dimension $E_0$, the descendants have all possible Lorentz spins $[s]$ shown in (\ref{eq:MULT4BuildingBlock}), and lie in the SO(4) representations that result from tensoring row by row the representations listed in (\ref{eq:MULT4BuildingBlock}) with $(\ell_1 , \ell_2 )$. Finally, the dimension of the $p$-th descendant is $E_0+\frac{p}{2}$. 

For easy reference, the outcome of this exercise for all $(\ell_1 , \ell_2)$, with $\ell_1\geq\ell_2$ without loss of generality, is listed in tables~\ref{table: long00}--\ref{table: longhighl1l2}. The table entries show the spin and SO(4) charges, in the format $[s]^{(\ell_1^\prime,\, \ell_2^\prime)}$, of each possible state in the multiplet. The corresponding dimensions $\Delta$ are given next to each entry, and these are grouped as descendants of the superconformal primary at the top of each table. An entry of the form $[s]^{(\ell_1\pm a,\ell_2\pm b)}$ denotes four states in total (this differs from the convention adopted, in a different context, in the main text: see below (\ref{eq:KKFamIIIell=0nneq0})). Also, negative Dynkin labels are not allowed, and the corresponding states must be removed as they are actually absent. These spurious states only occur in tables \ref{table: longl0}--\ref{table: longhighl1l2}. Table \ref{table: longhighl1l2} is valid at face value for all  $\ell_1,\ell_2\geq2$, with all entries therein present. The same table is also valid for  $\ell_i=\tfrac32$ for either or both $i=1,2$, but the states at level $Q^4$ with negative Dynkin labels need to be discarded. Similar comments apply to tables \ref{table: longl0}--\ref{table: longl1}. Tables~\ref{table: long00}--\ref{table: longl1} involve fewer states compared to  table~\ref{table: longhighl1l2} and, without going through the constructive algorithm specified above, it is not obvious which states must be crossed out in table \ref{table: longhighl1l2} to recover tables \ref{table: long00}--\ref{table: longl1}. Only tables \ref{table: long00}, \ref{table: long10}, \ref{table: long11}, \ref{table: longl0}, \ref{table: longl1} and \ref{table: longhighl1l2} play a role in the KK spectra described in this paper. The remaining tables necessarily involve strictly half-integer SO(4) labels for the superconformal primary and are only included for completeness.

The dimensions $E_0$ and Dynkin labels $(\ell_1 , \ell_2 ) $ of the (superconformal primary of the) OSp$(4|4)$  multiplets constructed with the above algorithm must respect the unitarity bound
\begin{equation} \label{eq: unitaritybound}
	E_0\geq s_0+\ell_1+\ell_2+1\;,
\end{equation}
with $s_0 =0$ for the graviton multiplets listed in the tables. The multiplets undergo shortening when the bound is saturated, in which case they split into short graviton and gravitino multiplets as:
\begin{equation} \label{eq: recombN4}
	\lgrav_4[\ell_1+\ell_2+1 ,\ell_1,\,\ell_2]  \rightarrow \sgrav_4[ \ell_1+\ell_2+1 , \ell_1,\,\ell_2]+\sgino_4[\ell_1+\ell_2+3, \ell_1+1,\,\ell_2+1]\;.
\end{equation}
See \cite{Cordova:2016emh} for the state contents of these $\cN=4$ short multiplets for specific values of $(\ell_1 , \ell_2)$. 

It is also useful to give the splitting of the above $\cN=4$ graviton multiplets under the supergroup embedding (\ref{eq:OSp44OSp22U(1)F}) into $\cN=2$ multiplets of definite $\textrm{U}(1)_F$ flavour charge. The $\textrm{U}(1)_R$ R-symmetry group of $ \textrm{OSp}(4|2)$ and $\textrm{U}(1)_F$ are the subgroups of the SO(4) R-symmetry of $\textrm{OSp}(4|4)$ specified in (\ref{eq:Embedding1}) and below that equation. Branching accordingly the SO(4) representations of the states in tables ~\ref{table: long00}--\ref{table: longhighl1l2}, recombining them into $\textrm{OSp}(4|2)$ multiplets using the $\cN=2$ tables of appendix A of \cite{Klebanov:2008vq}, and keeping track of the flavour charges, we obtain

\newpage

{\setlength\arraycolsep{1.5pt}
\begin{eqnarray}	\label{eq: N4toN2branching}
	\lgrav_4&&\!\!\!\!\!\big[E_0,\ell_1,\ell_2\big]= \qquad \qquad  \nonumber \\
	\qquad	\qquad &&\bigoplus_{m_1=-\ell_1}^{\ell_1}\bigoplus_{m_2=-\ell_2}^{\ell_2}
	\Big\{
	\underline{ \lgrav}_2\big[E_0+1,y_{m_1m_2}; f_{m_1 m_2} \big] \nonumber\\[5pt]
     &\oplus& \underline{\lgino}_2\big[E_0+\tfrac12, y_{m_1m_2} ; f_{m_1 m_2} +1\big] 
     \oplus \underline{\lgino}_2\big[E_0+\tfrac12, y_{m_1m_2} ; f_{m_1 m_2}  -1\big] \nonumber \\[5pt]
&\oplus&\lgino_2\big[E_0+\tfrac32, y_{m_1m_2} ; f_{m_1 m_2} +1\big] 
\oplus\lgino_2\big[E_0+\tfrac32, y_{m_1m_2} ; f_{m_1 m_2}  -1\big] \nonumber \\[5pt]
	&\oplus& 		\underline{\lvec}_2 \big[E_0,y_{m_1m_2}; f_{m_1 m_2} \big]  \nonumber \\[5pt]
	&\oplus& \lvec_2\big[E_0+1, y_{m_1m_2} ; f_{m_1 m_2} +2 \big]
	\oplus \lvec_2\big[E_0+1, y_{m_1m_2} ; f_{m_1 m_2}  \big]\nonumber\\[5pt]
	&\oplus&\lvec_2\big[E_0+1,y_{m_1m_2} ; f_{m_1 m_2} -2 \big]
	\oplus\lvec_2\big[E_0+2, y_{m_1m_2} ; f_{m_1 m_2} \big]
  \Big\}\; ,
\end{eqnarray}
}with $y_{m_1m_2}$ and $f_{m_1m_2}$ given in (\ref{eqRFCharges}). When $E_0$ saturates the $\cN=4$ unitarity bound (\ref{eq: unitaritybound}), the $\cN=4$ multiplet on the l.h.s.~of (\ref{eq: N4toN2branching}) becomes short as in (\ref{eq: recombN4}), and the underlined $\cN=2$ multiplets on the r.h.s.~at $(m_1 = -\ell_1 , m_2 = -\ell_2)$ and $(m_1 = \ell_1 , m_2 = \ell_2)$ shorten as well. The case of interest to this paper has, in particular, $\ell_1$ and $\ell_2$ further restricted to be equal, as in (\ref{eq:ShorteningQNs}), in the short multiplets. With this further restriction, the relevant $\cN=2$ multiplets in (\ref{eq: N4toN2branching}) that undergo shortening are thus
{\setlength\arraycolsep{0pt}
\begin{eqnarray}\label{eq:shorteningpattern}
&&\lgrav_2\big[ \ell+2+\epsilon  , \pm \ell ; 0 \big] \rightarrow
\sgrav_2\big[\ell+2  , \pm \ell ; 0 \big] \oplus \sgino_2\big[\ell+\tfrac52  , \pm ( \ell + 1 ) ; 0 \big]\, ,	\nonumber \\[7pt]
&&\lgino_2\big[ \ell+ \tfrac32 +\epsilon , \pm \ell ; + 1  \big]\rightarrow
\sgino_2\big[ \ell+ \tfrac32  , \pm \ell ; + 1  \big] \oplus \svec_2\big[ \ell+ 2  , \pm ( \ell + 1 ) ; + 1  \big]\,,\nonumber \\[7pt]
&&\lgino_2\big[ \ell+ \tfrac32 +\epsilon , \pm \ell ; - 1  \big]\rightarrow
\sgino_2\big[ \ell+ \tfrac32  , \pm \ell ; - 1  \big] \oplus \svec_2\big[ \ell+ 2  , \pm ( \ell  + 1 ) ; - 1  \big]\,,\nonumber \\[7pt]
&&\lvec_2\big[ \ell+1 +\epsilon , \pm \ell ; 0 \big] \rightarrow
\svec_2\big[\ell+1  , \pm \ell ; 0 \big]\oplus \text{HYP}_2\big[\ell+2  , \pm ( \ell + 2 ) ; 0 \big] \, ,
\end{eqnarray}
}%
as in \eqref{eq:ShorteningQNsCM}.
Only the flavour-neutral short multiplets here make it to the list of protected multiplets in table \ref{tab:ShortN=2Spectrum}. The short flavoured multiplets appear accidentally in the spectra of the SO(4) and SU$(2)_F$ points, joining other multiplets into SO(4) and SU$(2)_F$ representations. An extreme case of the shortening conditions occurs when the graviton becomes massless. In this case, we have the following splitting of a massless $\cN=4$ graviton multiplet into $\cN=2$ massless ones:
\begin{equation} \label{eqMGRAV4and2}
			\text{MGRAV}_4[1,0,0]=\text{MGRAV}_2[2,0;0]\oplus\text{MGINO}_2[\tfrac32,0;\pm1]\oplus\text{MVEC}_2[1,0;0] \; . 
		\end{equation}

We conclude with the observation that the multiplicities, the (superconformal primary) U$(1)_R$ charge,  and the (overall) U$(1)_F$ flavour charge of the $\cN=2$ multiplets that compose  $\lgrav_4\big[E_0,\ell_1,\ell_2\big]$ according to (\ref{eq: N4toN2branching}) can be also retrieved in the following manner. Introducing fugacites $u$ and $x$ for U$(1)_R$ and U$(1)_F$, define for each multiplet on the r.h.s.~of (\ref{eq: N4toN2branching}) the functions

\newpage 

{\setlength\arraycolsep{2pt}
\begin{eqnarray}	\label{eq:fugacities}
	\nu_{\lgrav_2}^{E_0+1}&=&\nu_{\lvec_2}^{E_0}=\nu_{\lvec_2}^{E_0+2}=\frac{\big[1-\big(u x\big)^{2\ell_1+1}\big]\big[1-\big(\tfrac{u}{x}\big)^{2\ell_2+1}\big]}{\big(ux\big)^{\ell_1}\big(\tfrac{u}{x}\big)^{\ell_2}\big(1-ux\big)\big(1-\tfrac{u}{x}\big)}\;,\nonumber\\[8pt]
	\nu_{\lgino_2}^{E_0+\frac12}&=&\nu_{\lgino_2}^{E_0+\frac32}=\frac{(x+1)}{x}\frac{\big[1-\big(ux\big)^{2\ell_1+1}\big]\big[1-\big(\tfrac{u}{x}\big)^{2\ell_2+1}\big]}{\big(ux\big)^{\ell_1}\big(\tfrac{u}{x}\big)^{\ell_2}\big(1-ux\big)\big(1-\tfrac{u}{x}\big)}\;,\nonumber\\[8pt]
	\nu_{\lvec_2}^{E_0+1}&=&\frac{(x^2+x+1)}{x}\frac{\big[1-\big(ux\big)^{2\ell_1+1}\big]\big[1-\big(\tfrac{u}{x}\big)^{2\ell_2+1}\big]}{\big(ux\big)^{\ell_1}\big(\tfrac{u}{x}\big)^{\ell_2}\big(1-ux\big)\big(1-\tfrac{u}{x}\big)}\; ,
\end{eqnarray}
}with $\nu_{\lgrav_2}^{E_0+1}$ corresponding to the ${\lgrav_2}$'s with dimension $E_0+1$, etc. Expanding these functions at fixed $\ell_1$ and $\ell_2$ in powers of $u$ and $x$, the multiplicity $m$, R-charge $y_0$ and flavour charge $f$ of a multiplet can be read off from the term $m \, u^{y_0} x^f$ in the expansion of its associated function $\nu$.

\newpage

 \begin{table}[h]

\centering

\resizebox{0.56\textwidth}{!}{

\begin{tabular}{ccl}
\hline\\[-7pt]
		&	$E_0$			&	$[0]^{(0,\,0)}$			\\[9pt]
$Q$		&	$E_0+\tfrac12$		&	$[\tfrac12]^{\big(\tfrac12,\,\tfrac12\big)}$		\\[9pt]
$Q^2$	&	$E_0+1$			&	$[1]^{(1,\,0)}$	+ $[1]^{(0,\,1)}$			\\[5pt]
		&					&	$[0]^{(1,\,1)}$+$[0]^{(0,\,0)}$			\\[9pt]
$Q^3$	&	$E_0+\tfrac32$		&	$[\tfrac32]^{\big(\tfrac12,\,\tfrac12\big)}$		\\[5pt]
		&					&	$[\tfrac12]^{\big(\tfrac32,\,\tfrac12\big)}$+$[\tfrac12]^{\big(\tfrac12,\,\tfrac32\big)}$+$[\tfrac12]^{\big(\tfrac12,\,\tfrac12\big)}$	\\[9pt]
$Q^4$	&	$E_0+2$			&	$[2]^{(0,\,0)}$		\\[5pt]
		&					&	$[1]^{(1,\,1)}$+\,$[1]^{(1,\,0)}$+$[1]^{(0,\,1)}$	\\[5pt]
		&					&	$[0]^{(2,\,0)}$+ $[0]^{(0,\,2)}$+$[0]^{(1,\,1)}$+$[0]^{(0,\,0)}$\\[9pt]
$Q^5$	&	$E_0+\tfrac52$		&	$[\tfrac32]^{\big(\tfrac12,\,\tfrac12\big)}$		\\[5pt]
		&					&	$[\tfrac12]^{\big(\tfrac32,\,\tfrac12\big)}$+$[\tfrac12]^{\big(\tfrac12,\,\tfrac32\big)}$+$[\tfrac12]^{\big(\tfrac12,\,\tfrac12\big)}$	\\[9pt]
$Q^6$	&	$E_0+3$			&	$[1]^{(1,\,0)}$	+ $[1]^{(0,\,1)}$			\\[5pt]
		&					&	$[0]^{(1,\,1)}$+$[0]^{(0,\,0)}$			\\[9pt]
$Q^7$	&	$E_0+\tfrac72$		&	$[\tfrac12]^{\big(\tfrac12,\,\tfrac12\big)}$		\\[9pt]
$Q^8$	&	$E_0+4$			&	$[0]^{(0,\,0)}$		\\\hline
\end{tabular}
}

\caption{\footnotesize{States in the long graviton supermultiplet LGRAV$_4[E_0,0,0]$.}\normalsize}
\label{table: long00}
\end{table}
%
%
 \begin{table}[H]

\centering

\resizebox{0.82\textwidth}{!}{

\begin{tabular}{ccl}
\hline\\[-7pt]
		&	$E_0$			&	$[0]^{\big(\tfrac12,\,0\big)}$			\\[9pt]
$Q$		&	$E_0+\tfrac12$		&	$[\tfrac12]^{\big(1,\,\tfrac12\big)}$+$[\tfrac12]^{\big(0,\,\tfrac12\big)}$		\\[9pt]
$Q^2$	&	$E_0+1$			&	$[1]^{\big(\tfrac32,\,0\big)}$+$[1]^{\big(\tfrac12,\,0\big)}$+ $[1]^{\big(\tfrac12,\,1\big)}$			\\[5pt]
		&					&	$[0]^{\big(\tfrac32,\,1\big)}$+$[0]^{\big(\tfrac12,\,1\big)}$+$[0]^{\big(\tfrac12,\,0\big)}$			\\[9pt]
$Q^3$	&	$E_0+\tfrac32$		&	$[\tfrac32]^{\big(1,\,\tfrac12\big)}$+$[\tfrac32]^{\big(0,\,\tfrac12\big)}$		\\[5pt]
		&					&	$[\tfrac12]^{\big(2,\,\tfrac12\big)}$+2\,$[\tfrac12]^{\big(1,\,\tfrac12\big)}$+$[\tfrac12]^{\big(1,\,\tfrac32\big)}$+$[\tfrac12]^{\big(0,\,\tfrac32\big)}$+$[\tfrac12]^{\big(0,\,\tfrac12\big)}$	\\[9pt]
$Q^4$	&	$E_0+2$			&	$[2]^{\big(\tfrac12,\,0\big)}$		\\[5pt]
		&					&	$[1]^{\big(\tfrac32,\,1\big)}$+2\,$[1]^{\big(\tfrac12,\,1\big)}$+$[1]^{\big(\tfrac32,\,0\big)}$+$[1]^{\big(\tfrac12,\,0\big)}$	\\[5pt]
		&					&	$[0]^{\big(\tfrac52,\,0\big)}$+ $[0]^{\big(\tfrac32,\,0\big)}$+$[0]^{\big(\tfrac12,\,2\big)}$+$[0]^{\big(\tfrac32,\,1\big)}$+$[0]^{\big(\tfrac12,\,1\big)}$+$[0]^{\big(\tfrac12,\,0\big)}$\\[9pt]
$Q^5$	&	$E_0+\tfrac52$		&	$[\tfrac32]^{\big(1,\,\tfrac12\big)}$+$[\tfrac32]^{\big(0,\,\tfrac12\big)}$		\\[5pt]
		&					&	$[\tfrac12]^{\big(2,\,\tfrac12\big)}$+2\,$[\tfrac12]^{\big(1,\,\tfrac12\big)}$+$[\tfrac12]^{\big(1,\,\tfrac32\big)}$+$[\tfrac12]^{\big(0,\,\tfrac32\big)}$+$[\tfrac12]^{\big(0,\,\tfrac12\big)}$	\\[9pt]
$Q^6$	&	$E_0+3$			&	$[1]^{\big(\tfrac32,\,0\big)}$+$[1]^{\big(\tfrac12,\,0\big)}$+ $[1]^{\big(\tfrac12,\,1\big)}$			\\[5pt]
		&					&	$[0]^{\big(\tfrac32,\,1\big)}$+$[0]^{\big(\tfrac12,\,1\big)}$+$[0]^{\big(\tfrac12,\,0\big)}$			\\[9pt]
$Q^7$	&	$E_0+\tfrac72$		&	$[\tfrac12]^{\big(1,\,\tfrac12\big)}$+$[\tfrac12]^{\big(0,\,\tfrac12\big)}$		\\[9pt]
$Q^8$	&	$E_0+4$			&	$[0]^{\big(\tfrac12,\,0\big)}$		\\\hline
\end{tabular}
}

\caption{\footnotesize{States in the long graviton supermultiplet LGRAV$_4[E_0,\tfrac12,0]$.}\normalsize}

\label{table: long120}
\end{table}
%
%
 \begin{table}[H]

\centering

\resizebox{\textwidth}{!}{

\begin{tabular}{ccl}
\hline\\[-7pt]
		&	$E_0$			&	$[0]^{\big(\tfrac12,\,\tfrac12\big)}$			\\[9pt]
$Q$		&	$E_0+\tfrac12$		&	$[\tfrac12]^{(1,\,1)}$+$[\tfrac12]^{(1,\,0)}$+$[\tfrac12]^{(0,\,1)}$+$[\tfrac12]^{(0,\,0)}$		\\[9pt]
$Q^2$	&	$E_0+1$			&	$[1]^{\big(\tfrac32,\,\tfrac12\big)}$	+ $[1]^{\big(\tfrac12,\,\tfrac32\big)}$+ 2\,$[1]^{\big(\tfrac12,\,\tfrac12\big)}$			\\[5pt]
		&					&	$[0]^{\big(\tfrac32,\,\tfrac32\big)}$+$[0]^{\big(\tfrac32,\,\tfrac12\big)}$+ $[0]^{\big(\tfrac12,\,\tfrac32\big)}$+ 2\,$[0]^{\big(\tfrac12,\,\tfrac12\big)}$			\\[9pt]
$Q^3$	&	$E_0+\tfrac32$		&	$[\tfrac32]^{(1,\,1)}$+$[\tfrac32]^{(1,\,0)}$+$[\tfrac32]^{(0,\,1)}$+$[\tfrac32]^{(0,\,0)}$		\\[5pt]
		&					&	$[\tfrac12]^{(2,\,1)}$+$[\tfrac12]^{(1,\,2)}$+$[\tfrac12]^{(2,\,0)}$+$[\tfrac12]^{(0,\,2)}$+	3$[\tfrac12]^{(1,\,1)}$+2$[\tfrac12]^{(1,\,0)}$+2$[\tfrac12]^{(0,\,1)}$+$[\tfrac12]^{(0,\,0)}$	\\[9pt]
$Q^4$	&	$E_0+2$			&	$[2]^{\big(\tfrac12,\,\tfrac12\big)}$		\\[5pt]
		&					&	$[1]^{\big(\tfrac32,\,\tfrac32\big)}$+2\,$[1]^{\big(\tfrac32,\,\tfrac12\big)}$+ 2\,$[1]^{\big(\tfrac12,\,\tfrac32\big)}$+ 3\,$[1]^{\big(\tfrac12,\,\tfrac12\big)}$	\\[5pt]
		&					&	$[0]^{\big(\tfrac52,\,\tfrac12\big)}$+ $[0]^{\big(\tfrac12,\,\tfrac52\big)}$+$[0]^{\big(\tfrac32,\,\tfrac32\big)}$+2\,$[0]^{\big(\tfrac32,\,\tfrac12\big)}$+ 2\,$[0]^{\big(\tfrac12,\,\tfrac32\big)}$+ 2\,$[0]^{\big(\tfrac12,\,\tfrac12\big)}$\\[9pt]
$Q^5$	&	$E_0+\tfrac52$		&	$[\tfrac32]^{(1,\,1)}$+$[\tfrac32]^{(1,\,0)}$+$[\tfrac32]^{(0,\,1)}$+$[\tfrac32]^{(0,\,0)}$		\\[5pt]
		&					&	$[\tfrac12]^{(2,\,1)}$+$[\tfrac12]^{(1,\,2)}$+$[\tfrac12]^{(2,\,0)}$+$[\tfrac12]^{(0,\,2)}$+	3$[\tfrac12]^{(1,\,1)}$+2$[\tfrac12]^{(1,\,0)}$+2$[\tfrac12]^{(0,\,1)}$+$[\tfrac12]^{(0,\,0)}$	\\[9pt]
$Q^6$	&	$E_0+3$			&	$[1]^{\big(\tfrac32,\,\tfrac12\big)}$	+ $[1]^{\big(\tfrac12,\,\tfrac32\big)}$+ 2\,$[1]^{\big(\tfrac12,\,\tfrac12\big)}$			\\[5pt]
		&					&	$[0]^{\big(\tfrac32,\,\tfrac32\big)}$+$[0]^{\big(\tfrac32,\,\tfrac12\big)}$+ $[0]^{\big(\tfrac12,\,\tfrac32\big)}$+ 2\,$[0]^{\big(\tfrac12,\,\tfrac12\big)}$			\\[9pt]
$Q^7$	&	$E_0+\tfrac72$		&	$[\tfrac12]^{(1,\,1)}$+$[\tfrac12]^{(1,\,0)}$+$[\tfrac12]^{(0,\,1)}$+$[\tfrac12]^{(0,\,0)}$		\\[9pt]
$Q^8$	&	$E_0+4$			&	$[0]^{\big(\tfrac12,\,\tfrac12\big)}$		\\\hline
\end{tabular}
}

\caption{\footnotesize{States in the long graviton supermultiplet LGRAV$_4[E_0,\tfrac12,\,\tfrac12]$.}\normalsize}
\label{table: long1212}
\end{table}
%
%
 \begin{table}[H]

\centering

\resizebox{\textwidth}{!}{

\begin{tabular}{ccl}
\hline\\[-7pt]
		&	$E_0$			&	$[0]^{(1,\,0)}$			\\[9pt]
$Q$		&	$E_0+\tfrac12$		&	$[\tfrac12]^{\big(\tfrac32,\,\tfrac12\big)}$ + $[\tfrac12]^{\big(\tfrac12,\,\tfrac12\big)}$		\\[9pt]
$Q^2$	&	$E_0+1$			&	$[1]^{(2,\,0)}$ + $[1]^{(1,\,1)}$ + $[1]^{(1,\,0)}$ + $[1]^{(0,\,0)}$			\\[5pt]
		&					&	$[0]^{(2,\,1)}$ + $[1]^{(1,\,1)}$ + $[1]^{(1,\,0)}$ + $[1]^{(0,\,1)}$			\\[9pt]
$Q^3$	&	$E_0+\tfrac32$		&	$[\tfrac32]^{\big(\tfrac32,\,\tfrac12\big)}$ + $[\tfrac32]^{\big(\tfrac12,\,\tfrac12\big)}$		\\[5pt]
		&					&	$[\tfrac12]^{\big(\tfrac52,\,\tfrac12\big)}$ + $[\tfrac12]^{\big(\tfrac32,\,\tfrac32\big)}$ + 2\,$[\tfrac12]^{\big(\tfrac32,\,\tfrac12\big)}$ + 
								$[\tfrac12]^{\big(\tfrac12,\,\tfrac32\big)}$ + 2\,$[\tfrac12]^{\big(\tfrac12,\,\tfrac12\big)}$		\\[9pt]
$Q^4$	&	$E_0+2$			&	$[2]^{(1,\,0)}$		\\[5pt]
		&					&	$[1]^{(2,\,1)}$ + $[1]^{(2,\,0)}$ + 2\,$[1]^{(1,\,1)}$ + $[1]^{(1,\,0)}$ + $[1]^{(0,\,1)}$ + $[1]^{(0,\,0)}$	\\[5pt]
		&					&	$[0]^{(3,\,0)}$ + $[0]^{(2,\,1)}$ + $[0]^{(2,\,0)}$ + $[0]^{(1,\,2)}$ + $[0]^{(1,\,1)}$ + 2\,$[0]^{(1,\,0)}$ + $[0]^{(0,\,1)}$		\\[9pt]
$Q^5$	&	$E_0+\tfrac52$		&	$[\tfrac32]^{\big(\tfrac32,\,\tfrac12\big)}$ + $[\tfrac32]^{\big(\tfrac12,\,\tfrac12\big)}$		\\[5pt]
		&					&	$[\tfrac12]^{\big(\tfrac52,\,\tfrac12\big)}$ + $[\tfrac12]^{\big(\tfrac32,\,\tfrac32\big)}$ + 2\,$[\tfrac12]^{\big(\tfrac32,\,\tfrac12\big)}$ + 
								$[\tfrac12]^{\big(\tfrac12,\,\tfrac32\big)}$ + 2\,$[\tfrac12]^{\big(\tfrac12,\,\tfrac12\big)}$		\\[9pt]
$Q^6$	&	$E_0+3$			&	$[1]^{(2,\,0)}$ + $[1]^{(1,\,1)}$ + $[1]^{(1,\,0)}$ + $[1]^{(0,\,0)}$			\\[5pt]
		&					&	$[0]^{(2,\,1)}$ + $[1]^{(1,\,1)}$ + $[1]^{(1,\,0)}$ + $[1]^{(0,\,1)}$			\\[9pt]
$Q^7$	&	$E_0+\tfrac72$		&	$[\tfrac12]^{\big(\tfrac32,\,\tfrac12\big)}$ + $[\tfrac12]^{\big(\tfrac12,\,\tfrac12\big)}$		\\[9pt]
$Q^8$	&	$E_0+4$			&	$[0]^{(1,\,0)}$		\\\hline
\end{tabular}
}

\caption{\footnotesize{States in the long graviton supermultiplet LGRAV$_4[E_0,1,0]$.}\normalsize}
\label{table: long10}
\end{table}
%
%
 \begin{table}[H]

\centering

\resizebox{\textwidth}{!}{

\begin{tabular}{ccl}
\hline\\[-7pt]
		&	$E_0$			&	$[0]^{\big(1,\,\tfrac12\big)}$			\\[9pt]
$Q$		&	$E_0+\tfrac12$		&	$[\tfrac12]^{\big(\tfrac32,\,1\big)}$ + $[\tfrac12]^{\big(\tfrac32,\,0\big)}$ + $[\tfrac12]^{\big(\tfrac12,\,1\big)}$ + $[\tfrac12]^{\big(\tfrac12,\,0\big)}$		\\[9pt]
$Q^2$	&	$E_0+1$			&	$[1]^{\big(2,\,\tfrac12\big)}$ + $[1]^{\big(1,\,\tfrac32\big)}$ + 2\,$[1]^{\big(1,\,\tfrac12\big)}$ + $[1]^{\big(0,\,\tfrac12\big)}$		\\[5pt]
		&					&	$[0]^{\big(2,\,\tfrac32\big)}$ + $[0]^{\big(2,\,\tfrac12\big)}$ + $[0]^{\big(1,\,\tfrac32\big)}$ + 2\,$[0]^{\big(1,\,\tfrac12\big)}$ 
								+ $[0]^{\big(0,\,\tfrac32\big)}$ + $[0]^{\big(0,\,\tfrac12\big)}$		\\[9pt]
$Q^3$	&	$E_0+\tfrac32$		&	$[\tfrac32]^{\big(\tfrac32,\,1\big)}$ + $[\tfrac32]^{\big(\tfrac32,\,0\big)}$ + $[\tfrac32]^{\big(\tfrac12,\,1\big)}$ + $[\tfrac32]^{\big(\tfrac12,\,0\big)}$		\\[5pt]
		&					&	$[\tfrac12]^{\big(\tfrac52,\,1\big)}$ + $[\tfrac12]^{\big(\tfrac52,\,0\big)}$ + $[\tfrac12]^{\big(\tfrac32,\,2\big)}$ + 3\,$[\tfrac12]^{\big(\tfrac32,\,1\big)}$ 
								+ 2\,$[\tfrac12]^{\big(\tfrac32,\,0\big)}$ + $[\tfrac12]^{\big(\tfrac12,\,2\big)}$		\\[5pt]
		&					&	+ 3\,$[\tfrac12]^{\big(\tfrac12,\,1\big)}$ + 2\,$[\tfrac12]^{\big(\tfrac12,\,0\big)}$		\\[9pt]
$Q^4$	&	$E_0+2$			&	$[2]^{\big(1,\,\tfrac12\big)}$		\\[5pt]
		&					&	$[1]^{\big(2,\,\tfrac32\big)}$ + 2\,$[1]^{\big(2,\,\tfrac12\big)}$ + 2\,$[1]^{\big(1,\,\tfrac32\big)}$ + 3\,$[1]^{\big(1,\,\tfrac12\big)}$ 
								+ $[1]^{\big(0,\,\tfrac32\big)}$ + 2\,$[1]^{\big(0,\,\tfrac12\big)}$		\\[5pt]
		&					&	$[0]^{\big(3,\,\tfrac32\big)}$ + $[0]^{\big(2,\,\tfrac32\big)}$ + 2\,$[0]^{\big(2,\,\tfrac12\big)}$ + $[0]^{\big(1,\,\tfrac52\big)}$ 
								+ 2\,$[0]^{\big(1,\,\tfrac32\big)}$ + 3\,$[0]^{\big(1,\,\tfrac12\big)}$		\\[5pt]
		&					&	+ $[0]^{\big(0,\,\tfrac32\big)}$ + $[0]^{\big(0,\,\tfrac12\big)}$		\\[9pt]
$Q^5$	&	$E_0+\tfrac52$		&	$[\tfrac12]^{\big(\tfrac52,\,1\big)}$ + $[\tfrac12]^{\big(\tfrac52,\,0\big)}$ + $[\tfrac12]^{\big(\tfrac32,\,2\big)}$ + 3\,$[\tfrac12]^{\big(\tfrac32,\,1\big)}$ 
								+ 2\,$[\tfrac12]^{\big(\tfrac32,\,0\big)}$ + $[\tfrac12]^{\big(\tfrac12,\,2\big)}$		\\[5pt]
		&					&	+ 3\,$[\tfrac12]^{\big(\tfrac12,\,1\big)}$ + 2\,$[\tfrac12]^{\big(\tfrac12,\,0\big)}$		\\[9pt]
$Q^6$	&	$E_0+3$			&	$[1]^{\big(2,\,\tfrac12\big)}$ + $[1]^{\big(1,\,\tfrac32\big)}$ + 2\,$[1]^{\big(1,\,\tfrac12\big)}$ + $[1]^{\big(0,\,\tfrac12\big)}$		\\[5pt]
		&					&	$[0]^{\big(2,\,\tfrac32\big)}$ + $[0]^{\big(2,\,\tfrac12\big)}$ + $[0]^{\big(1,\,\tfrac32\big)}$ + 2\,$[0]^{\big(1,\,\tfrac12\big)}$ 
								+ $[0]^{\big(0,\,\tfrac32\big)}$ + $[0]^{\big(0,\,\tfrac12\big)}$		\\[9pt]
$Q^7$	&	$E_0+\tfrac72$		&	$[\tfrac12]^{\big(\tfrac32,\,1\big)}$ + $[\tfrac12]^{\big(\tfrac32,\,0\big)}$ + $[\tfrac12]^{\big(\tfrac12,\,1\big)}$ + $[\tfrac12]^{\big(\tfrac12,\,0\big)}$		\\[9pt]
$Q^8$	&	$E_0+4$			&	$[0]^{\big(1,\,\tfrac12\big)}$		\\\hline
\end{tabular}
}

\caption{\footnotesize{States in the long graviton supermultiplet LGRAV$_4[E_0,1,\tfrac12]$.}\normalsize}
\label{table: long112}
\end{table}
%
%
 \begin{table}[H]

\centering

\resizebox{\textwidth}{!}{

\begin{tabular}{ccl}
\hline\\[-7pt]
		&	$E_0$			&	$[0]^{(1,\,1)}$		\\[9pt]
$Q$		&	$E_0+\tfrac12$		&	$[\tfrac12]^{\big(\tfrac32,\,\tfrac32\big)}$ + $[\tfrac12]^{\big(\tfrac32,\,\tfrac12\big)}$ + $[\tfrac12]^{\big(\tfrac12,\,\tfrac32\big)}$ 
								+ $[\tfrac12]^{\big(\tfrac12,\,\tfrac12\big)}$		\\[9pt]
$Q^2$	&	$E_0+1$			&	$[1]^{(2,\,1)}$ + $[1]^{(1,\,2)}$ + 2\,$[1]^{(1,\,1)}$ + $[1]^{(1,\,0)}$ + $[1]^{(0,\,1)}$		\\[5pt]
		&					&	$[0]^{(2,\,2)}$ + $[0]^{(2,\,1)}$ + $[0]^{(2,\,0)}$ + $[0]^{(1,\,2)}$ + 2\,$[0]^{(1,\,1)}$ + $[0]^{(1,\,0)}$ \\[5pt]
		&					&	+ $[0]^{(0,\,2)}$+ $[0]^{(0,\,1)}$+ $[0]^{(0,\,0)}$		\\[9pt]
$Q^3$	&	$E_0+\tfrac32$		&	$[\tfrac32]^{\big(\tfrac32,\,\tfrac32\big)}$ + $[\tfrac32]^{\big(\tfrac32,\,\tfrac12\big)}$ + $[\tfrac32]^{\big(\tfrac12,\,\tfrac32\big)}$ 
								+ $[\tfrac32]^{\big(\tfrac12,\,\tfrac12\big)}$		\\[5pt]
		&					&	$[\tfrac12]^{\big(\tfrac52,\,\tfrac32\big)}$ + $[\tfrac12]^{\big(\tfrac52,\,\tfrac12\big)}$ + $[\tfrac12]^{\big(\tfrac32,\,\tfrac52\big)}$ 
								+ 3\,$[\tfrac12]^{\big(\tfrac32,\,\tfrac32\big)}$ + 3\,$[\tfrac12]^{\big(\tfrac32,\,\tfrac12\big)}$ \\[5pt]
		&					&	+ $[\tfrac12]^{\big(\tfrac12,\,\tfrac52\big)}$ + 3\,$[\tfrac12]^{\big(\tfrac32,\,\tfrac12\big)}$ + $[\tfrac12]^{\big(\tfrac12,\,\tfrac12\big)}$		\\[9pt]
$Q^4$	&	$E_0+2$			&	$[2]^{(1,\,1)}$		\\[5pt]
		&					&	$[1]^{(2,\,2)}$ + 2\,$[1]^{(2,\,1)}$ + $[1]^{(2,\,0)}$ + 2\,$[1]^{(1,\,2)}$ + 3\,$[1]^{(1,\,1)}$ + 2\,$[1]^{(1,\,0)}$ 		\\[5pt]
		&					&	+ $[1]^{(0,\,2)}$ + 2\,$[1]^{(0,\,1)}$ + $[1]^{(0,\,0)}$		\\[5pt]
		&					&	$[0]^{(3,\,1)}$ + $[0]^{(2,\,2)}$ + 2\,$[0]^{(2,\,1)}$ + $[0]^{(2,\,0)}$ + $[0]^{(1,\,3)}$ + 2\,$[0]^{(1,\,2)}$ + 4\,$[0]^{(1,\,1)}$		\\[5pt]
		&					&	+ $[0]^{(1,\,0)}$ + $[0]^{(0,\,2)}$ + $[0]^{(0,\,1)}$ + $[0]^{(0,\,0)}$		\\[9pt]
$Q^5$	&	$E_0+\tfrac52$		&	$[\tfrac32]^{\big(\tfrac32,\,\tfrac32\big)}$ + $[\tfrac32]^{\big(\tfrac32,\,\tfrac12\big)}$ + $[\tfrac32]^{\big(\tfrac12,\,\tfrac32\big)}$ 
								+ $[\tfrac32]^{\big(\tfrac12,\,\tfrac12\big)}$		\\[5pt]
		&					&	$[\tfrac12]^{\big(\tfrac52,\,\tfrac32\big)}$ + $[\tfrac12]^{\big(\tfrac52,\,\tfrac12\big)}$ + $[\tfrac12]^{\big(\tfrac32,\,\tfrac52\big)}$ 
								+ 3\,$[\tfrac12]^{\big(\tfrac32,\,\tfrac32\big)}$ + 3\,$[\tfrac12]^{\big(\tfrac32,\,\tfrac12\big)}$ \\[5pt]
		&					&	+ $[\tfrac12]^{\big(\tfrac12,\,\tfrac52\big)}$ + 3\,$[\tfrac12]^{\big(\tfrac32,\,\tfrac12\big)}$ + $[\tfrac12]^{\big(\tfrac12,\,\tfrac12\big)}$		\\[9pt]
$Q^6$	&	$E_0+3$			&	$[1]^{(2,\,1)}$ + $[1]^{(1,\,2)}$ + 2\,$[1]^{(1,\,1)}$ + $[1]^{(1,\,0)}$ + $[1]^{(0,\,1)}$		\\[5pt]
		&					&	$[0]^{(2,\,2)}$ + $[0]^{(2,\,1)}$ + $[0]^{(2,\,0)}$ + $[0]^{(1,\,2)}$ + 2\,$[0]^{(1,\,1)}$ + $[0]^{(1,\,0)}$ \\[5pt]
		&					&	+ $[0]^{(0,\,2)}$+ $[0]^{(0,\,1)}$+ $[0]^{(0,\,0)}$		\\[9pt]
$Q^7$	&	$E_0+\tfrac72$		&	$[\tfrac12]^{\big(\tfrac32,\,\tfrac32\big)}$ + $[\tfrac12]^{\big(\tfrac32,\,\tfrac12\big)}$ + $[\tfrac12]^{\big(\tfrac12,\,\tfrac32\big)}$ 
								+ $[\tfrac12]^{\big(\tfrac12,\,\tfrac12\big)}$		\\[9pt]
$Q^8$	&	$E_0+4$			&	$[0]^{(1,\,1)}$		\\\hline
\end{tabular}
}

\caption{\footnotesize{States in the long graviton supermultiplet LGRAV$_4[E_0,1,1]$.}\normalsize}
\label{table: long11}
\end{table}
%
%
 \begin{table}[H]

\centering

\resizebox{\textwidth}{!}{

\begin{tabular}{ccl}
\hline\\[-7pt]
		&	$E_0$			&	$[0]^{(\ell_1,\,0)}$			\\[9pt]
$Q$		&	$E_0+\tfrac12$		&	$[\tfrac12]^{\big(\ell_1\pm\tfrac12,\,\tfrac12\big)}$		\\[9pt]
$Q^2$	&	$E_0+1$			&	$[1]^{(\ell_1\pm1,\,0)}$ + $[1]^{(\ell_1,\,1)}$ + $[0]^{(\ell_1,\,0)}$		\\[5pt]
		&					&	$[0]^{(\ell_1\pm1,\,1)}$ + $[0]^{(\ell_1,\,1)}$ + $[0]^{(\ell_1,\,0)}$		\\[9pt]
$Q^3$	&	$E_0+\tfrac32$		&	$[\tfrac32]^{\big(\ell_1\pm\tfrac12,\,\tfrac12\big)}$		\\[5pt]
		&					&	$[\tfrac12]^{\big(\ell_1\pm\tfrac32,\,\tfrac12\big)}$ + $[\tfrac12]^{\big(\ell_1\pm\tfrac12,\,\tfrac32\big)}$	 + 2\,$[\tfrac12]^{\big(\ell_1\pm\tfrac12,\,\tfrac12\big)}$		\\[9pt]
$Q^4$	&	$E_0+2$			&	$[2]^{(\ell_1,\,0)}$		\\[5pt]
		&					&	$[1]^{(\ell_1\pm1,\,1)}$ + $[1]^{(\ell_1\pm1,\,0)}$ + 2\,$[1]^{(\ell_1,\,1)}$ + $[1]^{(\ell_1,\,0)}$		\\[5pt]
		&					&	$[0]^{(\ell_1\pm2,\,0)}$ + $[0]^{(\ell_1\pm1,\,1)}$ + $[0]^{(\ell_1\pm1,\,0)}$ + $[0]^{(\ell_1,\,2)}$ + $[0]^{(\ell_1,\,1)}$ + 2\,$[0]^{(\ell_1,\,0)}$		\\[9pt]
$Q^5$	&	$E_0+\tfrac52$		&	$[\tfrac32]^{\big(\ell_1\pm\tfrac12,\,\tfrac12\big)}$		\\[5pt]
		&					&	$[\tfrac12]^{\big(\ell_1\pm\tfrac32,\,\tfrac12\big)}$ + $[\tfrac12]^{\big(\ell_1\pm\tfrac12,\,\tfrac32\big)}$	 + 2\,$[\tfrac12]^{\big(\ell_1\pm\tfrac12,\,\tfrac12\big)}$		\\[9pt]
$Q^6$	&	$E_0+3$			&	$[1]^{(\ell_1\pm1,\,0)}$ + $[1]^{(\ell_1,\,1)}$ + $[0]^{(\ell_1,\,0)}$		\\[5pt]
		&					&	$[0]^{(\ell_1\pm1,\,1)}$ + $[0]^{(\ell_1,\,1)}$ + $[0]^{(\ell_1,\,0)}$		\\[9pt]
$Q^7$	&	$E_0+\tfrac72$		&	$[\tfrac12]^{\big(\ell_1\pm\tfrac12,\,\tfrac12\big)}$		\\[9pt]
$Q^8$	&	$E_0+4$			&	$[0]^{(\ell_1,\,0)}$		\\\hline
\end{tabular}
}

\caption{\footnotesize{States in the long graviton supermultiplet LGRAV$_4[E_0,\ell_1,0]$, with $\ell_1\geq\tfrac32$. For $\ell_1=\tfrac32$, the negative Dynkin label at the $Q^4$ level is absent.}\normalsize}
\label{table: longl0}
\end{table}
%
%
 \begin{table}[H]

\centering

\resizebox{\textwidth}{!}{

\begin{tabular}{ccl}
\hline\\[-7pt]
		&	$E_0$			&	$[0]^{\big(\ell_1,\,\tfrac12\big)}$		\\[9pt]
$Q$		&	$E_0+\tfrac12$		&	$[\tfrac12]^{\big(\ell_1\pm\tfrac12,\,1\big)}$ + $[\tfrac12]^{\big(\ell_1\pm\tfrac12,\,0\big)}$		\\[9pt]
$Q^2$	&	$E_0+1$			&	$[1]^{\big(\ell_1\pm1,\,\tfrac12\big)}$	+ $[1]^{\big(\ell_1,\,\tfrac32\big)}$+ 2\,$[1]^{\big(\ell_1,\,\tfrac12\big)}$		\\[5pt]
		&					&	$[0]^{\big(\ell_1\pm1,\,\tfrac32\big)}$ + $[0]^{\big(\ell_1\pm1,\,\tfrac12\big)}$ + $[0]^{\big(\ell_1,\,\tfrac32\big)}$ + 2\,$[0]^{\big(\ell_1,\,\tfrac12\big)}$		\\[9pt]
$Q^3$	&	$E_0+\tfrac32$		&	$[\tfrac32]^{\big(\ell_1\pm\tfrac12,\,1\big)}$ + $[\tfrac32]^{\big(\ell_1\pm\tfrac12,\,0\big)}$		\\[5pt]
		&					&	$[\tfrac12]^{\big(\ell_1\pm\tfrac32,\,1\big)}$ + $[\tfrac12]^{\big(\ell_1\pm\tfrac32,\,0\big)}$ + $[\tfrac12]^{\big(\ell_1\pm\tfrac12,\,2\big)}$ 
								+ 3\,$[\tfrac12]^{\big(\ell_1\pm\tfrac12,\,1\big)}$ + $[\tfrac12]^{\big(\ell_1\pm\tfrac12,\,0\big)}$		\\[9pt]
$Q^4$	&	$E_0+2$			&	$[2]^{\big(\ell_1,\,\tfrac12\big)}$		\\[5pt]
		&					&	$[1]^{\big(\ell_1\pm1,\,\tfrac32\big)}$ + $[1]^{\big(\ell_1\pm1,\,\tfrac12\big)}$ + 2\,$[1]^{\big(\ell_1,\,\tfrac32\big)}$ 
								+ 3\,$[1]^{\big(\ell_1,\,\tfrac12\big)}$		\\[5pt]
		&					&	$[0]^{\big(\ell_1\pm2,\,\tfrac12\big)}$ + $[0]^{\big(\ell_1\pm1,\,\tfrac32\big)}$ + 2\,$[0]^{\big(\ell_1\pm1,\,\tfrac12\big)}$ 
								+ $[0]^{\big(\ell_1,\,\tfrac52\big)}$ + 2\,$[0]^{\big(\ell_1,\,\tfrac32\big)}$ + 3\,$[0]^{\big(\ell_1,\,\tfrac12\big)}$		\\[9pt]
$Q^5$	&	$E_0+\tfrac52$		&	$[\tfrac32]^{\big(\ell_1\pm\tfrac12,\,1\big)}$ + $[\tfrac32]^{\big(\ell_1\pm\tfrac12,\,0\big)}$		\\[5pt]
		&					&	$[\tfrac12]^{\big(\ell_1\pm\tfrac32,\,1\big)}$ + $[\tfrac12]^{\big(\ell_1\pm\tfrac32,\,0\big)}$ + $[\tfrac12]^{\big(\ell_1\pm\tfrac12,\,2\big)}$ 
								+ 3\,$[\tfrac12]^{\big(\ell_1\pm\tfrac12,\,1\big)}$ + $[\tfrac12]^{\big(\ell_1\pm\tfrac12,\,0\big)}$		\\[9pt]
$Q^6$	&	$E_0+3$			&	$[1]^{\big(\ell_1\pm1,\,\tfrac12\big)}$	+ $[1]^{\big(\ell_1,\,\tfrac32\big)}$+ 2\,$[1]^{\big(\ell_1,\,\tfrac12\big)}$		\\[5pt]
		&					&	$[0]^{\big(\ell_1\pm1,\,\tfrac32\big)}$ + $[0]^{\big(\ell_1\pm1,\,\tfrac12\big)}$ + $[0]^{\big(\ell_1,\,\tfrac32\big)}$ + 2\,$[0]^{\big(\ell_1,\,\tfrac12\big)}$		\\[9pt]
$Q^7$	&	$E_0+\tfrac72$		&	$[\tfrac12]^{\big(\ell_1\pm\tfrac12,\,1\big)}$ + $[\tfrac12]^{\big(\ell_1\pm\tfrac12,\,0\big)}$		\\[9pt]
$Q^8$	&	$E_0+4$			&	$[0]^{\big(\ell_1,\,\tfrac12\big)}$		\\\hline
\end{tabular}
}

\caption{\footnotesize{States in the long graviton supermultiplet LGRAV$_4[E_0,\ell_1,\tfrac12]$, with $\ell_1\geq\tfrac32$. For $\ell_1=\tfrac32$, the negative Dynkin label at the $Q^4$ level is absent.}\normalsize}
\label{table: longl12}
\end{table}
%
%
 \begin{table}[H]

\centering

\resizebox{\textwidth}{!}{

\begin{tabular}{ccl}
\hline\\[-7pt]
		&	$E_0$			&	$[0]^{(\ell_1,\,1)}$		\\[9pt]
$Q$		&	$E_0+\tfrac12$		&	$[\tfrac12]^{\big(\ell_1\pm\tfrac12,\,\tfrac32\big)}$ + $[\tfrac12]^{\big(\ell_1\pm\tfrac12,\,\tfrac12\big)}$		\\[9pt]
$Q^2$	&	$E_0+1$			&	$[1]^{(\ell_1\pm1,\,1)}$ + $[1]^{(\ell_1,\,2)}$ + 2\,$[1]^{(\ell_1,\,1)}$ + $[1]^{(\ell_1,\,0)}$		\\[5pt]
		&					&	$[0]^{(\ell_1\pm1,\,2)}$ + $[0]^{(\ell_1\pm1,\,1)}$ + $[0]^{(\ell_1\pm1,\,0)}$ + $[0]^{(\ell_1,\,2)}$ + 2\,$[0]^{(\ell_1,\,1)}$	 + $[0]^{(\ell_1,\,0)}$	\\[9pt]
$Q^3$	&	$E_0+\tfrac32$		&	$[\tfrac32]^{\big(\ell_1\pm\tfrac12,\,\tfrac32\big)}$ + $[\tfrac32]^{\big(\ell_1\pm\tfrac12,\,\tfrac12\big)}$		\\[5pt]
		&					&	$[\tfrac12]^{\big(\ell_1\pm\tfrac32,\,\tfrac32\big)}$ + $[\tfrac12]^{\big(\ell_1\pm\tfrac32,\,\tfrac12\big)}$ 
								+ $[\tfrac12]^{\big(\ell_1\pm\tfrac12,\,\tfrac52\big)}$ + 3\,$[\tfrac12]^{\big(\ell_1\pm\tfrac12,\,\tfrac32\big)}$ 
								+ 2\,$[\tfrac12]^{\big(\ell_1\pm\tfrac12,\,\tfrac12\big)}$		\\[9pt]
$Q^4$	&	$E_0+2$			&	$[2]^{(\ell_1,\,1)}$		\\[5pt]
		&					&	$[1]^{(\ell_1\pm2,\,1)}$ + 2\,$[1]^{(\ell_1\pm1,\,1)}$ + $[1]^{(\ell_1\pm1,\,0)}$ + 2\,$[1]^{(\ell_1,\,2)}$ + 3\,$[1]^{(\ell_1,\,1)}$ + 2\,$[1]^{(\ell_1,\,0)}$		\\[5pt]
		&					&	$[0]^{(\ell_1\pm2,\,1)}$ + $[0]^{(\ell_1\pm1,\,2)}$ + 2\,$[0]^{(\ell_1\pm1,\,1)}$ + $[0]^{(\ell_1\pm1,\,0)}$ + $[0]^{(\ell_1,\,3)}$ + 2\,$[0]^{(\ell_1,\,2)}$ 		\\[5pt]
		&					&	+  4\,$[0]^{(\ell_1,\,1)}$ + $[0]^{(\ell_1,\,0)}$		\\[9pt]
$Q^5$	&	$E_0+\tfrac52$		&	$[\tfrac32]^{\big(\ell_1\pm\tfrac12,\,\tfrac32\big)}$ + $[\tfrac32]^{\big(\ell_1\pm\tfrac12,\,\tfrac12\big)}$		\\[5pt]
		&					&	$[\tfrac12]^{\big(\ell_1\pm\tfrac32,\,\tfrac32\big)}$ + $[\tfrac12]^{\big(\ell_1\pm\tfrac32,\,\tfrac12\big)}$ 
								+ $[\tfrac12]^{\big(\ell_1\pm\tfrac12,\,\tfrac52\big)}$ + 3\,$[\tfrac12]^{\big(\ell_1\pm\tfrac12,\,\tfrac32\big)}$ 
								+ 2\,$[\tfrac12]^{\big(\ell_1\pm\tfrac12,\,\tfrac12\big)}$		\\[9pt]
$Q^6$	&	$E_0+3$			&	$[1]^{(\ell_1\pm1,\,1)}$ + $[1]^{(\ell_1,\,2)}$ + 2\,$[1]^{(\ell_1,\,1)}$ + $[1]^{(\ell_1,\,0)}$		\\[5pt]
		&					&	$[0]^{(\ell_1\pm1,\,2)}$ + $[0]^{(\ell_1\pm1,\,1)}$ + $[0]^{(\ell_1\pm1,\,0)}$ + $[0]^{(\ell_1,\,2)}$ + 2\,$[0]^{(\ell_1,\,1)}$	 + $[0]^{(\ell_1,\,0)}$	\\[9pt]
$Q^7$	&	$E_0+\tfrac72$		&	$[\tfrac12]^{\big(\ell_1\pm\tfrac12,\,\tfrac32\big)}$ + $[\tfrac12]^{\big(\ell_1\pm\tfrac12,\,\tfrac12\big)}$		\\[9pt]
$Q^8$	&	$E_0+4$			&	$[0]^{(\ell_1,\,1)}$		\\\hline
\end{tabular}
}

\caption{\footnotesize{States in the long graviton supermultiplet LGRAV$_4[E_0,\ell_1,1]$, with $\ell_1\geq\tfrac32$. For $\ell_1=\tfrac32$, the negative Dynkin label at the $Q^4$ level is absent.}\normalsize}
\label{table: longl1}
\end{table}
%
%
 \begin{table}[H]

\centering

\resizebox{\textwidth}{!}{

\begin{tabular}{ccl}
\hline\\[-7pt]
		&	$E_0$			&	$[0]^{(\ell_1,\,\ell_2)}$			\\[9pt]
$Q$		&	$E_0+\tfrac12$		&	$[\tfrac12]^{\big(\ell_1\pm\tfrac12,\,\ell_2\pm\tfrac12\big)}$		\\[9pt]
$Q^2$	&	$E_0+1$			&	$[1]^{(\ell_1\pm1,\,\ell_2)}$	+ $[1]^{(\ell_1,\,\ell_2\pm1)}$+ 2\,$[1]^{(\ell_1,\,\ell_2)}$			\\[5pt]
		&					&	$[0]^{(\ell_1\pm1,\,\ell_2\pm1)}$+$[0]^{(\ell_1\pm1,\,\ell_2)}$+ $[0]^{(\ell_1,\,\ell_2\pm1)}$+ 2\,$[0]^{(\ell_1,\,\ell_2)}$			\\[9pt]
$Q^3$	&	$E_0+\tfrac32$		&	$[\tfrac32]^{\big(\ell_1\pm\tfrac12,\,\ell_2\pm\tfrac12\big)}$		\\[5pt]
		&					&	$[\tfrac12]^{\big(\ell_1\pm\tfrac32,\,\ell_2\pm\tfrac12\big)}$+$[\tfrac12]^{\big(\ell_1\pm\tfrac12,\,\ell_2\pm\tfrac32\big)}$+3\,$[\tfrac12]^{\big(\ell_1\pm\tfrac12,\,\ell_2\pm\tfrac12\big)}$		\\[9pt]
$Q^4$	&	$E_0+2$			&	$[2]^{(\ell_1,\,\ell_2)}$		\\[5pt]
		&					&	$[1]^{(\ell_1\pm1,\,\ell_2\pm1)}$+2\,$[1]^{(\ell_1\pm1,\,\ell_2)}$+ 2\,$[1]^{(\ell_1,\,\ell_2\pm1)}$+ 3\,$[1]^{(\ell_1,\,\ell_2)}$	\\[5pt]
		&					&	$[0]^{(\ell_1\pm2,\,\ell_2)}$+ $[0]^{(\ell_1,\,\ell_2\pm2)}$+$[0]^{(\ell_1\pm1,\,\ell_2\pm1)}$+2\,$[0]^{(\ell_1\pm1,\,\ell_2)}$+ 2\,$[0]^{(\ell_1,\,\ell_2\pm1)}$+ 4\,$[0]^{(\ell_1,\,\ell_2)}$\\[9pt]
$Q^5$	&	$E_0+\tfrac52$		&	$[\tfrac32]^{\big(\ell_1\pm\tfrac12,\,\ell_2\pm\tfrac12\big)}$		\\[5pt]
		&					&	$[\tfrac12]^{\big(\ell_1\pm\tfrac32,\,\ell_2\pm\tfrac12\big)}$+$[\tfrac12]^{\big(\ell_1\pm\tfrac12,\,\ell_2\pm\tfrac32\big)}$+3\,$[\tfrac12]^{\big(\ell_1\pm\tfrac12,\,\ell_2\pm\tfrac12\big)}$		\\[9pt]
$Q^6$	&	$E_0+3$			&	$[1]^{(\ell_1\pm1,\,\ell_2)}$	+ $[1]^{(\ell_1,\,\ell_2\pm1)}$+ 2\,$[1]^{(\ell_1,\,\ell_2)}$			\\[5pt]
		&					&	$[0]^{(\ell_1\pm1,\,\ell_2\pm1)}$+$[0]^{(\ell_1\pm1,\,\ell_2)}$+ $[0]^{(\ell_1,\,\ell_2\pm1)}$+ 2\,$[0]^{(\ell_1,\,\ell_2)}$			\\[9pt]
$Q^7$	&	$E_0+\tfrac72$		&	$[\tfrac12]^{\big(\ell_1\pm\tfrac12,\,\ell_2\pm\tfrac12\big)}$		\\[9pt]
$Q^8$	&	$E_0+4$			&	$[0]^{(\ell_1,\,\ell_2)}$		\\\hline
\end{tabular}
}

\caption{\footnotesize{States in the long graviton supermultiplet LGRAV$_4[E_0,\ell_1,\,\ell_2]$, with $\ell_1,\,\ell_2\geq\tfrac32$. For $\ell_i=\tfrac32$, negative Dynkin labels at the $Q^4$ level are absent.}\normalsize}
\label{table: longhighl1l2}
\end{table}

\newpage

\bibliography{references}

\end{document}